\documentclass[aps,superscriptaddress,nofootinbib]{revtex4}
\usepackage{hyperref}
\usepackage{epsfig,rotating}
\usepackage{amsmath,amssymb}
\usepackage{dsfont}

\numberwithin{equation}{section}

\newcommand{\vx}{\vec{x}}

\newcommand{\vq}{\vec{q}}

\newcommand{\ga}{\gamma}

\newcommand{\uvq}{\widehat{\bf{q}}}

\newcommand{\be}{\begin{equation}}
\newcommand{\ee}{\end{equation}}
\newcommand{\bea}{\begin{eqnarray}}
\newcommand{\eea}{\end{eqnarray}}

\begin{document}

\title{Production  of heavy sterile neutrinos from vector boson decay
at electroweak temperatures}
\author{Louis Lello}
\email{lal81@pitt.edu}
\affiliation{Department of Physics and
Astronomy, University of Pittsburgh, Pittsburgh, PA 15260}
\affiliation{Department of Physics, Brookhaven National Laboratory,
Upton, NY 11973}
\author{Daniel Boyanovsky}
\email{boyan@pitt.edu} \affiliation{Department of Physics and
Astronomy, University of Pittsburgh, Pittsburgh, PA 15260}
\author{Robert D. Pisarski}
\email{pisarski@bnl.gov}
\affiliation{Department of Physics, Brookhaven National Laboratory,
Upton, NY 11973}
\affiliation{RIKEN/BNL Research Center, Brookhaven National Laboratory,
Upton, NY 11973}

\date{\today}

\begin{abstract}

In the standard model extended with a see-saw mass matrix,
we study the production of sterile neutrinos
from the decay of vector bosons at temperatures near the masses of the electroweak bosons.

We derive a general quantum kinetic equation for the
production of sterile neutrinos and their
effective mixing angles, which is applicable over a wide range of temperature,
to all orders in interactions of the standard model,
and to leading order in a small mixing angle for the neutrinos.

We emphasize the relation between the production rate and
Landau damping at one loop order, and show that
production rates and effective mixing angles depend sensitively
upon the neutrino's helicity.
Sterile neutrinos with positive helicity interact more
weakly with the medium than those with negative
helicity, and their effective mixing angle is not modified significantly.
Negative helicity states   couple more strongly to the vector bosons, but
their mixing angle is strongly suppressed by the medium.
Consequently,
if the mass of the sterile neutrino is $\lesssim 8.35\,\mathrm{MeV}$,
there are fewer states with negative helicity produced than those with
positive helicity.
There is an Mikheyev-–Smirnov-–Wolfenstein type resonance in
the absence of lepton asymmetry, but due to screening by the damping rate,
the production rate is not enhanced.
Sterile neutrinos with negative helicity freeze-out at
$T^-_f\simeq 5\,\mathrm{GeV}$ whereas positive helicity neutrinos
freeze-out at $T^+_f \simeq 8\,\mathrm{GeV}$, with
both distributions far from thermal.
As the temperature decreases, due to competition between a decreasing
production rate and an increasing mixing angle,
the distribution function for states
with negative helicity  is broader
in momentum and hotter than that for those with positive helicity.
Sterile neutrinos produced via vector boson decay do not
satisfy the abundance, lifetime and cosmological
constraints to be the sole dark matter component in the universe.
Massive sterile neutrinos produced via vector boson decay might solve the
$^{7}Li$ problem, albeit at the very edge of the possible parameter space.
A heavy sterile neutrino with a mass of a few MeV
could decay into light sterile neutrinos, of a few keV in mass,
that contribute to warm dark matter. We argue that heavy sterile neutrinos with
lifetime  $\leq 1/H_0$ reach local thermodynamic equilibrium.

\end{abstract}


\maketitle

\section{Introduction}\label{sec:intro}
The paradigm of standard cosmology
is inflation plus cold dark matter, $\Lambda CDM$.
While it succeeds in describing the formation of structure at
large scales, there are discrepancies at smaller scales,
especially galactic and sub-galactic.  There is the core-cusp problem:
N-body simulations of cold dark matter produce dark matter
profiles that generically feature cusps, yet observations
suggest that the profile has a smooth core
\cite{corecusp,dalcantonhogan}. There is also the
missing satellites problem, where simulations also predict that
typical galazies are surrounded by satellites dominated by
dark matter, which is inconsistent with current observations
\cite{toobig}.
Both of these problems can be solved by allowing
some fraction of the dark matter to be
warm dark matter, WDM \cite{wdm1,wdm2,wdm3,wdm4,wdm5,wdm6}.
A possible candidate for warm dark matter is a massive ``sterile" neutrino
\cite{warmdm,dodwid,dodwid2,dolgovhansen,abazajian3,sterilesexperiment,kusenkorev,sterilereview}.
Whether dark matter is hot or cold depends upon its free streaming length,
$\lambda_{fs}$, which is the cut-off scale
in the linear power spectrum of density perturbations.
Cold dark matter with
$\lambda_{fs} \lesssim$ pc produces cuspy profiles,
while warm dark matter with
$\lambda_{fs} \sim \mbox{few kpc}$ gives cored profiles.

One important question is whether these disrepancies can be explained
with the standard $\Lambda CDM$ model by including the effects of
baryons in the simulations.
Recent simulations of warm dark matter which include
velocity dispersion suggest that cores do form, but do
not yet reliably constrain the mass of the WDM candidate
in a model independent manner \cite{padu}.

In order to evade cosmological bounds
the distribution functions of warm dark matter cannot be thermal \cite{planck}.
How sterile neutrinos are produced in the early universe was studied
originally in Ref. \cite{dolgovenqvist, dolgovreview}.
In Refs. \cite{dodwid,shifuller,asaka,laine1,laine,laine2,abacir}
it was argued that sterile neutrinos are a viable candidate
for warm dark matter, which is  produced out of local
equilibrium either non-resonantly \cite{dolgovenqvist,dodwid,asaka,laine}
or resonantly in the presence of a lepton asymmetry \cite{shifuller}.
Models in which a scalar decays into a pair of sterile neutrinos
at the electroweak scale (or higher) also yields
a non-thermal distribution
\cite{boyan1,kusenko1,petraki,merle,drewescalar}.
Observations of the X-ray emission spectrum of the Andromeda galaxy
with Chandra led to tight constraints on the non-resonant
production of sterile neutrinos
\cite{casey}.  More recently, the report of observation of a 3.5 keV
signal from the XMM Newton X-ray telescope has been argued to be
due to a 7 keV sterile neutrino\cite{bulbul,boyarsky},
although this interpretation has been challenged
\cite{jeltema,malysh,anderson,sekiya}.
The prospect of a keV sterile dark matter candidate
continues to motivate studies in both theory and observation
\cite{merle,abazajian0,abazajian,abazajian2,kaplinghat2,kaplinghat3,kusenko1,lellolightsterile,added2,abacir,merleini}.

Neutrino masses, mixing and oscillations are
uncontroversial evidence of physics beyond the standard model.
A robust experimental program has brought measurements of most of the
parameters associated with light neutrino masses
\cite{pdg,neutrinoexperiments} with several relevant questions
poised to be answered in the near future \cite{upcomingexperiments}.
Short baseline neutrino oscillation experiments such as LSND and MiniBooNE
\cite{lsnd,miniboone} present a picture of the neutrino sector which
may require an additional sterile neutrino species of mass $\sim 1 eV$
\cite{sterilesexperiment,giunti2,mirizzi}
but there remains tension with other experiments
\cite{nosterilesexperiment} and a definitive resolution of these anomalies
requires further experiments
\cite{lasserre,decay,lello1,shrock,barger,giunti}.
An interpretation of short baseline experimental anomalies
as a signal for sterile neutrinos leads to a relatively light mass
$\sim eV$ which rules out this putative new sterile neutrino as a
candidate for dark matter.  Howevever, many well motivated
extensions beyond the standard model posit the existence of heavy neutrinos.
It has been argued that sterile neutrinos with a mass on the order
of $\mathrm{MeV}$ or larger \cite{gninenko} could decay
and explain the short baseline anomalies.
Alternatively, heavy sterile neutrinos produced through
rare decay channels could also explain the anomaly \cite{heavysterile}.
Recent proposals make the case for a program to  search for
heavy neutrinos \cite{heavyoscil,boyancascade} in a
wide range of experiments including  hadron colliders
\cite{goran,added1,added3,han,pilafsis,bonivento}.
Furthermore, it has been argued that heavy sterile neutrinos
in the mass range $100-500 \,\mathrm{MeV}$
can decay non thermally and so evade bounds from cosmology
and accelerator experiments \cite{fullkuse}.
Sterile neutrinos with mass $\simeq \mathrm{MeV}$
can be of cosmological relevance
in models where the reheating temperature is low \cite{gelmini}.

A heavy sterile neutrino with mass $\simeq 14 \,\mathrm{MeV}$,
a mixing angle $\theta \simeq 10^{-3}$, and a lifetime
$\tau_s \simeq 1.8\times 10^{5}\,s$ has been proposed \cite{ishida}
as a novel solution to the ``Lithium-problem''.
This is the nearly threefold discrepancy between
the standard big-bang nucleosynthesis (BBN) prediction and
observed abundance of $^{7}Li$
\cite{fields,serpico,pospelovbbn,ishida,poulin}.
This solution relies on the energy injected by the decay of the
sterile neutrino to destroy part of $^{7}Be$ prior to its
conversion into $^{7}Li$ in the late stages of BBN
\cite{poulin,ishida}. This mechanism  has been recently re-analyzed
and confirmed in Ref.\cite{salvati} with a sterile neutrino mass
$\simeq 4.35 \,\mathrm{MeV}$, mixing angle $\theta \leq 10^{-5}-10^{-4}$
and   lifetime $\simeq 1.8\,\times 10^{5}\, s$.
An important bonus of this mechanism is that the decay of
the heavy sterile neutrino, $\simeq \mathrm{MeV}$ in mass,
yields an \emph{increase} in the effective number
of relativistic species $\Delta N_{eff} \simeq 0.34$
at the $95\%\,\mathrm{CL}$\cite{salvati}. The energy injection
from the decay of heavy sterile neutrinos with longer lifetime
may also contribute to early ionization \cite{haiman}.
Although there is no experimental evidence for such heavy
sterile neutrinos, there are stringent accelerator and cosmological
bounds on their possible masses and mixing angles with active
 neutrinos \cite{kusbounds,gelmini,oleg,aaron}.

There is a hierarchy of masses for the light active neutrinos,
with nearly two orders of magnitude between the mass squared differences for the explanation of solar and atmospheric neutrino mixing.

Possible extensions beyond the standard model may also accommodate
a hierarchy of \emph{heavy} neutrinos \cite{sterilereview,asaka2,drewes}.

Current and future underground neutrino detectors may be able to
probe dark matter candidates with $\simeq \mathrm{few}\,\mathrm{MeV}$
\cite{pospelov}.  The possibility of a hierarchy of \emph{heavy}
sterile neutrinos offers novel production mechanisms for warm (and hot)
dark matter, from the cascade decays of heavy neutrinos to lighter
ones.  This possibility is similar to models of many dark matter components
proposed recently \cite{dienes}, where the decay of
a heavy field seeds the production of a light one.
This leads to a scenario of \emph{mixed dark matter}
described by several species of massive neutrinos
with non-equilibrium distribution functions, and thereby
evade Lyman-$\alpha$ constraints\cite{lymanboyar}.
A recent article argued on various possible production mechanisms of sterile neutrino directly
 from standard model processes available throughout the thermal history of the universe
 and analyzed in detail the scenario of production
 of mixed dark matter (colder, warmer and hotter)
 from pion decay shortly after the QCD crossover\cite{mixedlou}.

This analysis, along with previous work \cite{lellolightsterile},
also suggests that the decay of a heavy sterile neutrino into a light active neutrino
increases the effective number of neutrinos, $N_{eff}$.
This has been studied recently in Ref.\cite{salvati} in the context of
energy injection post (BBN) from the decay of a heavy sterile neutrino with lifetime $\simeq 10^5\,\mathrm{secs}$.

\vspace{5mm}

\textbf{Motivation and Goals:}

Sterile neutrinos with masses in the range
$\mathrm{KeV}-\mathrm{few}\,\mathrm{MeV}$
may  play an important role in cosmology.

Most of the studies of their production and freeze-out have focused
on the well motivated mass range of $\mathrm{few}\,\mathrm{KeV}$
as possible warm dark matter candidates. However, if the
hierarchy of masses and mixing of light active neutrinos
is of any guide in extensions beyond the standard model,
a possible hierarchy of heavier sterile neutrinos that mix with
the light active neutrinos, with very small mixing angles, may emerge.

In this scenario, the possibility that heavier neutrinos yield
a mixture of dark matter components, from cold, heavy species
to warm, light ones, with important cosmological impact, and the possibility that $\simeq \mathrm{MeV}$ sterile neutrinos \emph{may} yield a solution to the $^{7}Li$ problem\cite{poulin,ishida,salvati}
motivates our study of the production and freeze out
of sterile neutrinos in a wider range of masses and temperatures.
In this article we study the production of sterile neutrinos
solely from \emph{standard model interactions}.  Ref. \cite{mixedlou}
identified several possible processes available
throughout the thermal history of the Universe that \emph{may}
lead to the production of a sterile species
from its mixing with active neutrinos.


Recently, Ref. \cite{laine2} studied the damping rate of $\mathrm{GeV}$ sterile neutrinos at high temperature within the context of the washout rate of leptonic densities.

In contrast, we focus on sterile
neutrinos with masses $ \lesssim \mathrm{few}\,\mathrm{MeV}$,
which is appropriate both to dark matter and the
possible solution of the $^7Li$ problem.
We highlight the important role which different helicity  channels play
for the production rate and mixing angles, including
cosmological expansion.  We also compute the
non-equilibrium distribution functions for
different helicities, along with various cosmological constraints.


In this article our   goals are twofold:
\begin{itemize}
\item{ \textbf{i:)}
Using quantum kinetics, we seek to provide a consistent
description of both production and freeze-out, valid in a wide
range of temperature, under a minimal set of assumptions.
These are:
\textbf{a)} except for the coupling
between sterile and active neutrinos
via a see-saw type mass matrix, we only consider the interactions of the
standard model.
\textbf{b)} consistent with bounds from accelerator experiments
and cosmology,
\cite{kusbounds,gelmini,casey,bulbul,boyarsky,kaplinghat2,oleg,aaron}
we assume that the vacuum mixing angle, $\theta \ll 1$.
Taken together, these bounds suggest that $\theta^2\lesssim 10^{-5}$
for a wide range of masses $M_s \lesssim 300 \,\mathrm{MeV}$.
\textbf{c):}
Interactions in the standard model can be treated perturbatively,
and that the relevant degrees of freedom, including active neutrinos,
are in local thermal equilibrium (LTE) during both production and freeze-out
of sterile species.  The latter is consistent with a small mixing angle.
}

\item{\textbf{ii:)}
We work to leading order in the electro-weak coupling $\alpha_W$,
without any assumption on the mass scales of the sterile neutrinos.
To leading order in $\alpha_W$, the production of
sterile neutrinos occurs from the decay of $W$ and $Z$ bosons in
the thermal bath.

We focus on the temperatures at the electroweak scale,
$T \simeq M_{W}, M_Z$.  This is sufficiently below the temperature
for the electroweak phase transition, which is a
crossover at $T_{ew} \simeq 160 \mathrm{GeV}$ \cite{laine},
so that the $W$ and $Z$ bosons are in local thermal equilibrium,
with masses close to those at zero temperature.
}

\end{itemize}

\vspace{2mm}

\textbf{Brief summary of results:}
For simplicity we consider a model with one active and one sterile
neutrino.  Our main results are:

\vspace{2mm}

\begin{itemize}
\item{We obtain the mass eigenstates,
effective mixing angles and damping rates directly
from the equations of motion in the medium in terms of the full
self-energy \emph{to all orders in weak interactions}.
We give an expression for the effective mixing angles
which is broadly valid for $\theta \ll 1$ and to all orders
in perturbation theory in standard model couplings, at any temperature.
The mixing angle in the medium  depends strongly on helicity:
negative helicity neutrinos (and positive helicity antineutrinos)
feature mixing angles which are strongly suppressed at high temperature.
In contrast, for positive helicity neutrinos
(and negative helicity anti-neutrinos),
the corrections to the mixing angle are subleading, so that the
effective mixing angle is nearly the same as that in vacuum.
This happens because the interaction of neutrinos with positive helicity
is helicity suppressed.
Damping rates are also helicity dependent
and suppressed for those with positive helicity, however because the effective mixing angle is \emph{larger} than that of the negative helicity states, the resulting production rate is comparable in a wide range of masses.
We obtain the general form of the quantum kinetic equation
that describes the production and freeze out of sterile-like neutrinos
The production rate is determined by the damping rate of
sterile-like  \emph{mass eigenstates} and the mixing angle in the medium.
Although the  production rate of positive helicity  states is
suppressed by helicity, it is comparable to the
rate for those with negative helicity, because over a wide regime of masses
it is compensated by the suppression of the mixing angle for neutrinos
with negative helicity.}

    \item{For sterile-like masses $M_s \ll M_W$ we find a
Mikheyev-Smirnov-Wolfenstein (MSW) \cite{msw} resonance in
the \emph{absence of a leptonic asymmetry}.  However,
it is screened by the damping rate and so does \emph{not}
lead to enhanced production.
For this mass range of $M_s$ negative helicity states
freeze out at $T^-_f \simeq 5 \,\mathrm{GeV}$
whereas positive helicity states freeze-out at
$T^+_f \simeq \,8 \mathrm{GeV}$. Both feature highly
non-thermal distribution functions, where   for the negative
helicity states the distribution function is broader and hotter
than that for positive helicity.

Paradoxically, this is a consequence of a
\emph{longer freeze-out time} for the negative helicity states,
despite the fact that their coupling to the environment is stronger.
This is a surprising result, stemming from  a competition
between a diminishing damping rate and an \emph{increasing}
effective mixing angle as the temperature decreases.
We argue that this leading order production mechanism establishes
a \emph{lower bound} for the abundance.
We find however, that sterile-like neutrinos produced via
vector boson decay do not satisfy the various bounds on
lifetimes and mixing angles to be viable $\mathrm{keV}$
dark matter candidates. However they can be suitable as
the $\mathrm{MeV}$ sterile neutrinos that are conjectured
to provide a solution to the $^{7}Li$ problem.  This is
with the caveat that there seems to be tension among the
various bounds available in the literature \cite{salvati,aaron}.
Just as these heavier neutrinos may decay by injecting energy
into the plasma as the solution to this problem, we also conjecture
that they may also decay into lighter $\simeq \,\mathrm{KeV}$
sterile neutrinos, with a much smaller branching ratio,
that could be suitable candidates for warm dark matter.  }

\end{itemize}

To the best of our knowledge there has not
yet been a systematic study
of the \emph{full dynamics of production and
freeze-out obtaining the non equilibrium distribution functions}
of heavy sterile neutrinos with $M_s \lesssim \mathrm{few}\,\mathrm{MeV}$
at the scale $T \simeq 100 \mathrm{GeV}$ with cosmological expansion.
Our analysis is motivated by the possible cosmological
relevance of sterile neutrinos in a wide range of masses,
and complements previous studies that focus on lower temperature regimes.


\section{ Mass eigenstates, damping rates    and mixing angles in the medium:}

We consider the Standard Model with only one leptonic generation: one active neutrino and its charged lepton partner  and one $SU(2)$ singlet Dirac sterile neutrino within a type-I see-saw scenario\cite{minkowski} to discuss the main aspects in the simplest setting.  This choice differs from other versions of the type-I see-saw that include only a right handed Majorana neutrino. The important difference is that in the Majorana case there is no   antineutrino contribution, quantitatively this results in an overall factor $1/2$ in the final abundances.

With one lepton doublet $L_L = ( \nu_{aL}, l_L )^T $  and one $SU(2)$ singlet $\nu_s$ the Lagrangian density is

 \bea\label{Lagunseesaw} \mathcal{L} = \mathcal{L}_{SM}+\overline{\nu}_s   \,i
{\not\!{\partial}} \nu_s - Y \overline{L}_L\,\tilde{\Phi} \, \nu_{sR} +h.c. - M_s \overline{\nu_s}\, \nu_s   \eea where $\tilde{\Phi} = i \tau_2 \Phi^*$ and $\Phi$ is the usual Higgs doublet. We consider temperatures sufficiently below the electroweak crossover at $T_{ew} \simeq 160 \,\mathrm{GeV}$ that the Higgs expectation value is very nearly the zero temperature value. After
spontaneuous symmetry breaking and in unitary gauge the Yukawa coupling yields an off-diagonal Dirac mass term $m = Y \langle \Phi^0 \rangle$  that mixes $\nu_s$ and $\nu_a$,   neglecting the Higgs active-sterile coupling, since we will focus solely on production via vector boson interactions,  the Lagrangian density becomes
 \bea\label{Laguni} \mathcal{L} = \mathcal{L}_{SM}+\overline{\nu}_s   \,i
{\not\!{\partial}} \nu_s - \overline{\nu}_\alpha ~\mathds{M}_{\alpha \beta}~  \nu_{\beta}
  + \mathrm{h.c}~; ~\alpha,\beta = a,s ~,\eea where $a,s$ refer to active and sterile respectively and

\be \label{massmatrix} \mathds{M}=\left(
\begin{array}{cc}
 0 & m   \\
  m   & M_s \\
\end{array}%
\right)\,.  \ee  Introducing the ``flavor'' doublet $  ( {\nu}_a,
 {\nu}_s)$ the diagonalization
 of the mass term $\mathds{M}$ is achieved by a unitary transformation to the mass basis
  $( {\nu}_1,  {\nu}_2)$, namely \be \Big( \begin{array}{c}
                                                          \nu_a \\
                                                          \nu_s
                                                        \end{array}
\Big) = U(\theta) \Big( \begin{array}{c}
                                                          \nu_1 \\
                                                          \nu_2
                                                        \end{array} \Big) ~~;~~
                                                         U(\theta) = \Bigg( \begin{array}{cc}
                                                                              \cos(\theta) & \sin(\theta) \\
                                                                              -\sin(\theta) & \cos(\theta)
                                                                            \end{array}
                                                          \Bigg) ~,\label{unitrafo}\ee
where \be \cos(2\theta) = \frac{M_s}{\left[M^2_s + 4 m^2 \right]^\frac{1}{2}} ~~;~~   \sin(2\theta) = \frac{2 m}{\left[M^2_s + 4 m^2 \right]^\frac{1}{2}}~.
 \label{mixingangles}\ee In the mass basis \be  {\mathds{M}}_{diag}= U^{-1}(\theta)\,\mathds{M} \, U(\theta) = \Bigg( \begin{array}{cc}
                                                M_1 & 0 \\
                                               0 & M_2
                                              \end{array}
 \Bigg)~~;~~M_1 = \frac{1}{2}\left[M_s - \left[M^2_s + 4 m^2 \right]^\frac{1}{2}\right] ~~;~~
  M_2 = \frac{1}{2}\left[M_s + \left[M^2_s + 4 m^2 \right]^\frac{1}{2}\right]~. \label{diagmass}\ee We focus on the case $m\ll M_s$, therefore
  \be M_1 \simeq -\frac{m^2}{M_s} ~~;~~M_2 \simeq M_s ~~;~~
\sin(2\theta)\simeq 2\theta  \simeq \frac{2m}{M_s}   \ll 1 ~.\label{parameters}\ee

We refer to the heavier mass eigenstate with $M_2\simeq M_s$ as \emph{sterile-like} and the lighter mass eigenstate with $M_1 \propto m^2/M_s$ as \emph{active-like}, since these are primarily composed of the sterile and active ``flavors'' in the flavor basis.

We work in unitary gauge which exhibits the physical degrees of freedom  of massive vector bosons in thermodynamic equilibrium. The equations of motion were derived previously in references\cite{boywu,boyho1,boyho2}. In particular, ref.\cite{boywu} also includes contributions from Yukawa couplings between the sterile neutrino and scalar fields, but we will \emph{not} consider such extension here as this implies a particular model for the origin of the mass matrix. Our focus here is to study the sterile neutrino production solely from standard model interactions (charged and neutral currents) under the minimal set of assumptions discussed above.

\begin{figure}[ht!]
\begin{center}
\includegraphics[height=6cm,width=8cm,keepaspectratio=true]{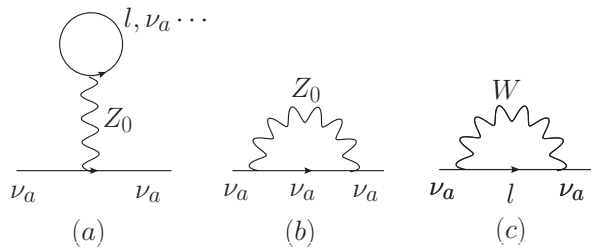}
\caption{One loop contributions to the self-energy. The neutral current tadpole is proportional to the lepton   (and quark) asymmetry.}
\label{fig:self}
\end{center}
\end{figure}

 Introducing the flavor doublet $\Psi^T = ( {\nu}_a,  {\nu}_s)$ the equation of motion \emph{in the flavor basis} is\cite{boywu,boyho1,boyho2}(for details see the appendix in ref.\cite{boyho1}).

\be \left(\,i\not\!{\partial}\,\mathds{1}-
{\mathds{M}}+{\mathbf{\Sigma}}^{t}\,\mathbb{L}\,\right)\,\Psi (\vx,t) + \int
d^3 x'   dt' \;  \,\mathbf{\Sigma}^{r}(\vx-\vx',t-t')
\;\mathbb{L} \, \Psi(\vx',t') = 0 \label{eqnofmot} \ee where $\mathds{1}$ is the identity matrix in flavor space and   $\mathbb{L}=(1-\gamma^5)/2$
is the left-handed chiral projection operator.  The full one-particle irreducible self-energy includes local tadpole ($\Sigma^t$) and non-local dispersive ($\Sigma^r(\vx-\vx',t-t')$)  contributions. It is solely arising from \emph{standard model interactions  to all orders} and is diagonal in the flavor basis, namely
\be \mathbf{\Sigma} \equiv \Sigma \, \Bigg( \begin{array}{cc}
                                                                              1 & 0 \\
                                                                             0 & 0
                                                                            \end{array}
                                                          \Bigg) \,.\label{sigmamtx}\ee Furthermore in factoring
out the projector $\mathbb{L}$, the remaining self-energy is calculated in the vector-like theory.  For example, the one loop contributions to the self-energy are shown in fig.(\ref{fig:self}), this is the leading order contribution to the self energy.

 Introducing the
space-time Fourier transform in a spatial volume $V$ \be
\Psi(\vec{x},t) = \frac{1}{\sqrt{V}} \sum_{\vec{q}}\int d\omega \,
e^{i\vec{q}\cdot\vec{x}}\,e^{-i\omega t} \tilde{\Psi}(\omega, \vec{q})
\label{FT}\ee and similarly for the self-energy kernels, the effective Dirac equation   in the flavor basis becomes\cite{boywu,boyho1,boyho2} (see the appendix in \cite{boyho1}) \be
\Bigg[\big(\gamma_0 \omega - \vec{\gamma}\cdot\vec{q}\big)\mathds{1} -
 {\mathds{M}} +   \Big(\mathbf{\Sigma}^{t}+\mathbf{\Sigma} (\omega,\vec{q})\Big) \,\mathbb{L}  \Bigg]\tilde{\Psi}(\omega,\vec{q}) =0 \,,
\label{eqnofmotFT}\ee this equation of motion is \emph{exact}, since the self-energy includes all order contributions in standard model couplings. The bracket in (\ref{eqnofmotFT}) is the inverse or the retarded propagator whose poles in the complex plane determine the dispersion relations and damping rates of \emph{the mass eigenstates in the medium}.

The tadpole contribution $\mathbf{\Sigma}^{t}$ is local therefore it is independent of $\omega,\vq$, in (\ref{eqnofmotFT}) $\Sigma(\omega,\vq)$, is the space-time Fourier transform of $\Sigma^r(\vx-\vx',t-t')$ and    features a dispersive representation\cite{boywu,boyho1,boyho2}

\be \Sigma(\omega,\vec{q}) = \frac{1}{\pi}\int_{-\infty}^\infty dq_0
\; \frac{\mathrm{Im}\Sigma(q_0,\vec{q}\,)}{q_0-\omega -i\,0^+ } \;
 .\label{selfa}\ee
 From this dispersive form   it follows that
 \be  \Sigma(\omega,\vec{q}) = \mathrm{Re}\Sigma(\omega,\vec{q}) + i \,\mathrm{Im} \Sigma(\omega,\vec{q}) \label{reimsig}\ee with
\be \mathrm{Re}\Sigma(\omega,\vec{q}) = \frac{1}{\pi}\int_{-\infty}^\infty dq_0
\; \mathcal{P}\Bigg[\frac{\mathrm{Im}\Sigma(q_0,\vec{q}\,)}{q_0-\omega } \Bigg]\,.  \label{resigma} \ee The real part yields the ``index of refraction'' in the medium and the imaginary part determines the \emph{damping rate} of the single (quasi) particle excitations.
The tadpole term must be calculated separately and does not feature a dispersive representation.

 Although in this article we will focus on the one-loop contributions to the self-energy from standard model charged and neutral current interactions, the form of the equations of motion and the dispersive form of the self-energy (not the tadpole) is generally valid in principle to
 \emph{all orders} in standard model interactions which are of the $V-A$ form.

 A subtle but important conceptual issue arises in the neutral current contribution to the self-energy with internal loop propagators for neutrinos. The propagators correspond to mass eigenstates, therefore in principle the perturbative loop expansion should be carried out in the \emph{mass basis} rather than in the flavor basis. Furthermore, if the neutrino propagators describe neutrinos thermalized in the medium in terms of the equilibrium  Fermi-Dirac distribution function, not only these propagators correspond to mass (energy) eigenstates but also are \emph{assumed} to be in thermal equilibrium. We will assume the following: i) very small mixing angle $\theta \ll 1$ so that to leading order in this mixing angle, the active-like mass eigenstate can be taken to be the active flavor eigenstate, ii) in the temperature regime of interest in this article $T \simeq M_{W,Z}$ active (flavor) neutrinos are in (local) thermal equilibrium. Under these assumptions (the validity of which will be confirmed later) we consider the internal loop propagators in the neutral current contribution to be those of active neutrinos  in thermal equilibrium to leading order in the mixing angle.

 As a consequence of the $V-A$ nature of the standard model couplings of neutrinos, $\Sigma(\omega,\vq)$ has the general form of a vector-like theory
 \be   {\Sigma}^t+\Sigma(\omega,\vec{q}) \equiv  \gamma^0 \, {A}(\omega,\vec{q})-\vec{\ga}\,\cdot\uvq ~ {B}(\omega,\vec{q})\,, \label{generalsig}\ee and
 \be \mathbf{\Sigma}^t+\mathbf{\Sigma}(\omega,\vec{q}) = \gamma^0 \, \mathbb{{A}}(\omega,\vec{q})-\vec{\ga}\,\cdot\uvq ~ \mathbb{{B}}(\omega,\vec{q}) \label{sigmtx} \ee where in the flavor basis
    \be
\mathds{A}(\omega,\vec{q}) = \Bigg(\begin{array}{cc}
                               A (\omega,\vec{q}) & 0 \\
                               0 & 0
                             \end{array} \Bigg)~~;~~\mathds{B}(\omega,\vec{q})  =   \Bigg(\begin{array}{cc}
                               B(\omega,\vec{q}) & 0 \\
                               0 &  0
                             \end{array} \Bigg)\,, \label{ABmatx}
                             \ee

 The equations of motion simplify by projecting with $\mathbb{L}  = (1-\gamma^5)/2; \mathbb{R} = (1+\gamma^5)/2$ and expanding in helicity eigenstates. Following the steps of ref.\cite{boywu} we write for the left ($L$) and right $(R)$ fields
  \be \Psi_L = \sum_{h=\pm 1} v^{h}
\otimes \Psi^h_L ~~;~~\Psi^h_L = \Bigg( \begin{array}{c}
                                            \nu^h_a \\
                                            \nu^h_s
                                          \end{array}
\Bigg)_L~, \label{varfiL} \ee and \be \Psi_R = \sum_{h=\pm 1} v^{h}
\otimes  \Psi^h_R ~~;~~\Psi^h_R  = \Bigg( \begin{array}{c}
                                            \nu^h_a \\
                                            \nu^h_s
                                          \end{array}
\Bigg)_R~, \label{varfiR} \ee where the left   and right
  handed doublets are written in the \emph{flavor} basis, and
$v^h$ are eigenstates of the  helicity operator \be
\widehat{h}(\uvq) = \gamma^0 \vec{\gamma}\cdot\,\uvq\,\gamma^5 =
\vec{\sigma}\cdot\,\uvq ~\Bigg( \begin{array}{cc}
                                  \mathds{1} & 0 \\
                                  0 & \mathds{1}
                                \end{array}
\Bigg)\label{helicity}\ee namely, \be \vec{\sigma}\cdot\,\uvq ~ v^h
= h ~ v^h~~;~~h=\pm 1 \,.\label{vhel}\ee

We find in the \emph{flavor} basis the equation of motion for the left and
right-handed component doublets  \be   \Big[ ( \omega^2 - q^2)
\mathds{1} + \big(\omega-h q\big)\big(\mathds{A} + h
\mathds{B}\big)  -\mathds{M}^2 \Big]\Psi^h_L    =    0 \label{psiLh} \ee and \be    \Big[ \omega -h\,q\Big] \Psi^h_R   =  \mathds{M} \,\gamma^0\, \Psi^h_L \label{psiRh}
 \,,  \ee  where $\mathds{M}$ is given by (\ref{massmatrix}) and
  \be  {\mathds{M}}^2 = {\overline{M}}^{\,2}
\mathds{1} + \frac{\delta M^2}{2}  \, \Bigg(
                            \begin{array}{cc}
                             -\cos(2\theta) & \sin(2\theta) \\
                              \sin(2\theta) & \cos(2\theta) \\
                            \end{array}
                            \Bigg) \,.\label{M2flav}\ee with \be \overline{M}^{\,2} \equiv
                            \frac{1}{2}\big(M^2_1+M^2_2\big)~~;~~\delta
                            M^2 \equiv  M^2_2-M^2_1  \,,
                            \label{massdefs}\ee and $M_{1,2}$ are
  given by eqn. (\ref{diagmass}). The results (\ref{psiLh},\ref{psiRh}) are \emph{general} for standard model couplings of the active (flavor) neutrinos and sterile neutrinos that only interact with active ones via a see-saw type mass matrix. Before discussing in detail the one-loop contribution from charged and neutral currents, we want to establish   general results for the effective mixing angle in the medium and damping rates. The operator on the left hand side of (\ref{psiLh})
  \be \Big[\mathbb{S}^h_L(\omega,q)\Big]^{-1} = \Big[ ( \omega^2 - q^2)
\mathds{1} + \big(\omega-h q\big)\big(\mathds{A} + h
\mathds{B}\big)  -\mathds{M}^2 \Big]\,, \label{inverseprop} \ee defines the inverse propagator in the flavor basis for the left handed component projected on helicity eigenstates. The correct ``mass eigenstates'' correspond to the (complex) poles of $\mathbb{S}$, the real part describes the correct propagating frequencies and the imaginary parts describe the \emph{damping rate} of single (quasi) particle excitations.
  We will extract these ``mass eigenstates'' from the (complex) zeroes of  $[\mathbb{S}^h_L]^{-1}$ by invoking the following approximations whose validity will be assessed below:
\begin{itemize}
\item{Ultrarelativistic approximation: $q \gg M_s$, this entails that the produced sterile like neutrinos freeze out while relativistic.}
    \item{ $\theta \ll 1$, in particular   we will assume that the self-energy correction is larger in magnitude than the vacuum mixing angle. The precise condition will be discussed below.}
  \item{Validity of the perturbative expansion, in particular that the self-energy corrections are smaller than the unperturbed dispersion relations. This assumption will be clarified and discussed in detail in the analysis that follows.}
\end{itemize}
Introducing
\be \Omega^h \equiv (\omega-h\,q)\,\big(A(\omega,q)+h\,B(\omega,q)\big) \,, \label{omegah}\ee and using (\ref{M2flav}) we obtain
\be \Big[\mathbb{S}^h_L(\omega,q)\Big]^{-1} =   \Bigg( \omega^2 - q^2-\overline{M}^{\,2}+\frac{\Omega^h}{2}\Bigg)
\mathds{1} - \frac{1}{2} \, \sqrt{\Big(\delta M^2 \cos(2\theta)+\Omega^h\Big)^2+\Big(\delta M^2 \sin(2\theta) \Big)^2}~~\Bigg(
                            \begin{array}{cc}
                             -\mathbb{C}^h_m(\omega,q) & \mathbb{S}^h_m(\omega,q) \\
                              \mathbb{S}^h_m(\omega,q) & \mathbb{C}^h_m(\omega,q) \\
                            \end{array}
                            \Bigg)  \label{inverseS}\ee
where
\bea \mathbb{C}^h_m(\omega,q) & = &  \frac{\delta M^2\cos(2\theta)+\Omega^h(\omega,q)}{ \sqrt{\Big(\Omega^h(\omega,q)+\delta M^2 \cos(2\theta)\Big)^2 + \Big( \delta M^2\, \sin(2\theta)\Big)^2  } } \label{Cm} \\ \mathbb{S}^h_m(\omega,q) & = &  \frac{\delta M^2 \sin(2\theta)}{ \sqrt{\Big(\Omega^h(\omega,q)+\delta M^2 \cos(2\theta)\Big)^2 + \Big( \delta M^2\, \sin(2\theta)\Big)^2  }} \,, \label{Sm} \eea with
\be (\mathbb{C}^h_m(\omega,q))^2 + (\mathbb{S}^h_m(\omega,q))^2 = 1 \,. \label{identity} \ee
\emph{If} the imaginary part of the self energy vanishes (or can be neglected) then
\be \mathbb{C}^h_m(\omega,q) = \cos(2\theta_m)~~;~~ \mathbb{S}^h_m(\omega,q)= \sin(2\theta_m) \,, \label{mixangmed}\ee and  $\theta_m$ would be the mixing angle in the medium. However, the absorptive (imaginary) part of the self-energy (related to the damping rate of quasi particle excitations) prevent such identification.

The matrix
\[ \Bigg(
                            \begin{array}{cc}
                             -\mathbb{C}^h_m(\omega,q) & \mathbb{S}^h_m(\omega,q) \\
                              \mathbb{S}^h_m(\omega,q) & \mathbb{C}^h_m(\omega,q) \\
                            \end{array}
                            \Bigg) \]
has null trace and determinant $(-1)$ as a consequence of (\ref{identity}), therefore  \emph{real} eigenvalues $\lambda  = \pm 1$ with the   following eigenvectors:
\bea  && \Bigg(
                            \begin{array}{c}
                             c^h(\omega,q) \\
                              -s^h(\omega,q)
                            \end{array}
                            \Bigg)    ~~;~~ \lambda = -1 \label{min1} \\
&&  \Bigg(
                            \begin{array}{c}
                             s^h(\omega,q) \\
                              c^h(\omega,q)
                            \end{array}
                            \Bigg)    ~~;~~ \lambda =  1 \label{plus1} \eea
where
\bea c^h(\omega,q) & = & \Bigg[\frac{1+\mathbb{C}^h_m(\omega,q)}{2}\Bigg]^{1/2} \label{cosi} \\
s^h(\omega,q) & = & \Bigg[\frac{1-\mathbb{C}^h_m(\omega,q)}{2}\Bigg]^{1/2} \,.\label{sisi}\eea For vanishing absorptive part $s^h  \equiv \sin(\theta_m), c^h \equiv \cos(\theta_m)$ with $\theta_m$ the (real) mixing angle in the medium.

To leading order for $\theta \ll 1$ and $M_1 \ll M_2 \simeq M_s$ we obtain the following eigenvalues of $\Big[\mathbb{S}\Big]^{-1}$
\be S^{-1} \simeq \omega^2 -q^2 -M^2_2 - \frac{\theta^2 \,(M^2_2)^2}{\Big(M^2_2+\Omega^h(\omega,q) \Big)}+\theta^2 M^2_2~~;~~\mathrm{for}~ \lambda = +1~~;~~\mathrm{eigenvector}~~\Bigg(
                            \begin{array}{c}
                             s^h(\omega,q) \\
                              c^h(\omega,q)
                            \end{array}
                            \Bigg)  \label{invspl1}\ee

\be S^{-1} \simeq \omega^2 -q^2 -M^2_1 + \Omega^h(\omega,q)  + \frac{\theta^2 \,(M^2_2)^2}{\Big(M^2_2+\Omega^h(\omega,q) \Big)}-\theta^2 M^2_2~~;~~\mathrm{for}~ \lambda = -1~~;~~\mathrm{eigenvector}~~\Bigg(
                            \begin{array}{c}
                             c^h(\omega,q) \\
                              -s^h(\omega,q)
                            \end{array}
                            \Bigg)\,.  \label{invspl2}\ee
It is clear that the eigenvector corresponding to  eigenvalue $+1$ corresponds to a sterile-like neutrino in the medium: the radiative correction (self-energy) enters solely with the mixing angle and vanishes for vanishing mixing angle, whereas the eigenvector corresponding to eigenvalue $-1$ is active-like, with radiative correction ($\Omega^h$) even for $\theta=0$. The inverse of $ \Big[\mathbb{S}^h_L(\omega,q)\Big]^{-1}$ is the inverse of the propagator, therefore its complex zeroes   describe the complex poles. Under the validity of perturbation theory (discussed below in detail) we write
\be \omega =\omega_j(q) + \delta \omega^h_j ~~;~~ \omega_j(q) =  \sqrt{q^2+M^2_j}\simeq q+ \frac{M^2_j}{2q} ~~;~~ j=1,2 \,\label{freqs} \ee in the relativistic approximation and introducing  (in the relativistic limit)
\be \Delta^h_j(q) + i \gamma^h_j(q) = \frac{\Omega^{h}(\omega_j,q)}{2q} \simeq  \frac{(\omega_j-h\,q)}{2q}\,\Bigg[A(\omega=q,q)+h\,B(\omega=q,q)\Bigg] \,, \label{delplugam}\ee with $j=2$  for sterile-like (eigenvalue $\lambda=1$) and $j=1$ for active-like (eigenvalue $\lambda =-1$)  where both $\Delta$ and $\gamma$ are real, and introducing
\be \xi = \frac{M^2_s}{2q} \label{chi}\ee we find the position of the poles in the propagator (``mass eigenstates'') at
\be \delta \omega^h_2 = \frac{\theta^2~\Big(\xi+\Delta^h_2(q)-i\gamma^h_2(q)\Big)}{\Bigg[\Big( 1+\frac{\Delta^h_2(q)}{\xi}\Big)^2+\Big(\frac{\gamma^h_2(q)}{\xi} \Big)^2\Bigg]}-\theta^2 \xi~~;~~\mathrm{for}~ \lambda = +1~~;~~\mathrm{eigenvector}~~\Bigg(
                            \begin{array}{c}
                             s^h(\omega=q,q) \\
                              c^h(\omega=q,q)
                            \end{array}
                            \Bigg) \, \label{deloste} \ee for the ``sterile-like'' neutrino and

\be \delta \omega^h_1 = -\Big(\Delta^h_1(q)+i\gamma^h_1(q)\Big)-\frac{\theta^2~\Big(\xi+\Delta^h_1(q)-i\gamma^h_1(q)\Big)}{\Bigg[\Big( 1+\frac{\Delta^h_1(q)}{\xi}\Big)^2+\Big(\frac{\gamma^h_1(q)}{\xi} \Big)^2\Bigg]}+\theta^2 \xi~~;~~\mathrm{for}~ \lambda = -1~~;~~\mathrm{eigenvector}~~\Bigg(
                            \begin{array}{c}
                             c^h(\omega=q,q) \\
                              -s^h(\omega=q,q)
                            \end{array}
                            \Bigg) \,  \label{deloact} \ee for the ``active-like'' neutrino.

 We now introduce the \emph{effective mixing angle in the medium   }
 \be \theta^h_{eff}(q) = \frac{\theta}{\Bigg[\Big( 1+\frac{\Delta^h_j(q)}{\xi}\Big)^2+\Big(\frac{\gamma^h_j(q)}{\xi} \Big)^2\Bigg]^{1/2}}\,, \label{tetaeff}\ee   for each mass eigenstate $j=1,2$, in terms of which the position of the (quasi) particle poles  (\ref{deloste},\ref{deloact}) are written as
 \be \delta \omega^h_2 =  \Delta^h_2(q)~\Big(\theta^h_{eff}(q)\Big)^2+\xi \Big[ \Big(\theta^h_{eff}(q)\Big)^2-\theta^2\Big]  - i~\gamma^h_2(q)\Big(\theta^h_{eff}(q)\Big)^2 ~~;~~ j=2~~ (\mathrm{sterile-like}) \,, \label{delsteri}  \ee
\be \delta \omega^h_1 = -\Delta^h_1(q)~\Big[1+\Big(\theta^h_{eff}(q)\Big)^2\Big]-\xi \Big[ \Big(\theta^h_{eff}(q)\Big)^2-\theta^2\Big] - i~\gamma^h_1(q)\Big[1-\Big(\theta^h_{eff}(q)\Big)^2 \Big]~~;~~ j=1~~(\mathrm{active-like}) \,. \label{delactivo}  \ee

Writing
\be \delta \omega^h_j = \Delta E^h_j -i \frac{\Gamma^h_j}{2} \,,\label{breit} \ee for the corresponding helicity component, the imaginary part  $\Gamma$ yields the damping rate for the single (quasi) particle excitations in the medium, namely the ``mass eigenstates'' in the medium evolve in time as
\be \nu^h_j(q) \simeq e^{-i\omega_j t}\,e^{-i\Delta E^h_j t} \,e^{-\Gamma^h_j  t/2} \Rightarrow |\nu^h_j(q)|^2 \simeq  e^{-\Gamma^h_j  t}~~;~~ j=1,2 \,,\label{amps}\ee where the damping rates $\Gamma^h_{j}$ are given by
\bea \Gamma^h_2 & = & 2 \gamma^h(q)\, \Big(\theta^h_{eff}(q)\Big)^2 \simeq 2\gamma^h(q)\,\sin^2(\theta^h_{eff}) ~~\mathrm{sterile-like} \label{Gamaste}\\
\Gamma^h_1 & = &  2 \gamma^h(q)\, \Big[1-\Big(\theta^h_{eff}(q)\Big)^2\Big] \simeq 2\gamma^h(q)\,\cos^2(\theta^h_{eff}) ~~\mathrm{active-like} \,. \label{Gamaac} \eea

Even when a particle cannot decay in the vacuum, the spectral density may feature a width in the medium as a consequence of dissipative processes arising from the coupling to excitations in the medium. In this case the width describes the relaxation of the quasi-particle in linear response\cite{boydavey,boyeff,weldcut,lebellac}.

The coefficient $\omega -h q$ in (\ref{omegah}) is noteworthy: for positive $\omega$ the positive helicity component $h=1$ is \emph{helicity}  suppressed, on the mass shell of the (vacuum) mass eigenstates in the relativistic limit $\omega-q \simeq M^2_{1,2}/2q$. This is the usual helicity suppression from the $V-A$ form of the interaction and has important consequences: the damping rate for \emph{positive helicity} sterile-like neutrinos is much smaller than that for the negative helicity, and the medium corrections to the mixing angle are also \emph{suppressed} for the positive helicity component. This suppression will have important and unexpected  consequences on the rate of production of the sterile-like species as will be discussed below in detail.

 The functions $\gamma^\pm(q)$ require  the combinations $A \mp B$ that define the self energy (\ref{generalsig}), these combinations are handily extracted as follows: introduce the four-vectors
\be Q^-_\mu = \frac{1}{q}(q, -  \vec{q}) ~~;~~ Q^+_\mu = \frac{1}{q}(q,    \vec{q}) \label{fourvecs}\ee and  defining
\be   \Sigma^-(q) \equiv  A(\omega=q,q)-B(\omega=q,q)~~;~~ \Sigma^+(q) \equiv  A(\omega=q,q)+B(\omega=q,q)\, \label{sigplumin}\ee
 we obtain
\bea \Big(A(\omega,q)-B(\omega,q)\Big)_{\omega=q} & = & \frac{1}{4} \,\mathrm{Tr} {\not\!{Q^-}}\,\Sigma(q,q) \equiv \Sigma^-(q)\,,  \label{AminB}\\
\Big(A(\omega,q)+B(\omega,q)\Big)_{\omega=q} & = & \frac{1}{4} \,\mathrm{Tr} {\not\!{Q^+}}\,\Sigma(q,q) \equiv \Sigma^+(q) \,,\label{AplusB}
\eea
\be \gamma^-(q) =   \,\mathrm{Im}\, \Sigma^-(q)~~;~~(\mathrm{negative}\,\mathrm{helicity}) \label{gamneghel}\ee
\be \gamma^+(q) =  \Bigg[\frac{M_s}{2q}\Bigg]^2 \,  \mathrm{Im}\, \Sigma^+(q)~~;~~(\mathrm{positive}\,\mathrm{helicity}) \label{gamposhel}\ee
\be \Delta^-(q) =   \,\mathrm{Re}\, \,\Sigma^-(q)~~;~~(\mathrm{negative}\,\mathrm{helicity}) \label{delneghel}\ee
\be \Delta^+(q) =  \Bigg[\frac{M_s}{2q}\Bigg]^2 \,  \mathrm{Re}\, \Sigma^+(q)~~;~~(\mathrm{positive}\,\mathrm{helicity}) \label{delposhel}\ee

 The damping rates for negative ($-$) and positive ($+$)  sterile- like neutrinos respectively are given by
\be \Gamma^\mp_2(q) = 2~ \Big(\theta^\mp_{eff}(q)\Big)^2 ~ \gamma^\mp(q) \label{gamasterfin} \ee as discussed in the next section these rates determine the production rates of sterile neutrinos of each helicity.
   For the active-like neutrinos we find
\bea \Gamma^-_1(q) & = & 2~ \Bigg[1-\Big(\theta^-_{eff}(q)\Big)^2 \Bigg]~ ~\mathrm{Im}  \,\Sigma^-(q)    \label{gamaactnegfin}  \\ \Gamma^+_1(q) & = & 2~\Bigg[1-\Big(\theta^-_{eff}(q)\Big)^2 \Bigg]~\Bigg[\frac{M_1}{2q}\Bigg]^2 \, \mathrm{Im}\, \Sigma^+(q)     \, , \label{gamaactposfin}\eea   for negative ($-$) and positive $(+)$ helicity respectively, the latter rate can be safely neglected for the light active-like neutrinos. The effective mixing angles are given by
\be \theta^\pm_{eff}(q) = \frac{\theta}{\Bigg[\Big( 1+\frac{\Delta^\pm(q)}{\xi}\Big)^2+\Big(\frac{\gamma^\pm(q)}{\xi} \Big)^2\Bigg]^{1/2}}\,, \label{tetaeffpm}\ee where in the relativistic limit

\bea && \frac{\Delta^-(q)}{\xi} = \frac{2q}{M^2_s}\,\mathrm{Re}\Sigma^-(q) ~~;~~ \frac{\Delta^+(q)}{\xi} = \frac{\mathrm{Re}\Sigma^+(q)}{2q} \label{delplumi}\\ && \frac{\gamma^-(q)}{\xi} = \frac{2q}{M^2_s}\,\mathrm{Im}\Sigma^-(q) ~~;~~ \frac{\gamma^+(q)}{\xi} = \frac{\mathrm{Im}\Sigma^+(q)}{2q} \,, \label{gamplumi}  \eea where $\Sigma^\pm$ are given by (\ref{AminB},\ref{AplusB}).

The important observation is that the self-energy $\Sigma(\omega,q)$ is calculated in \emph{the standard model for massless flavor neutrinos}.

The result for the effective mixing angle (\ref{tetaeffpm}) is valid for $\theta \ll 1$ and (\ref{delplumi},\ref{gamplumi}) are valid in the relativistic limit $q \gg M_s$ but otherwise \emph{general} and \emph{valid to all orders in perturbation theory in standard model couplings}.

A Mikheyev-Smirnov-Wolfenstein\cite{msw} (MSW) resonance is available whenever
\be \frac{\Delta^\pm(q)}{\xi} = -1  \,, \label{mswres}\ee for $\cos(\theta) \simeq 1$. However, this resonance is \emph{screened} by the imaginary part (damping rate) (this phenomenon was also noticed in ref.\cite{laine2}), and under the condition that $\theta \ll 1$ and the validity of the approximations leading to the above results, the possible presence of this resonance will not yield a large enhancement in the effective mixing angle. This aspect will be discussed in detail in section (\ref{sec:wdecay}).

These expressions are one of the main results of this study and summarize the effective mixing angles and damping rates \emph{generically} for standard model interactions under the assumptions of the validity of the relativistic approximation, perturbative expansion and $\theta \ll 1$.

From the expressions (\ref{Cm},\ref{Sm})) and (\ref{cosi},\ref{sisi})  one finds that for $\theta \ll 1$
\be s^h(\omega=q,q) \simeq  \theta^h_{eff}(q)\,e^{-i\phi^h(q)}~~;~~\phi^h(q) = \tan^{-1}\Big[ \frac{\gamma^h(q)}{\xi + \Delta^h(q)} \Big] \,, \label{sincho}\ee where the phase is irrelevant for transition probabilities and the quantum kinetic description of sterile neutrino production (transition probabilities per unit time see below).

\section{Quantum kinetics: production rates}
In order to understand how to extract the rate of sterile-like neutrino production from the damping rate obtained from the self energy and the effective mixing angle in the medium, let us  consider first the quantum kinetics of production  of the sterile-like mass eigenstate from W-decay in the case of vacuum mixing angle. This analysis clearly shows how the mixing angle in the medium enters in the quantum kinetic equation with a straightforward generalization to more general production processes.

If the mass $M_2$ of the heavy, sterile-like neutrino is such that $M_{W} > M_2 + m_l$ with $m_l$ the mass of the charged lepton $l$, then the mass eigenstate corresponding to the sterile-like neutrino can be produced from W-decay, a similar argument applies to $Z$-decay  if $M_Z >M_2+M_1$.
The charged current interaction vertex for the case of one generation is
\be \mathcal{L}_{cc} = \frac{g_w}{\sqrt{2}} \Big[ \overline{l} \gamma^\mu \mathbb{L}~ \nu_l ~W_\mu+ h.c. \Big]\,. \label{chargedcurr}\ee Writing the flavor eigenstate $\nu_l$ in the mass basis as
\be \nu_l = \cos(\theta) \nu_1 + \sin(\theta) \nu_2 \label{flavormass} \ee with $\nu_1$ being the active-like and $\nu_2$ the sterile-like mass eigenstates with $M_1,M_2$ respectively, yielding an interaction vertex for the sterile-like mass eigenstate $\nu_2$
\be \mathcal{L}_{s,cc} = \frac{g_w}{\sqrt{2}}\,\sin(\theta) \Big[ \overline{l} \gamma^\mu \mathbb{L}~ \nu_2 ~W_\mu+ h.c. \Big]\,. \label{chargedcurrste}\ee The dynamics of the production of $\nu_2$ from the process $W  \rightarrow \overline{l} ~\nu_2$ is obtained via the quantum kinetic equation for the process $W  \rightleftharpoons \overline{l}  \nu_2$\cite{mixedlou}, namely \be \frac{d n_2(q;t)}{dt} = \frac{d n_2(q;t)}{dt}\big|_{\textrm{gain}}-\frac{d n_2(q;t)}{dt}\big|_{\textrm{loss}}\,,\label{quakin}\ee where $n_2(q;t)$ is the distribution function of the sterile-like mass eigenstate $\nu_2$ and the gain and loss terms are extracted from the usual transition probabilities per unit time,
\bea \frac{d n_2(q;t)}{dt}\big|_{\textrm{gain}} & = &  \frac{2\pi~\sin^2(\theta)}{2E_2(q)} \int \frac{d^3k\, |\mathcal{M}_{fi}|^2} {(2\pi)^3 2E_W(p)2E_l(k)} \,N_B(p)(1-n_l(k))(1-n_2(q;t)) \,\delta(E_W(p)-E_l(k)-E_2(q))  \nonumber  \\
\frac{d n_2(q;t)}{dt}\big|_{\textrm{loss}} & = &  \frac{2\pi ~\sin^2(\theta)}{2E_2(q)} \int \frac{d^3k\, {|\mathcal{M}_{fi}|^2}}{(2\pi)^3 2E_W(p)2E_l(k)} \,(1+N_B(p))n_l(k)n_2(q;t) \,\delta(E_W(p)-E_l(k)-E_2(q)) \,,\label{gainminlossW}\eea where $p= |\vec{k}+\vec{q}|$ and
\be N_B(p) = \frac{1}{e^{E_W(p)/T}-1} ~~;~~ n_l(k) = \frac{1}{e^{E_l(k)/T}+1} \,,\label{disLTE}\ee and
$ |\mathcal{M}_{fi}|^2$ is the usual transition matrix element for $W \rightarrow \overline{l} \nu $ and we have assumed that the $W$ vector boson and charged lepton $l$ are in thermal equilibrium and displayed explicity the factor $\sin^2(\theta)$ factored out of the $M_{fi}$.
Therefore the quantum kinetic equation (\ref{quakin}) becomes of the form
\be \frac{d n_2(q;t)}{dt} = \Gamma^<(q)(1-n_2(q;t)) - \Gamma^>(q) n_2(q;t) \label{gainminlos} \ee where the gain and loss \emph{rates} are
\be \Gamma^<(q) =  \frac{2\pi \,\sin^2(\theta)}{2E_2(q)} \int \frac{d^3k\, {|\mathcal{M}_{fi}|^2}}{(2\pi)^3 2E_W(p)2E_l(k)} \,N_B(p)(1-n_l(k))\,\delta(E_W(p)-E_l(k)-E_2(q)) \label{gainrate} \ee
\be \Gamma^>(q) =  \frac{2\pi \,\sin^2(\theta)}{2E_2(q)} \int \frac{d^3k\, {|\mathcal{M}_{fi}|^2}}{(2\pi)^3 2E_W(p)2E_l(k)} \,(1+N_B(p)) n_l(k)\,\delta(E_W(p)-E_l(k)-E_2(q)) \,.\label{lossrate} \ee Because the $W,l_\alpha$ are in thermal equilibrium the gain and loss rates obey the \emph{detailed balance condition}
\be \Gamma^<(q)\,e^{E_2(q)/T} = \Gamma^>(q)\,, \label{detailedbalance} \ee which can be confirmed straightforwardly from the explicit expressions (\ref{gainrate},\ref{lossrate})  using the energy conserving delta functions and the relations
\be 1+N_B(E) = e^{E/T} N_B(E)~~;~~ 1-n_l(E) = e^{E/T} n_l(E)\,. \label{relas}\ee Using (\ref{detailedbalance}) the quantum kinetic equation (\ref{gainminlos}) reads
\be \frac{d n_2(q;t)}{dt} = \Gamma_2(q) \Big[n_{eq}(q) -n_2(q;t)\Big]\,, \label{qukin} \ee  where
\be  n_{eq}(q) = \frac{1}{e^{\frac{E_2(q)}{T}}+1} \label{neq}\ee is the \emph{equilibrium} (Fermi-Dirac) distribution function and
\be \Gamma_2(q)  = \Gamma^>(q)+\Gamma^<(q) = \frac{2\pi \,\sin^2(\theta)}{2E_2(q)} \int \frac{d^3k\, {|\mathcal{M}_{fi}|^2}}{(2\pi)^3 2E_W(p)2E_l(k)} \,\Big[N_B(p)+n_l(k) \Big]\,\delta(E_W(p)-E_l(k)-E_2(q)) \,. \label{gamakineq} \ee The approach to equilibrium is studied by writing $n_2(q;t) = n_{eq}(q) + \delta n_2(q;t)  $ , it follows from (\ref{gainminlos}) that
\be \delta n_2(q;t)= \delta n_2(q;0)\,e^{-\Gamma_2(q)\,t} \,.\label{asyeq}\ee The relaxation rate $\Gamma_2(q)$ is precisely the damping rate of single (quasi) particle excitations  (\ref{amps}) as discussed in refs.\cite{boydavey,boyeff,weldcut,lebellac}. Neutral current interactions are treated similarly by passing to the mass basis and keeping only the linear term in $\sin(\theta) \simeq \theta$ for $\theta \ll 1$.
It is clear from (\ref{gamakineq}) that
\be \Gamma_2(q) = \sin^2(\theta) \,\Gamma_{sm}(q) \label{gamsm}\ee where $\Gamma_{sm}(q)$ is the rate calculated in the standard model for the production of a massive neutrino, furthermore,  it  is given by the \emph{imaginary part of the standard model flavor neutrino self energy} evaluated on the massive neutrino mass shell. In the limit of a relativistic sterile-like mass eigenstate, $\Gamma_{sm}(q)$ is \emph{identical} to the imaginary part of the self-energy for an active massless neutrino in the standard model. In fact in this limit the quantum kinetic equation for the active-like mass eigenstate in the relativistic limit is the same as (\ref{qukin}) but with $\sin^2(\theta)$ in (\ref{gamsm}) replaced by $\cos^2(\theta)$.

Fundamentally the heart of the argument is simply detailed balance, a consequence of the main assumption that the plasma degrees of freedom are in thermodynamic equilibrium: the damping rate of single quasiparticle excitations $\Gamma_2(q)$  determines the approach to equilibrium in linear response\cite{boydavey,boyeff,weldcut,lebellac} and for $\theta \ll 1$ the quantum kinetic equation is linear in the population $n_2$ to leading order in $\theta$,  therefore the gain term in the quantum kinetic equation is simply related to the relaxation rate by detailed balance. This argument is general for $\theta \ll 1$. Therefore, comparing with the damping rate for the sterile-like mass eigenstate (\ref{Gamaste}), this analysis makes clear that for $\theta \ll 1$ the medium effects on the mixing angle in the quantum kinetic equation are incorporated by the simple replacement $\sin(\theta) \rightarrow \theta_{eff}(q)$ in (\ref{gamakineq},\ref{gamsm}), in other words the \emph{full quantum kinetic equation for sterile-like production} is
\be \frac{d n^h_2(q;t)}{dt} = \Gamma^h_2(q)  \Big[n_{eq}(q) -n^h_2(q;t)\Big]\,, \label{qukinfin} \ee where $\Gamma^\mp_2(q)$ are given by (\ref{gamasterfin}) with (\ref{gamneghel},\ref{gamneghel}). Hence, the \emph{production} rate of sterile-like neutrinos is \be \Gamma^h_{prod}(q) = \Gamma^h_2(q)\,n_{eq}(q)\,. \label{gamaprodu}\ee

In summary, the production rates for  sterile-like neutrinos of  negative ($-$) and positive
 ($+$ )helicities are given by
\bea \Gamma^-_{prod}(q)  & = &  2 \Big(\theta^-_{eff}(q)\Big)^2\,\mathrm{Im}\Sigma^-(q) \,n_{eq}(q) \label{neghelprodu}\\\Gamma^+_{prod}(q)  & = &  2 \Big(\theta^+_{eff}(q)\Big)^2\,\Bigg[ \frac{M_s}{2q} \Bigg]^2\,\mathrm{Im}\Sigma^+(q)\,n_{eq}(q) \label{poshelprodu} \eea
where the mixing angles     $\theta^\mp_{eff}(q)$ is given by (\ref{tetaeff}) with the definitions (\ref{gamneghel}-\ref{delposhel}). In the production rates (\ref{neghelprodu},\ref{poshelprodu}) $\Sigma(q)$ is the \emph{standard model self-energy for flavor neutrinos} evaluated on the relativistic mass shell, and   $n^h_2$   refer to the population of the \emph{sterile-like mass eigenstate} of helicity $h$. Because $\theta_{eff}$ depends on helicity the matrix elements $M_{fi}$ \emph{should not} be averaged over helicity (spin) states.

From the expression (\ref{gamakineq}) we can also glean how the helicity suppression is manifest in the case of massive neutrinos. For this it is convenient to look at the positive frequency solutions of the massive Dirac equation in the chiral representation ($\gamma^5 = diag(\mathds{1},-\mathds{1}$)) and in the helicity basis:
\bea \mathcal{U}_+(\vq)  & = &  N \Bigg( \begin{array}{c}
                       v_+(\vq) \\
                        -\varepsilon(q)\, v_+(\vq)
                      \end{array}
 \Bigg) ~~;~~ \mathcal{U}_-(\vq) = N \Bigg( \begin{array}{c}
                       -\varepsilon(q)\,v_-(\vq) \\
                        v_-(\vq)
                      \end{array}
 \Bigg) \label{Uspi}\eea where $v_{\pm}(\vq)$ are helicity eigenvectors (Weyl spinors) for $h=\pm 1$, and
 \be N =\sqrt{E_s(q)+q} ~~;~~\varepsilon(q) = \frac{M_s}{E(q)+q}\,, \label{Neps}\ee then
 \be \mathbb{L}~\mathcal{U}_+(\vq)= \varepsilon(q) ~N \Bigg( \begin{array}{c}
                       0 \\
                         v_+(\vq)
                      \end{array}
 \Bigg)\label{projUplus}\ee in the relativistic limit $q\gg M_s$ , $\varepsilon \simeq M_s/2q$, this projected wave function enters in the
 matrix element $M_{fi}$ for a massive positive helicity neutrino in the final state, therefore
 \be |M_{fi}|^2 \propto \Big( \frac{M_s}{2q}\Big)^2 \label{poshelsupmfi}\ee in agreement with the helicity suppression for the damping rate discussed in the previous section.

\vspace{2mm}

\textbf{Generality:} Although in the above discussion we focused on the production process $W\rightarrow \overline{l}\nu_2$, the result (\ref{qukinfin}) is   general for $\theta \ll 1$. Consider the standard model charged and neutral current vertices, writing these in the basis of mass eigenstates the charged current vertex is linear in the mass eigenstate $\nu_2$ therefore the vertex is $\propto \theta$. The neutral current vertex would feature a term linear in $\theta$ ($ \propto \overline{\nu}_1 \,\nu_2$) and  another $\propto \theta^2$ ($\propto \overline{\nu}_2 \,\nu_2$), for $\theta \ll 1$ this last term can be neglected and both charged and neutral current vertices are \emph{linear} in $\theta$ and $\nu_2$. Furthermore $\theta \ll 1$ justifies taking the ``active-like'' mass eigenstate to be in local thermal equilibrium (LTE) in the medium for $T \geq 0.1\,\mathrm{MeV}$ as its relaxation rate is much larger than that of the ``sterile-like'' eigenstate which is suppressed by $\propto \theta^2 \ll 1$. Because the interaction vertices are linear in the neutrino field to leading order in $\theta$,  the quantum kinetic equation (gain - loss) is obviously of the form (\ref{gainminlos}) and because the degrees of freedom that lead to the gain and loss terms are all in (LTE) the gain ($\Gamma^<$) and loss ($\Gamma^>$) rates must obey the detailed balance condition (\ref{detailedbalance}). This analysis leads  directly to the quantum kinetic equation (\ref{qukin}) after replacing $\theta \rightarrow \theta_{eff}(q)$ where $q$ is the momentum of the sterile-like neutrino on its mass shell, the $|M_{fi}|^2$ matrix element for the gain and loss transition rates are insensitive to the phase in (\ref{sincho}). Analyzing the approach to equilibrium leads to the identification of $\Gamma_2$ with the damping rate of the sterile-like mass eigenstate. This argument is general and the analysis presented above for $W \rightarrow \overline{l} \nu_2$ provides a direct example, which will be the focus of a detailed analysis in the next section.

\section{Sterile production from vector boson decay.}\label{sec:wdecay}

We now focus on the description of sterile(like) neutrino production via vector boson decay
$ W  \rightarrow \overline{l}\,\nu_2 ~~;~~ Z^0 \rightarrow \overline{\nu}_1 \nu_2 $ at temperature  $T\simeq M_{w,z}$, this is the leading order production process at this temperature. This temperature scale is sufficiently lower than the electroweak crossover scale $T \simeq 160\,\mathrm{GeV}$ that the Higgs field is near its vacuum expectation value and the finite temperature corrections to the $W,Z$ masses can be safely neglected\cite{laine}.
At high temperature, the propagator of charged leptons receives substantial hard thermal loop corrections from electromagnetic interactions (and quarks from both photons and gluons) for momenta $\leq e T$\cite{htl,pisarskihtl,frenkel,weldon,lebellac}. However, the decay of a vector boson \emph{at rest} in the plasma yields particles with momenta $\simeq M_{W,Z}/2$, therefore  the typical momenta of the charged lepton is $\mathcal{O}(T)$ and in this regime the hard thermal loop corrections are subleading and will be neglected in the following analysis. By the same reason   a  sterile neutrino of mass   $M_2 \ll M_{W,Z}$   will be produced relativistically. Low momentum sterile neutrinos (and charged leptons) \emph{can} be produced for highly boosted vector bosons in the medium, but those excitations will be Boltzmann suppressed for $T\simeq M_{W,Z}$, this will be explicitly confirmed in the analysis below.

  We will take the charged lepton and the active-like neutrino to be massless and refer generically to the vector boson mass as $M$, adapting the general result to the $W,Z$ a posteriori.  Under this approximation (justified for $ M_2 \ll T\simeq M_{W,Z}$) the one-loop self energy is the same for both charged and neutral current interactions, in the latter case the loop in fig. (\ref{fig:self}-(b)) includes the active-like neutrino (self-consistently) assumed to be in LTE.

The one-loop tadpole contribution from neutral currents (fig. (\ref{fig:self})) is given by\cite{notzold,boywu}
\be \Sigma^{t} = -\gamma^0  \, {\pi}  \,\frac{\alpha_w}{M^2_W}\,\sum_{f}C^{(f)}_v\,\int  \frac{d^3k}{(2\pi)^3}\,\big[n_f(k)-\overline{n}_f(k)\big]  = -\gamma^0  \,\frac{\pi}{6} \,\alpha_w \,\Big(\frac{T}{M_w}\Big)^2\,\sum_{f}C^{(f)}_v\,\mu^f\,\Bigg[ 1 +\frac{\mu^2_f}{\pi^2\,T^2}\Bigg] \,,\label{sigtad}\ee where $f$ are all the ultrarelativistic fermionic species in thermal equilibrium at temperature $T$ and chemical equilibria with chemical potentials $\mu^f$ respectively.  The tadpole $\Sigma^t$ is independent of  frequency and momentum and contributes only to $A$ in (\ref{generalsig},\ref{ABmatx}). Although we quote this result as part of the general formulation, we will  neglect the lepton and quark asymmetries in the following analysis setting $\mu^f=0$ for all fermionic species, thereby neglecting the contribution $\Sigma^t$ to the self-energy.

We  obtain the imaginary part of the self-energy (for both helicities) in (\ref{reimsig})  from which we will obtain the real part from the dispersion relation (\ref{resigma}).

For both charged and neutral current contributions (fig. (\ref{fig:self},(b),(c)) for relativistic leptons,  the imaginary
part of the self energy is given by\cite{boywu}

\be Im\Sigma(q_0,\vec{q}) =   \pi  { g^2 } ~\int
\frac{d^3k}{(2\pi)^3}  \,\int^{\infty}_{-\infty} dk_0
\Big[1-n_{f}(k_0)+N_b(p_0)\Big] \gamma^\mu
\rho_f(k_0,\vec{k})\rho_b(p_0,\vec{p})\gamma^\nu\,P_{\mu
\nu}(p_0,\vec{p})~~;~~p^\mu = q^\mu-k^\mu \,, \label{imsigsm}\ee where $f$ stands for the
fermionic species,  either a charged lepton $l$ for the charged current or the active neutrino $\nu_a$ (assumed in thermal equilibrium) for the neutral current contributions and $b$ for either vector boson in the intermediate state. The couplings and masses for the charged and neutral current contributions are
   \be g  = \Bigg\{ \begin{array}{c}
                                          \displaystyle{\frac{g_w}{\sqrt{2}}} ~~~~~~~~\mathrm{CC} \\
                                         \displaystyle{\frac{g_w}{ 2\cos(\theta_w)}}
                                         ~~\mathrm{NC}
                                        \end{array} ~~~~;~~~~ M  = \Bigg\{ \begin{array}{l}
                                          \displaystyle{ {M_W} } ~~~~~~~~~~~~~~~\mathrm{CC} \\
                                         \displaystyle{ {M_Z} = \frac{M_W}{ \cos(\theta_w)}}
                                         ~~\mathrm{NC}
                                        \end{array}~~;~~ \sin^2(\theta_w) \simeq 0.23 ~~;~~ \alpha_w = \frac{g^2_w}{4\pi} \simeq \frac{1}{32} \label{gsm}\ee

                                         The
spectral densities are respectively (for massless fermions)
 \be \rho_f(k_0,\vec{k}) = \frac{{\not\!{k}}}{2k}\,\Big[\delta(k_0-k)-\delta(k_0+k) \Big]~~;~~ {\not\!{k}}= \gamma^0k^0-\vec{\gamma}\cdot\vec{k} \,,\label{ferspecdens} \ee
  \be \rho_b(p_0,\vec{p})   =
\frac{1}{2W_p}\Big[\delta(p^0-W_p)-\delta(p^0+W_p)\Big] ~~;~~W_p =
\sqrt{p^2+M^2}~~;~~p^\mu = q^\mu - k^\mu \,,  \label{bosspecdens}\ee
 and the projection operator \be
P_{\mu \nu}(p_0,\vec{p}) = -\Bigg[g_{\mu \nu}- \frac{p_\mu
p_\nu}{M^2} \Bigg]~~;~~p^\mu = (p^0, \vec{p})~;~M^2 \equiv M^2_{z,w}
\label{projector} \ee and \be n_{f}(k_0)    =
\frac{1}{e^{k_0/T}+1} ~~;~~     N_b(p_0)    =  \frac{1}{e^{p_0/T}-1}~.
\label{DFs} \ee As per the discussion in the previous sections (see eqns.(\ref{omegah}, \ref{delplugam})), we need the combinations $A(q_0,q) \pm B(q_0,q)$ which are obtained from the traces (\ref{AminB},\ref{AplusB}). We find
\be  \mathrm{Im}\Big[A(q_0,q)\mp B(q_0,q)\Big]  =   \pi  { g^2 } ~\int^{\infty}_{-\infty}dk_0~\int
\frac{d^3k}{(2\pi)^3~4k\,W_p}L^{\mu \nu}\Big[Q^\pm;k\Big]P_{\mu \nu}\big[p\big]~\overline{\rho}_f(k_0,k)\overline{\rho}_b(p_0,p)
\Big[1-n_{f}(k_0)+N_b(p_0)\Big] \,, \label{imapmb}\ee  where
\bea \overline{\rho}_f (k_0,k) & = &  \Big[\delta(k_0-k)-\delta(k_0+k) \Big] \label{rhotf} \\
\overline{\rho}_b (p_0,p)   & = & \Big[\delta(p_0-W_p)-\delta(p_0+W_p)\Big]~~;~~p^\mu = q^\mu-k^\mu \label{rhotb}\eea
and
\be L^{\mu \nu}\Big[Q;k\Big] = \Big[Q^\mu k^\nu + Q^\nu k^\mu -g^{\mu \nu} Q\cdot k \Big]\,. \label{lepten}\ee

Using the various delta functions from $\overline{\rho}_f;\overline{\rho}_b$ we find for the negative helicity component
\be L^{\mu\nu}\Big[Q^+,k\Big] P_{\mu \nu}\big[p\big] = -\frac{M^2}{q}\Big[ F_1(q_0,q)+ k^0\,F_2(q_0,q) \Big]\,, \label{LdotPplus}\ee with
\bea  F_1(q_0,q) & = & \Bigg[1- \Bigg(\frac{(q^0)^2 -q^2}{M^2} \Bigg) \Bigg]\,\Bigg[1-   \frac{\Big(q^0-q\Big)^2}{2M^2}   \Bigg] \label{F1} \\
F_2(q_0,q) & = & 2 \frac{(q^0-q)}{M^2}\Bigg[1- \Bigg(\frac{(q^0)^2 -q^2}{2M^2} \Bigg) \Bigg] \,.   \label{F2} \eea Similarly, for the positive helicity component
\be L^{\mu\nu}\Big[Q^-,k\Big] P_{\mu \nu}\big[p\big] =  \frac{M^2}{q}\Big[ G_1(q_0,q)+ k^0\,G_2(q_0,q) \Big]\,, \label{LdotPmin}\ee with
\bea  G_1(q_0,q) & = & \Bigg[1- \Bigg(\frac{(q^0)^2 -q^2}{M^2} \Bigg) \Bigg]\,\Bigg[1-   \frac{\Big(q^0+q\Big)^2}{2M^2}   \Bigg] \label{G1} \\
G_2(q_0,q) & = & 2 \frac{(q^0+q)}{M^2}\Bigg[1- \Bigg(\frac{(q^0)^2 -q^2}{2M^2} \Bigg) \Bigg] \,.   \label{G2} \eea

Note the relation
\be G_{1}(q^0,q) =  F_{1}(q^0,-q)~~;~~ G_{2}(q^0,q)= F_{2}(q^0,-q)\,. \label{relaFG}\ee

Using the results above, it is straightforward to show that
\be \mathrm{Im}\Big[A(q_0,q)+B(q_0,q)\Big]=\mathrm{Im}\Big[A(-q_0,q)- B(-q_0,q)\Big]\,.\label{idims}\ee This identity relates the imaginary parts for positive energy, negative helicity neutrinos to negative energy positive helicity (anti-neutrinos) (in absence of a chemical potential). This identity guarantees that the production rate for negative (positive) helicity neutrinos is the same as for positive (negative) helicity anti-neutrinos and is a consequence of the vanishing chemical potentials under the assumption of vanishing lepton  and baryon  asymmetry and, consequently vanishing   neutral current tadpole contribution.

 It is convenient to change integration variables, with
\be W \equiv W_p = \sqrt{q^2+k^2+M^2-2qk \cos(\varphi)} \Rightarrow \frac{dW}{d \cos(\varphi)} = -\frac{q k}{W_p} \label{wp}\ee therefore
 \be \frac{d^3k}{W_p} = (2\pi) k^2 dk \frac{d(\cos(\varphi)}{W_p} = -(2\pi) \frac{k dk \, dW}{q}  \,, \ee
yielding
 \be  \mathrm{Im}\Big[A(q_0,q)\mp B(q_0,q)\Big]  =      \frac{ g^2 }{16\pi\,q}  ~\int^{\infty}_{-\infty}dk_0~\int^\infty_0
dk \, \int^{W+}_{W_-}dW~L\Big[Q^\pm;k\Big]\cdot P\big[p\big]~\overline{\rho}_f(k_0,k)\overline{\rho}_b(p_0,p)
\Big[1-n_{f}(k_0)+N_b(p_0)\Big]  \,, \label{imapmbfin}\ee where $p^0 = q^0-k^0$. Now in terms of the integration variables $k^0,k, W$
\be \overline{\rho}_b = \Big[\delta(q^0-k^0-W)-\delta(q^0-k^0+W)\Big] \label{rhobb} \ee and  the integration limits in $W$ are
\be   W^\pm = \sqrt{(q \pm k)^2+M^2}\,. \label{limsW}\ee
The technical details of the calculation of the spectral densities is relegated to appendix (\ref{app:specdens}), we neglect the zero temperature contribution, assuming that it has been systematically absorbed by renormalization,  focusing solely on the finite temperature terms.

\subsection{Imaginary parts (damping rates)}\label{sub:imparts}
We can now obtain the imaginary parts evaluated on the relativistic mass shells $q_0 \simeq q$ (for positive energy neutrinos). The analysis of the support for the delta functions in the appendix shows that on the relativistic mass shell $q_0 = q$ the only contribution to the imaginary parts arises from the product (\ref{four}) with $k^+ = \infty, k^- = M^2/4q$ )
\[ -\delta(k^0+k) \delta(q-k^0-W_p) ~~;~~\vec{p}=\vec{q}+\vec{k} \] corresponding to the process $W\rightarrow \overline{l} \nu_2$ (the anti  lepton $\overline{l}$ is recognized in the delta function $\delta(k^0+k)$ which determines that the energy is $-k$).  This is the contribution labelled $\mathbf{4})$ (\ref{four}) in the appendix. The analysis of the regions of support in $q_0$ given by (\ref{fouri}-\ref{fouriii}) shows a remarkable phenomenon: the contributions (\ref{fouri},\ref{fouriii}) describe the process of Landau damping\cite{robLD}, namely the process of emission and reabsorption of thermal excitations in the medium that yields a branch cut below the light cone in the spectral density that vanishes at $T=0$. These processes are the equivalent of those described in ref.\cite{robLD} for quarks interacting with gluons or electrons with photons. In these cases, namely the exchange of massless vector bosons, the Landau damping cut ends at $q_0=q$, however, for \emph{massive} vector boson exchange the Landau damping cut below the light cone \emph{merges} with the branch cut above the light cone given by (\ref{fourii}). The mass shell for an ultrarelativistic sterile-like neutrino falls right at the end of the Landau damping cut and the beginning of the continuation above the light cone. It is precisely this point in the spectral density that contributes to the damping rate of the ultrarelativistic sterile-like neutrino and vanishes for $T=0$. Therefore, the end point of the Landau damping cut yields the damping rate of the ultrarelativistic sterile-like neutrino which, in turn, yields their \emph{production rate} as a consequence of detailed balance as discussed above.

 This  contribution to (\ref{imapmb})  yields
\be  \mathrm{Im}\Big[A(q,q)\mp B(q,q)\Big]  =   -\pi  { g^2 } ~ \int
\frac{d^3k}{(2\pi)^3~4k\,W_p}L^{\mu \nu}\Big[Q^\pm;k\Big]P_{\mu \nu}\big[p\big]~
\Big[ n_{f}(k)+N_b(W_p)\Big]\,\delta(W_p-q-k) \,, \label{imapmbshell}\ee which   is \emph{precisely} the expression for the rate $\Gamma_2$ in the quantum kinetic equation (\ref{gamakineq}) with ultrarelativistic neutrinos and charged leptons\footnote{The lepton tensor $L_{\mu \nu}$ is in terms of $Q^\pm$ that is divided by the energy of the relativistic neutrino (see the definitions (\ref{fourvecs})).  }(up to the prefactor $\sin^2(\theta)$). The helicity suppression factor arises similarly to the discussion after  (\ref{gamsm}).

\begin{figure}[h!]
\begin{center}
\includegraphics[height=4in,width=3.2in,keepaspectratio=true]{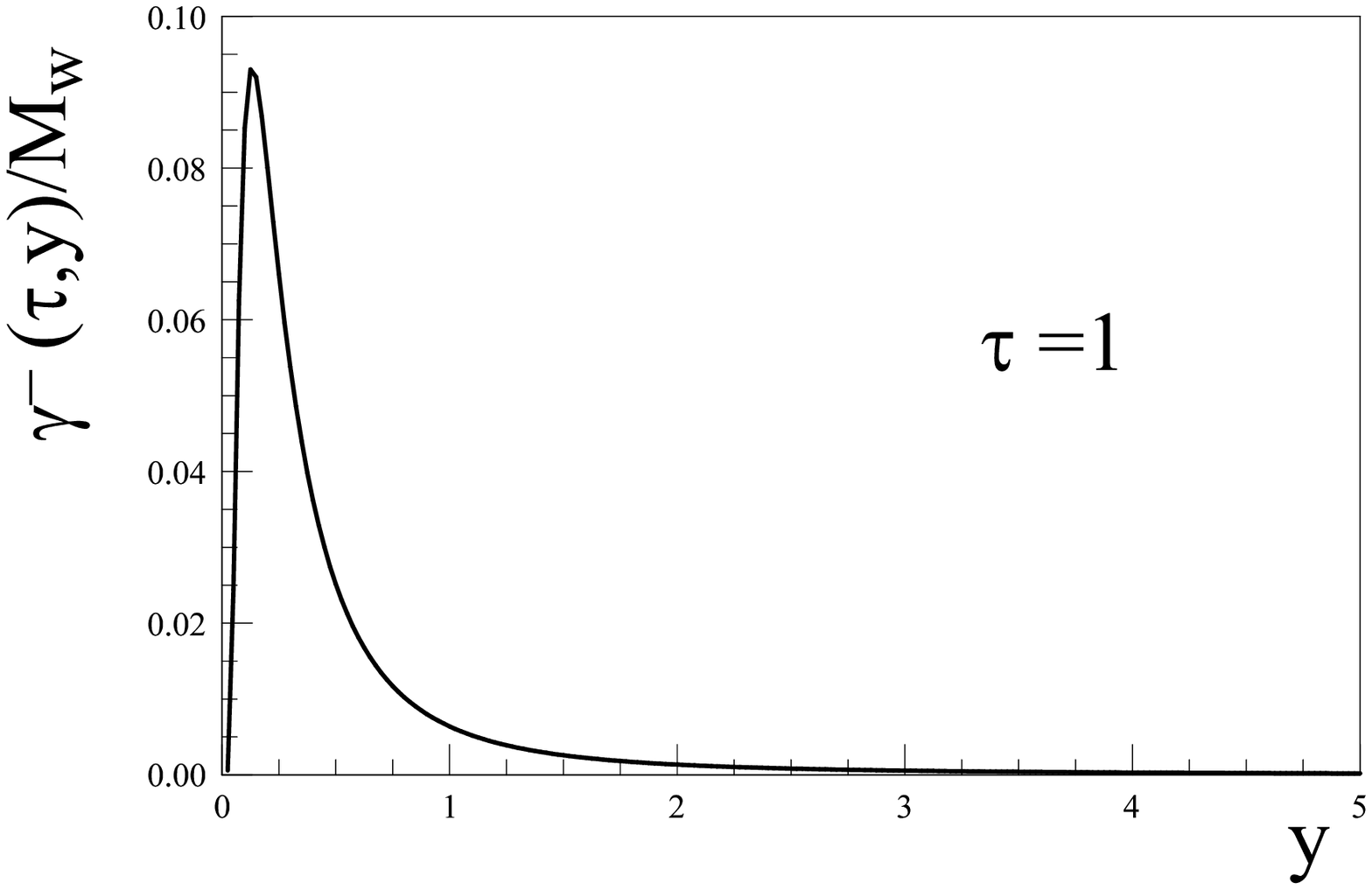}
\includegraphics[height=4in,width=3.2in,keepaspectratio=true]{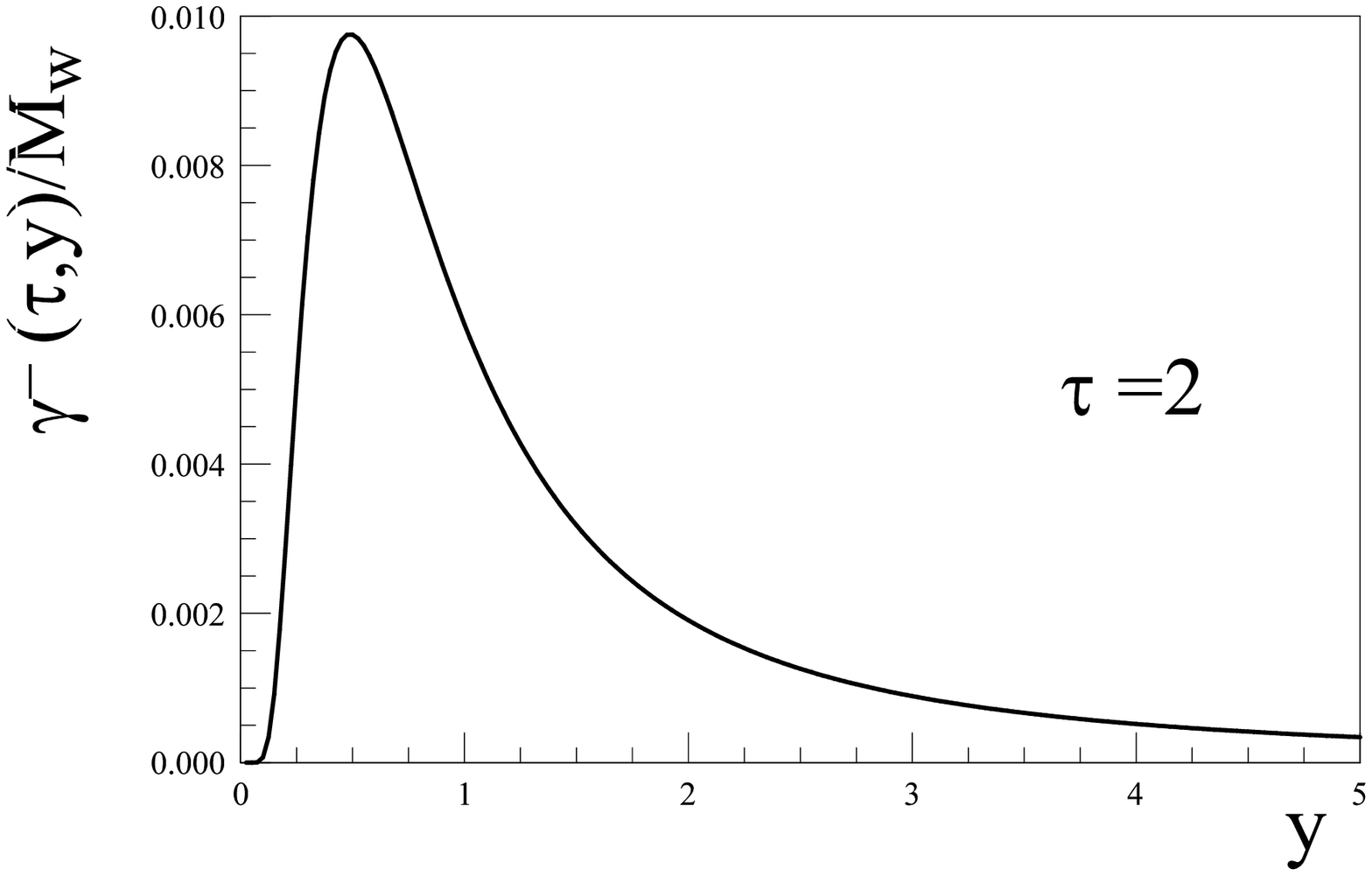}
\includegraphics[height=4in,width=3.2in,keepaspectratio=true]{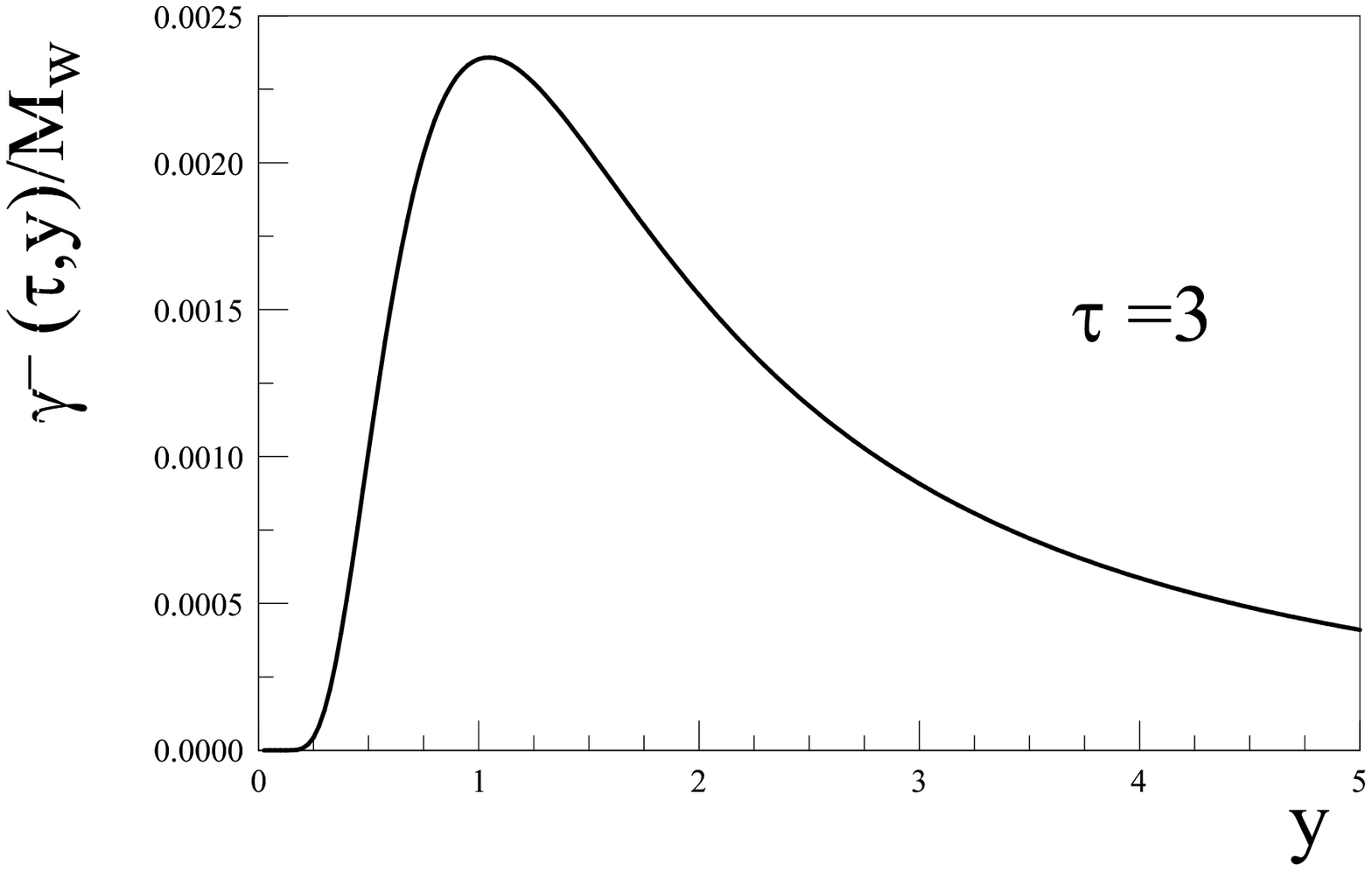}
\caption{  $\gamma^-(q)/M_W$ eqn. (\ref{gammamintot}) vs. $y=q/T$ for $\tau =M_W/T= 1,2,3$ respectively. }
\label{fig:gamaminus}
\end{center}
\end{figure}

\begin{figure}[h!]
\begin{center}
\includegraphics[height=4in,width=3.2in,keepaspectratio=true]{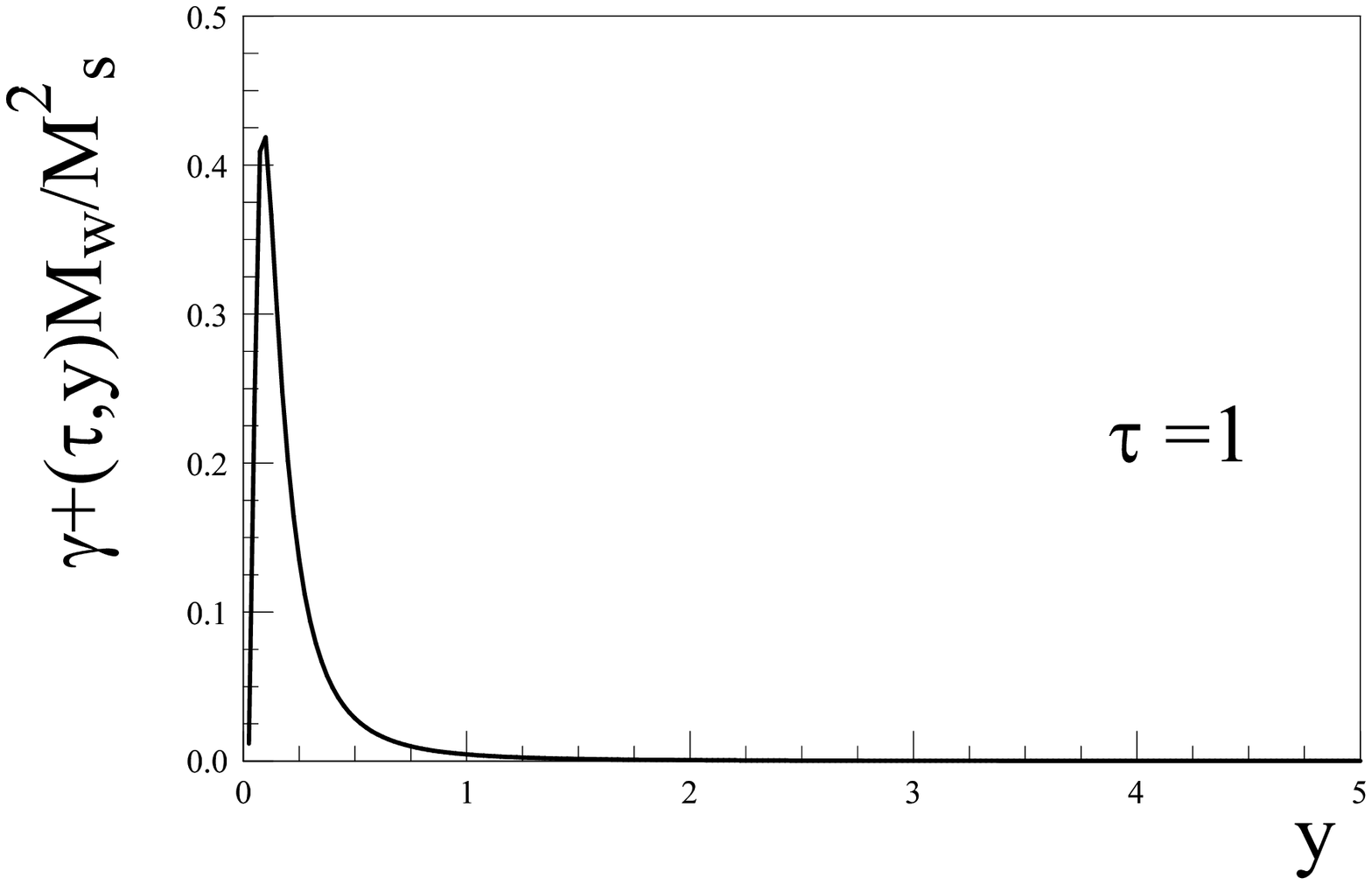}
\includegraphics[height=4in,width=3.2in,keepaspectratio=true]{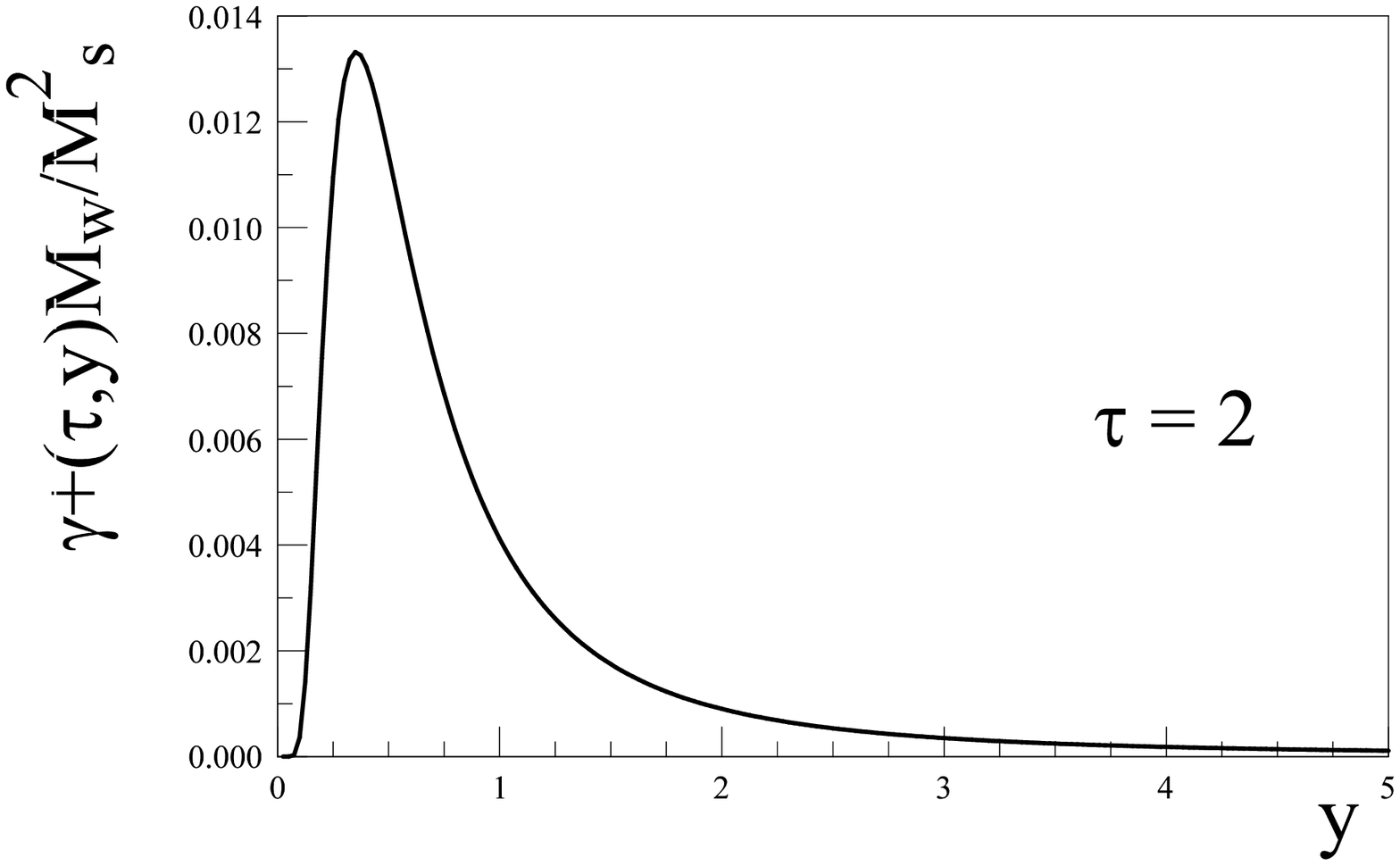}
\includegraphics[height=4in,width=3.2in,keepaspectratio=true]{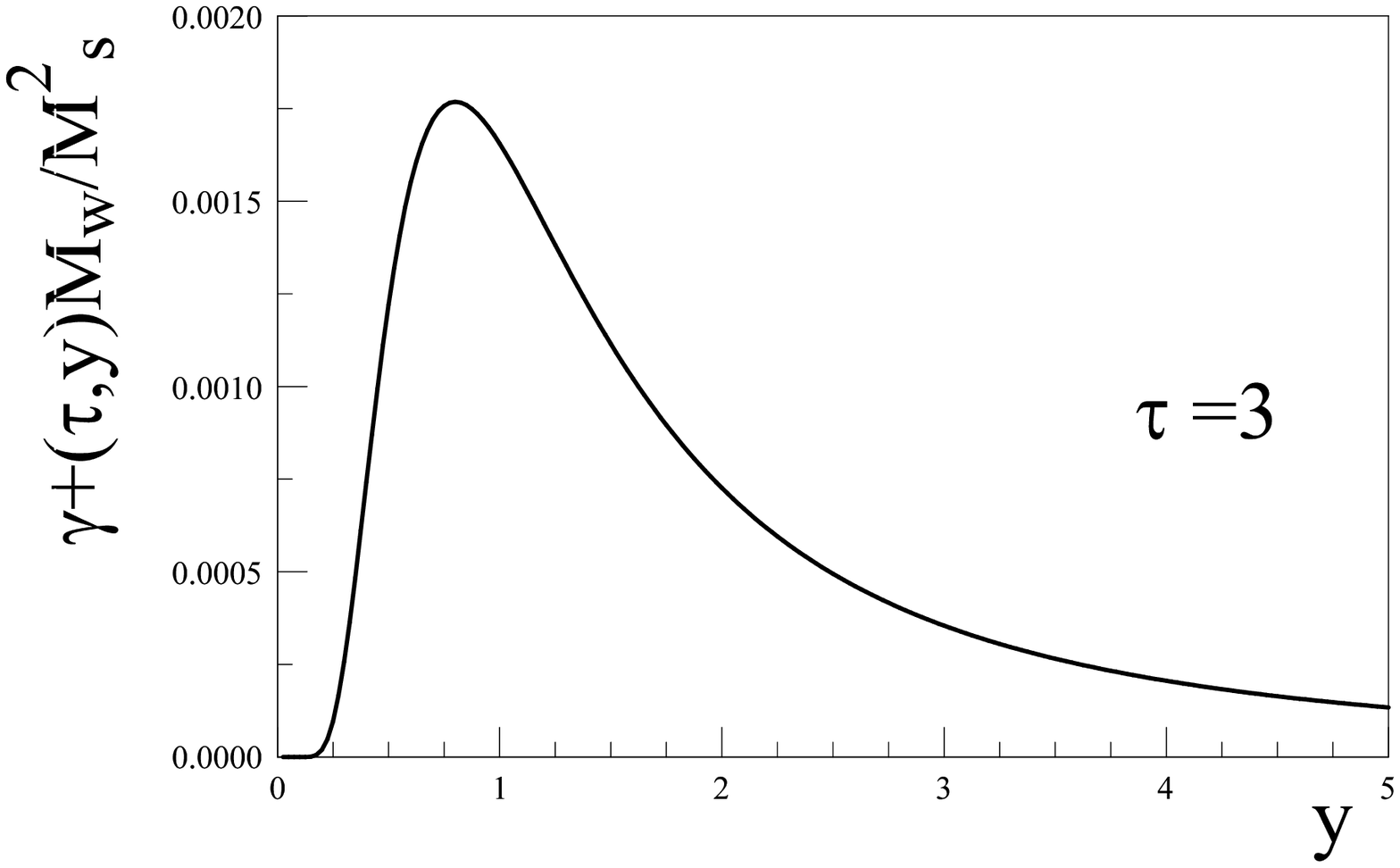}
\caption{  $\gamma^+(q)M_W/M^2_s$ eqn. (\ref{gammaplustot}) vs. $y=q/T$ for $\tau = M_W/T = 1,2,3$ respectively. }
\label{fig:gamaplus}
\end{center}
\end{figure}

 For negative helicity the terms $F_1(q^0=q,q)=1,F_2(q^0=q,q)=0$ and with the definitions (\ref{sigplumin}) we find for negative ($-$) and positive $(+)$ helicities respectively
\be \mathrm{Im}\Sigma^-(q)  = \frac{g^2 T}{16\pi} \frac{M^2}{q^2} \ln\Bigg[\frac{1+e^{-M^2/4qT}}{1-e^{-M^2/4qT}\,e^{-q/T}} \Bigg] \label{ImAminBshell}\ee

\be \mathrm{Im}\Sigma^+(q) = \frac{g^2 T}{16\pi}  \Bigg\{ \ln\Bigg[\frac{1+e^{-M^2/4qT}}{1-e^{-M^2/4qT}\,e^{-q/T}}\Bigg]  + \frac{2T}{q} \sum_{n=1}^{\infty} \frac{e^{-n\,M^2/4qT} }{n^2}\,\Big(e^{-n\,q/T}-(-1)^n \Big)  \Bigg\} \,. \label{ImAplusBshell}\ee These expressions clearly show the suppression for $q \ll M$ for $M \simeq T$ as a consequence of the fact that the decay products feature energy $\simeq M/2$.
These results pertain generically to a vector boson of mass M, we must add the contributions from the charged and neutral vector bosons with their respective masses and couplings. Anticipating the study with cosmological expansion in the next sections we take as a reference mass that of the $W$ vector boson $M_w$ and  introduce the dimensionless variables
\be \tau \equiv \frac{M_W}{T} ~~;~~ y = \frac{q}{T} \label{dimvars}\ee  with the standard model relations (\ref{gsm}) and defining
\be L[\tau,y] = \ln\Bigg[\frac{1+e^{-\tau^2/4y}}{1-e^{-\tau^2/4y}\,e^{-y}} \Bigg] ~~;~~ \sigma[\tau,y] = \frac{2}{y}   \sum_{n=1}^{\infty} \frac{e^{-n\,\tau^2/4y} }{n^2}\,\Big(e^{-n\,y}-(-1)^n \Big) ~~;~~ c \equiv \cos(\theta_w) \simeq 0.88 \label{Lsigcdefs}\ee  the sum of the contributions yield for $\gamma^\mp(q)$ (\ref{gamneghel},\ref{gamposhel})
\be \gamma^-(\tau;y)=   M_W \, \frac{\alpha_w\,\tau}{y^2} \, \Bigg[\frac{1}{8}\,L[\tau,y]+\frac{1}{16c^4}\,L\big[\frac{\tau}{c},y\big] \Bigg] \label{gammamintot}\ee
\be \gamma^+(\tau;y)=   {\alpha_w} \,M_W \Big( \frac{M_s}{M_W}\Big)^2\,\frac{\tau} {4y^2} \, \Bigg\{\frac{1}{8}\,\Big(L[\tau,y]+ \sigma(\tau,y)\Big)+\frac{1}{16c^2}\,\Big(L\big[\frac{\tau}{c},y\big] +\sigma[\frac{\tau}{c},y\big]\Big)   \Bigg\}\,. \label{gammaplustot}\ee The helicity suppression of the positive helicity rate $\gamma^+(q)$ is manifest in the ratio $M^2_s/M^2_W$, this is expected on the grounds that the typical momentum of the emitted neutrino is $\simeq M_W/2$. As a function of $y=q/T$ the rates feature a maximum at $\simeq \tau^2/8$, they are displayed in figs. (\ref{fig:gamaminus},\ref{fig:gamaplus}).

The suppression of the imaginary parts on-shell (damping rates) as $y\rightarrow 0$ has a simple explanation: for a   vector boson of mass $M$  decaying \emph{at rest} in the plasma into two relativistic leptons, energy conservation implies that each lepton carries a momentum $ M/2$, and for $\tau \simeq 1$ this implies $y \simeq 1/2$. For the neutrino to feature $y \ll 1$ it must be that the massive vector boson is\emph{ highly boosted} in the plasma but the probability of such state is exponentially suppressed, thus resulting in an exponential suppression of low momentum neutrinos.

For the mixing angles in the medium (\ref{tetaeff}) we need $\gamma^\pm(q)/\xi$ with $\xi$ given by (\ref{chi}), namely
\be \frac{\gamma^-(\tau;y)}{\xi} = \Big(\frac{M_W}{M_s}\Big)^2 ~I^-(\tau;y)~~;~~ I^-(\tau,y)= \frac{2\alpha_w }{y}\,\Bigg[\frac{1}{8}\,L[\tau,y]+\frac{1}{16c^4}\,L\big[\frac{\tau}{c},y\big] \Bigg]\label{gaminchi} \ee
\be \frac{\gamma^+(\tau;y)}{\xi}\equiv I^+(\tau;y)=    \frac{{\alpha_w}}{2y}  \, \Bigg\{\frac{1}{8}\,\Big(L[\tau,y]+ \sigma(\tau,y)\Big)+\frac{1}{16c^2}\,\Big(L\big[\frac{\tau}{c},y\big] +\sigma[\frac{\tau}{c},y\big]\Big)   \Bigg\}\,. \label{gapluschi}\ee

\begin{figure}[ht!]
\begin{center}
\includegraphics[height=4in,width=3.2in,keepaspectratio=true]{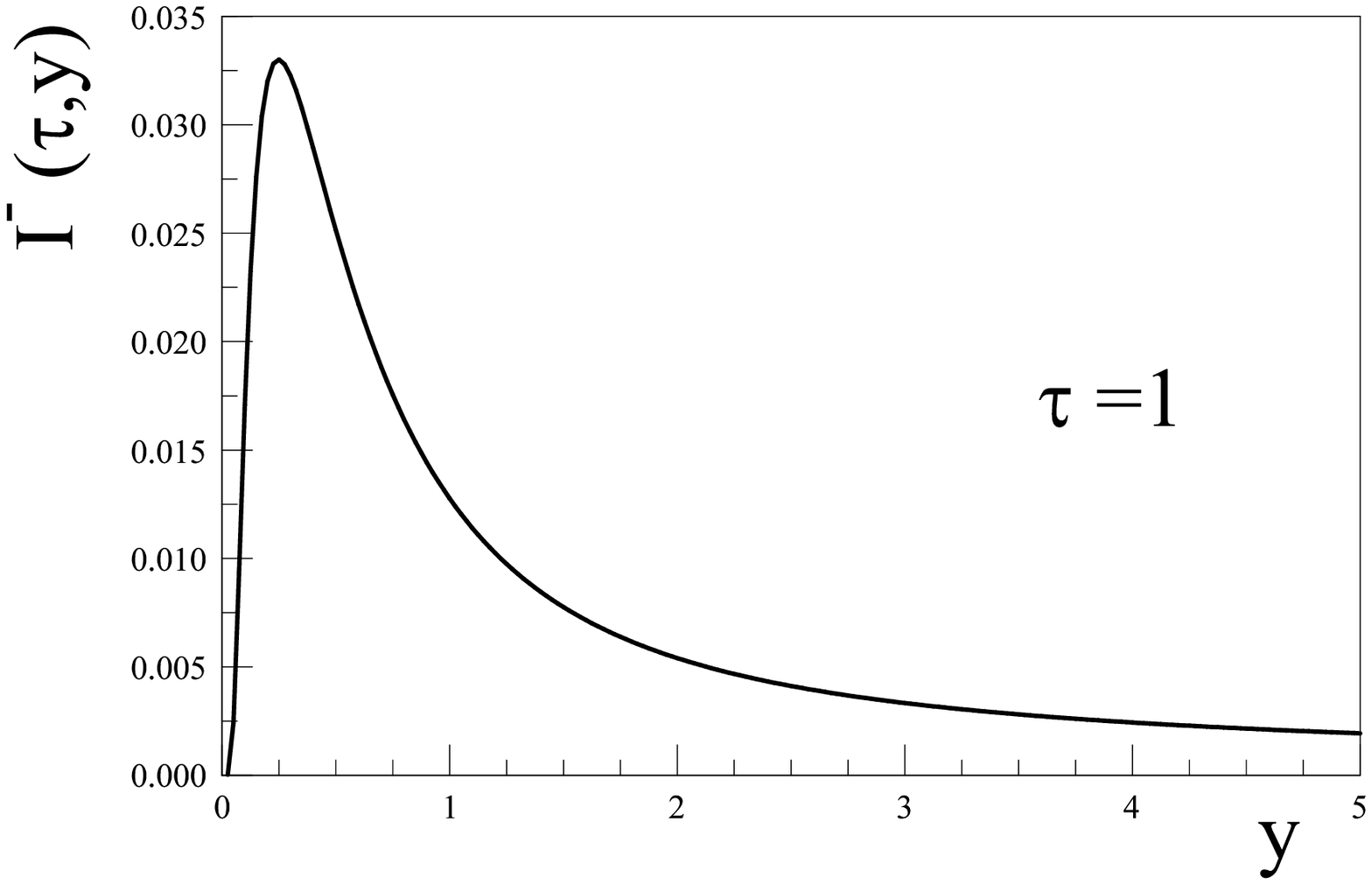}
\includegraphics[height=4in,width=3.2in,keepaspectratio=true]{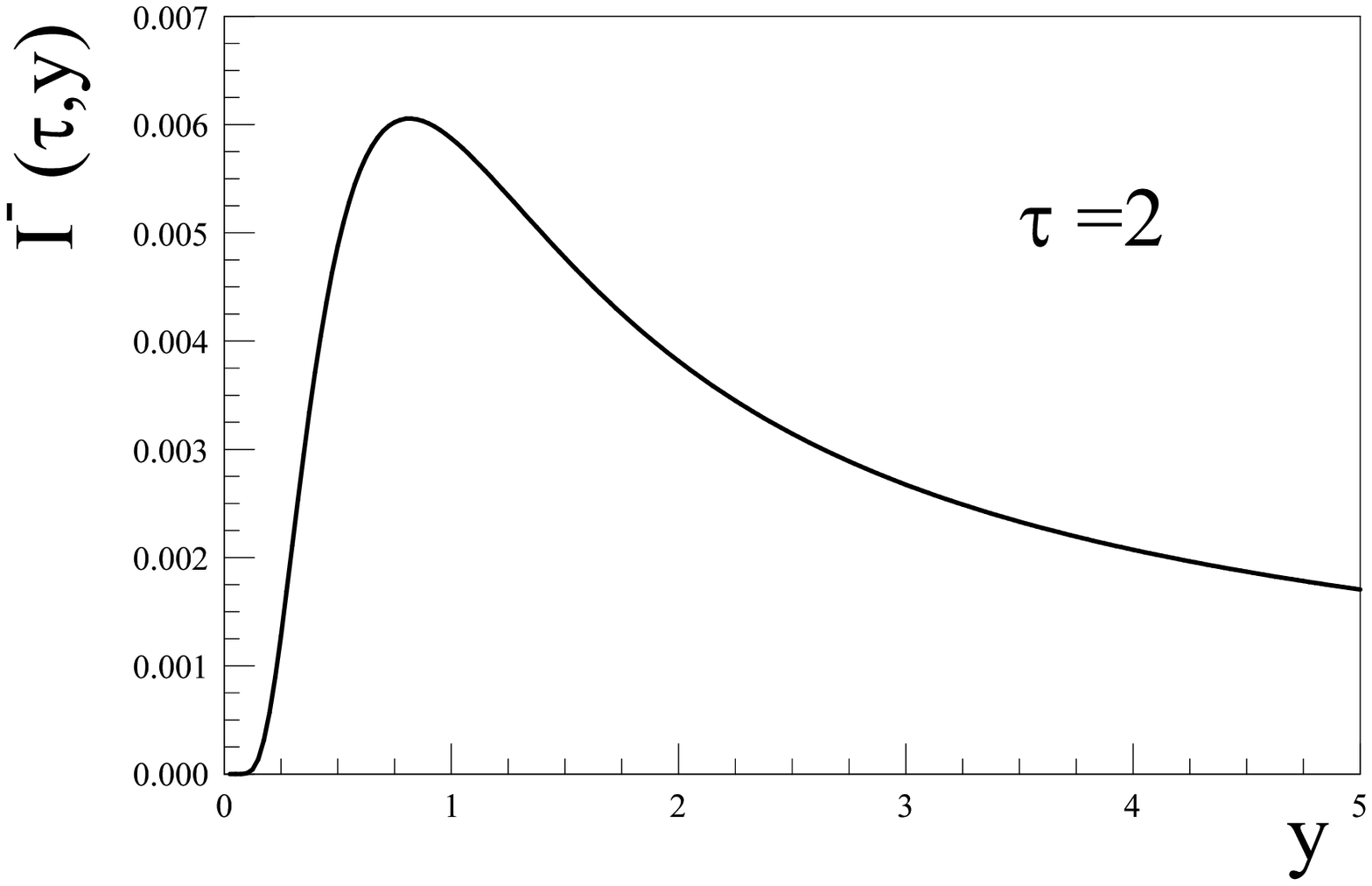}
\includegraphics[height=4in,width=3.2in,keepaspectratio=true]{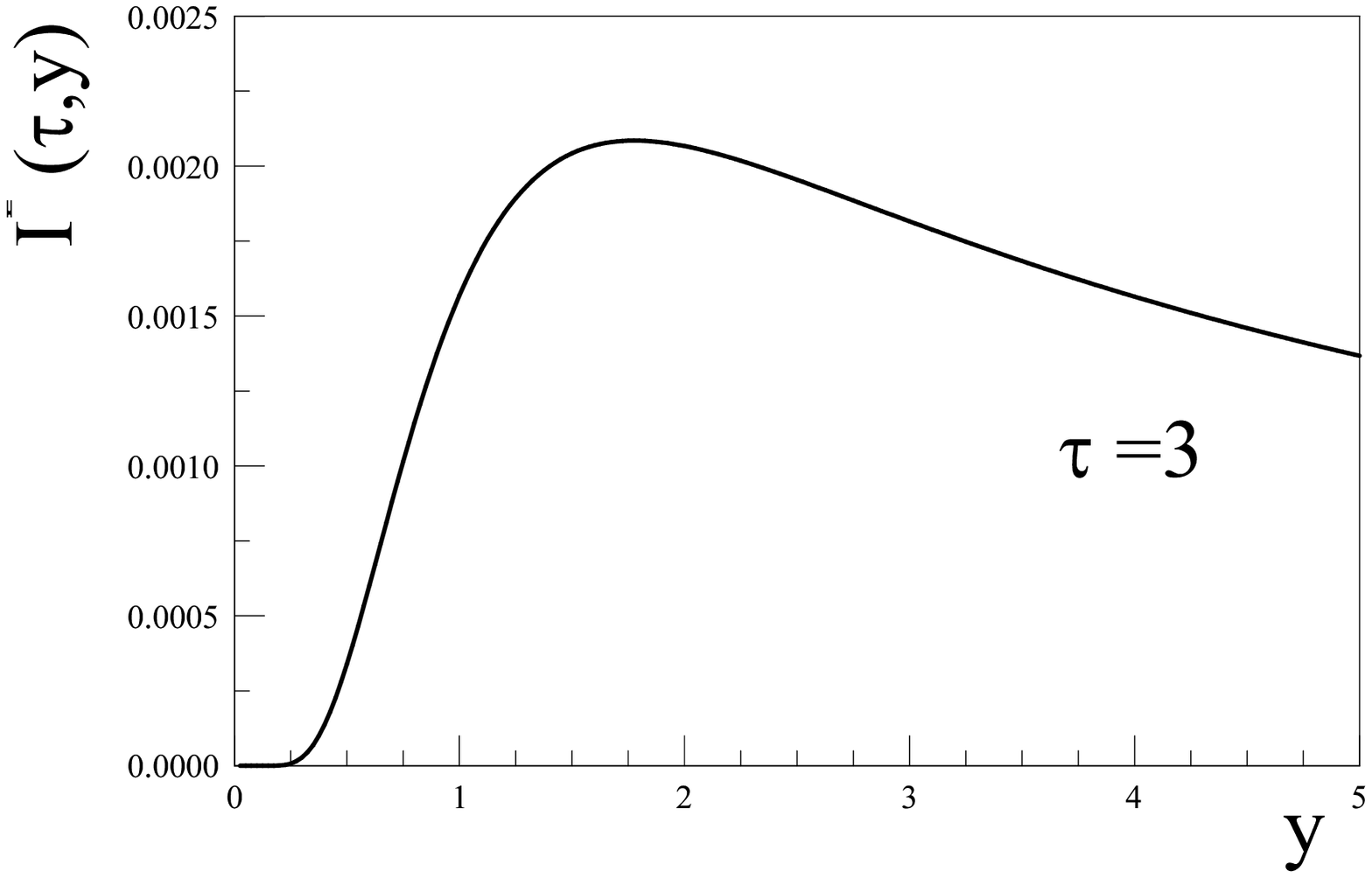}
\caption{  $I^-(\tau,y)$ eqn. (\ref{gaminchi}) vs. $y=q/T$ for $\tau =M_W/T= 1,2,3$ respectively. }
\label{fig:Iminus}
\end{center}
\end{figure}

\begin{figure}[ht!]
\begin{center}
\includegraphics[height=4in,width=3.2in,keepaspectratio=true]{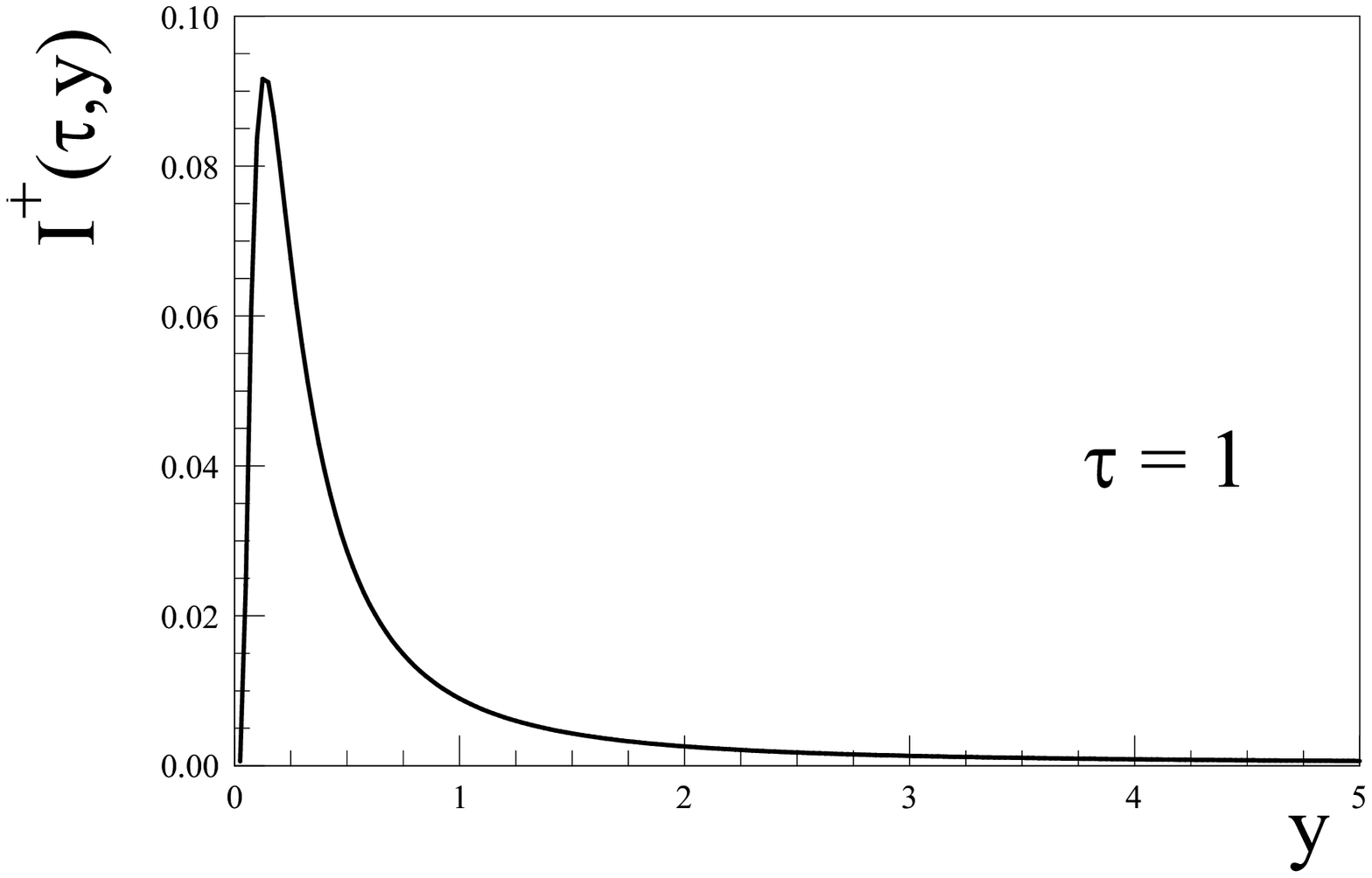}
\includegraphics[height=4in,width=3.2in,keepaspectratio=true]{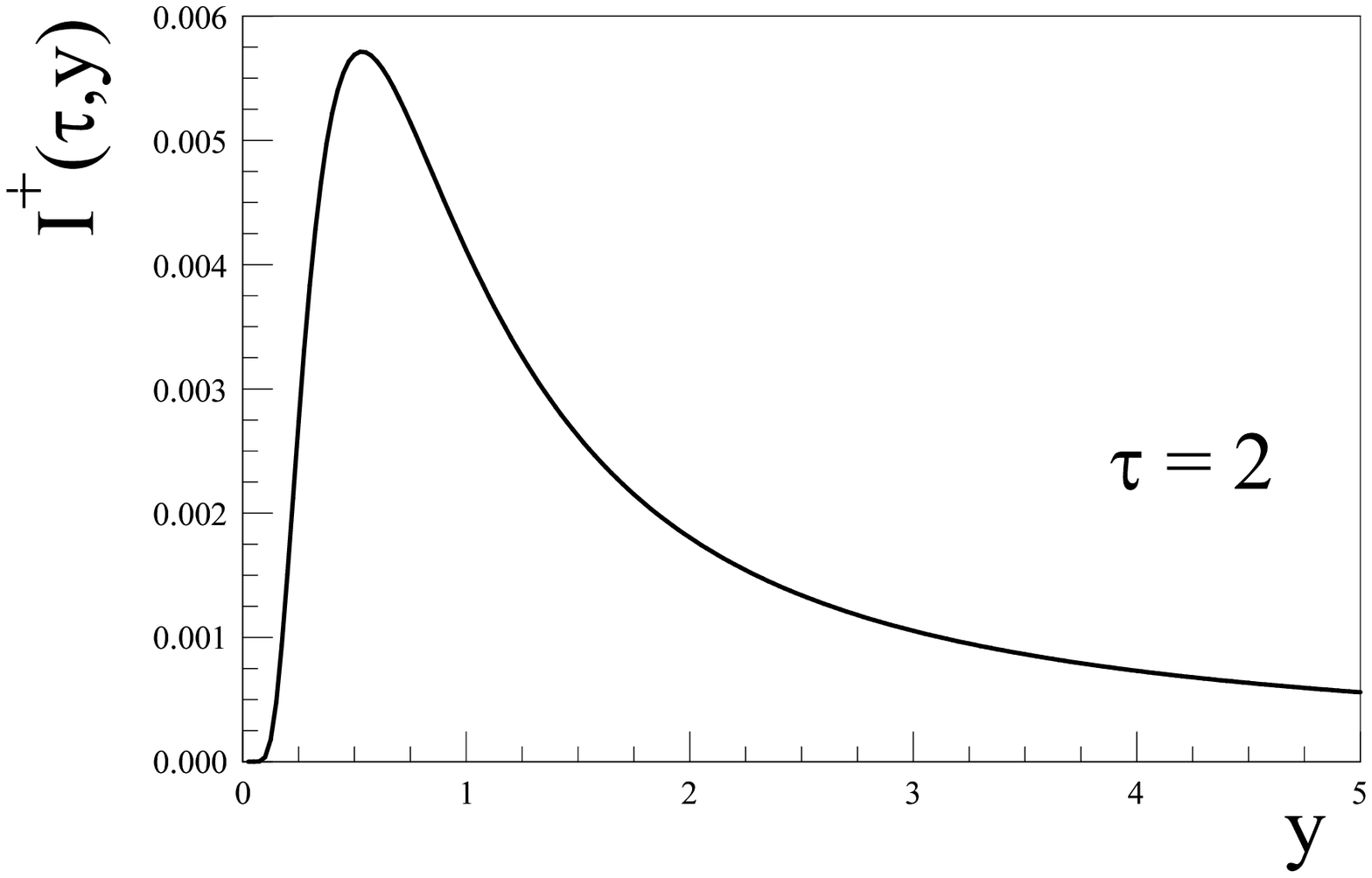}
\includegraphics[height=4in,width=3.2in,keepaspectratio=true]{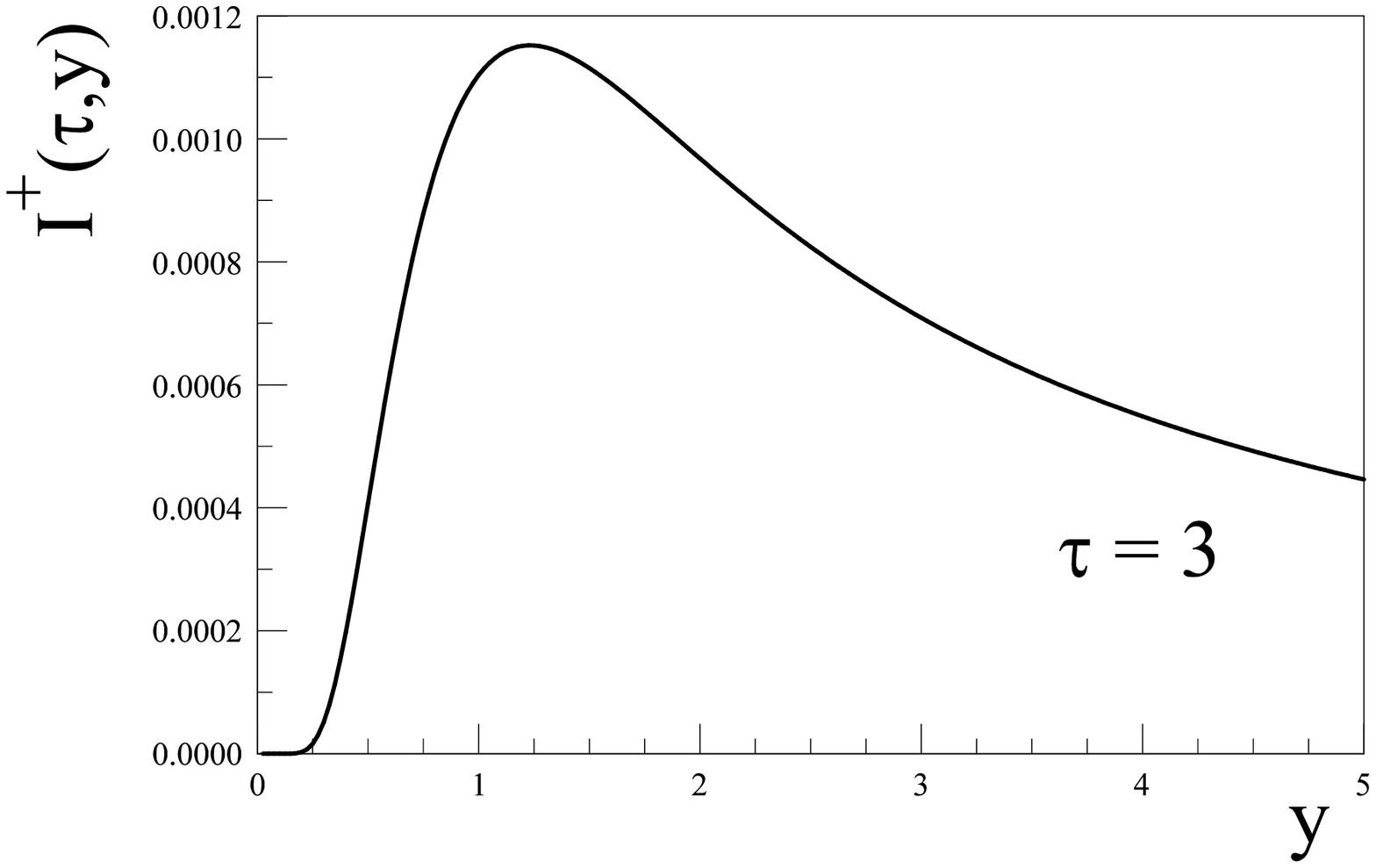}
\caption{  $I^+(\tau,y)$ eqn. (\ref{gapluschi}) vs. $y=q/T$ for $\tau =M_W/T= 1,2,3$ respectively. }
\label{fig:Iplus}
\end{center}
\end{figure}

Figs.(\ref{fig:Iminus},\ref{fig:Iplus}) display $I^\mp(\tau,y)$ for $\tau = 1,2,3$. The main observation is that $I^\mp \ll 1$ in the whole range of $y$ for $\tau \gtrsim 1$. This is important: note that $\gamma^-/\xi$ is \emph{enhanced} by the factor $M^2_w/M^2_s$ therefore for $M_W/M_s \gtrsim 10^2$  it follows that  $\gamma^-/\xi \gg 1$ for $ \tau \gtrsim 1 $ for $y \simeq 1$, this will result in a large suppression of the effective mixing angle in the medium. On the other hand the helicity suppression implies that $\gamma^+/\xi = I^+(\tau,y) \ll 1$ in the whole range of $y$ for $\tau \geq 1$, this will result in a vanishingly small correction to the effective mixing angle in the medium, which in this case will be nearly the
same as that for the vacuum. These points will be revisited again below when we discuss the corrections to the mixing angle \emph{vis-a-vis}  the production rate in the expanding cosmology.

\vspace{2mm}

\section{Real part: index of refraction}\label{sec:realpart}

The index of refraction or real part of the self-energy is obtained from the dispersive representation (\ref{resigma}). In appendix (\ref{app:specdens}) we provide the details of the calculation of $\mathrm{Re}\Sigma^\mp$, both are of the form
\be  \mathrm{Re}\Sigma^\mp (q) = \frac{g^2\, T}{16\pi^2}  \,\frac{M^2}{q^2} \, \mathcal{K}^\mp[\frac{M}{T},\frac{q}{T}]  \,, \label{Resigform}\ee where $\mathcal{K}^\mp[\tau,y]$ are dimensionless functions of the ratios $ M /T;y = q/T$ calculated numerically implementing the steps detailed in the appendix. Combining the contributions from charged and neutral currents we
find
\be  \mathrm{Re}\Sigma^\mp (\tau,y)= M_W \frac{\tau}{y^2}\frac{\alpha_w}{4\pi}\Big[\frac{1}{2}\,\mathcal{K}^\mp[\tau,y]+
\frac{1}{4c^4}\,\mathcal{K}^\mp[\frac{\tau}{c},y] \Big] \,.\label{totResigform}\ee

Of relevance for the in-medium mixing angle are the ratios $\Delta^\pm/\xi$ with $\Delta^\pm$ given by (\ref{delneghel},\ref{delposhel}) and $\xi$ by (\ref{chi}).

\vspace{2mm}

\textbf{Negative helicity:}   For negative helicity, with the definitions (\ref{chi},\ref{delneghel})  we find
\be \frac{\Delta^-(q)}{\xi}  =  \Big(\frac{M_W}{M_s} \Big)^2~J^-(\tau,y)\,, \label{deloverchimin} \ee where
\be J^-(\tau,y) =  \,\frac{\alpha_w}{2\pi\,y}\,\Big[\frac{1}{2}\,\mathcal{K}^-[\tau,y]+
\frac{1}{4c^4}\,\mathcal{K}^-[\frac{\tau}{c},y] \Big]\,. \label{relajkmin}\ee

\begin{figure}[ht!]
\begin{center}
\includegraphics[height=4in,width=3.5in,keepaspectratio=true]{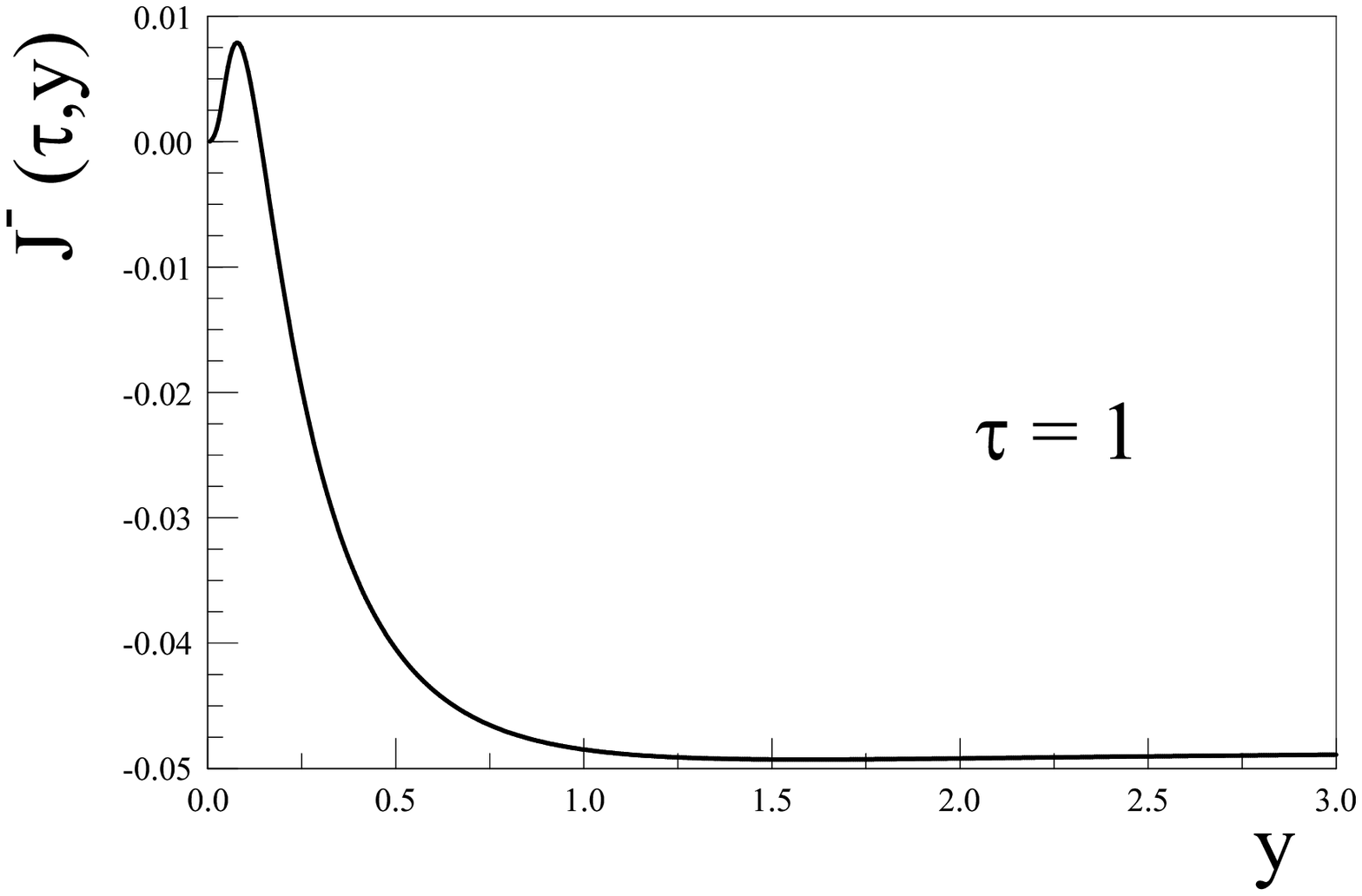}
\includegraphics[height=4in,width=3.5in,keepaspectratio=true]{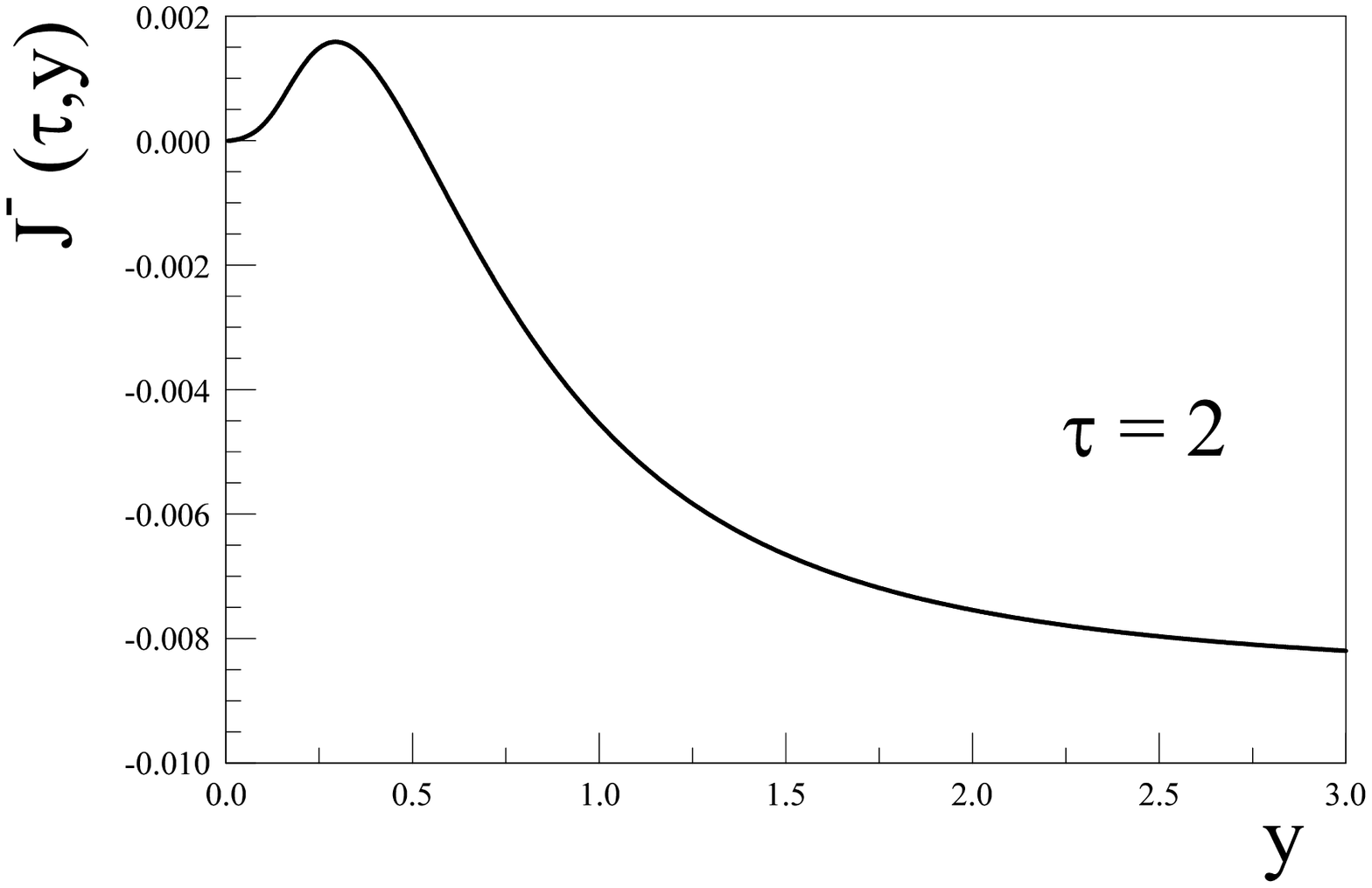}
\caption{  $J^-(\tau,y)$ eqn. (\ref{relajkmin}) vs. $y=q/T$ for $\tau =M_W/T= 1,2$ respectively.  }
\label{fig:Jmin}
\end{center}
\end{figure}

\vspace{2mm}

\textbf{Low temperature limit:} The limit $\tau \gg 1; \tau \gg y$ ($q \ll M_{w}$)  affords an analytic treatment the details of which are summarized in the appendix. In this limit we find that the general form of the real part of the self-energy is given by
\be \mathrm{Re}\Sigma^-(q) = \frac{14 \pi^2}{90}\, g^2 \,\Bigg(\frac{T}{M}\Bigg)^4\,q \label{lowTreSneghel}\ee  adding the charged and neutral current contributions we find in this limit
\be \mathrm{Re}\Sigma^-_{tot}(\tau,y) = M_W \, \frac{ 28\, \pi^3\,\alpha_w  }{90} \Big[1+\frac{1}{2}\,\cos^2(\theta_w)\Big]\, \frac{y}{\tau^5} \,,  \label{reStotneghel} \ee which agrees with those of refs.\cite{notzold,boywu}, and
\be \frac{\Delta^-}{\xi} =  \frac{ 28 \pi^3\,\alpha_w }{45} \, \Bigg(\frac{M_W}{M_s}\Bigg)^2\,\Big[1+\frac{1}{2}\,\cos^2(\theta_w)\Big]\, \frac{y^2}{\tau^6} \,,  \label{deltotneghel} \ee

\vspace{2mm}

\textbf{Positive Helicity} For positive helicity, with the definitions (\ref{chi},\ref{delposhel})  we find
\be \frac{\Delta^+(q)}{\xi}  =    J^+(\tau,y)\,, \label{deloverchimas} \ee where
\be J^+(\tau,y) =  \,\frac{\alpha_w\,\tau^2}{8\pi\,y^3}\,\Big[\frac{1}{2}\,\mathcal{K}^+[\tau,y]+
\frac{1}{4c^4}\,\mathcal{K}^+[\frac{\tau}{c},y] \Big]\,. \label{relajkmas}\ee

\begin{figure}[h!]
\begin{center}
\includegraphics[height=4in,width=3.2in,keepaspectratio=true]{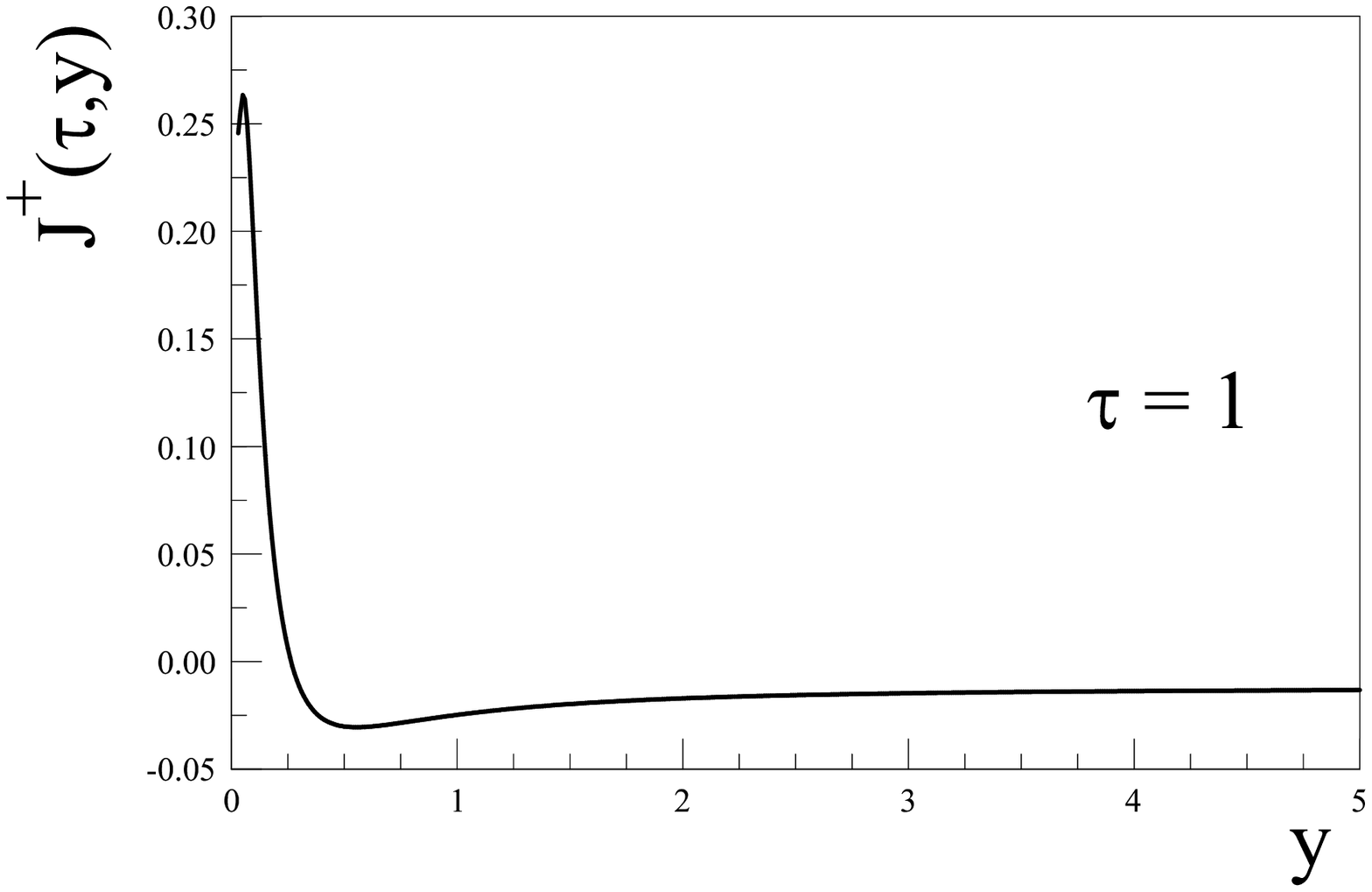}
\includegraphics[height=4in,width=3.2in,keepaspectratio=true]{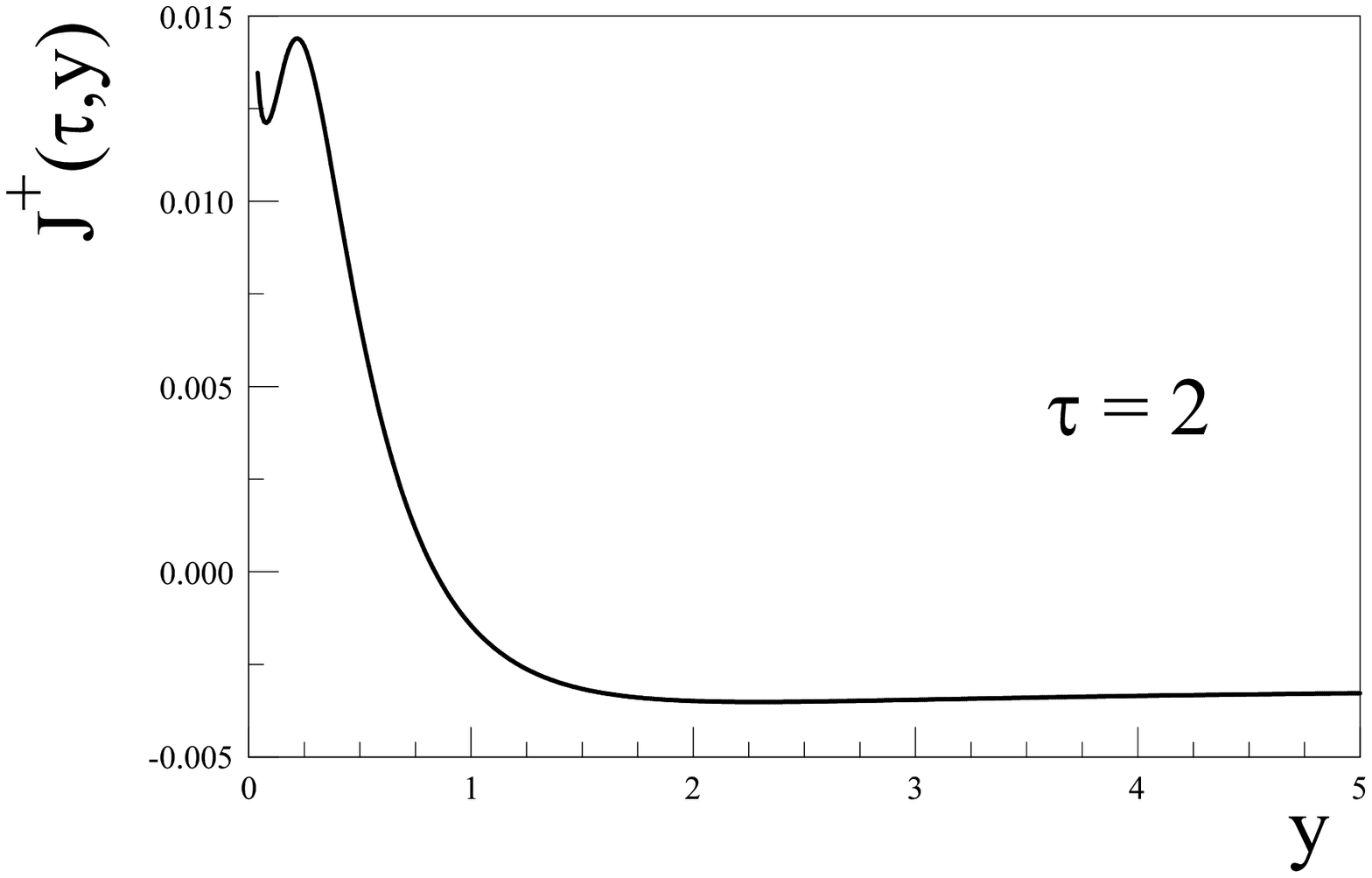}
\caption{  $J^+(\tau,y)$ eqn. (\ref{relajkmas}) vs. $y=q/T$ for $\tau =M_W/T= 1,2$ respectively. }
\label{fig:Jplus}
\end{center}
\end{figure}

Fig.(\ref{fig:Jplus}) displays $J^+(\tau,y)$ vs. $y$ for $\tau=1,2$, we see that $J^+ \ll 1$ for all values of $y$ diminishing rapidly as a function of $\tau$. This results in a small (and negligible) correction to the mixing angle in the medium.

\vspace{2mm}

\textbf{Low temperature limit:} As in the previous case, the limit $\tau \gg 1; \tau \gg y$ ($q \ll M_{w}$)  can be obtained analytically, see the appendix for details. The general form of the real part of the self-energy in this case is
\be \mathrm{Re}\Sigma^+(q) = \frac{14 \pi^2}{180}\, g^2 \,\Bigg(\frac{T}{M}\Bigg)^4\,q \label{lowTreSposhel}\ee  adding the charged and neutral current contributions we find in this limit
\be \mathrm{Re}\Sigma^+_{tot}(\tau,y) = M_W \, \frac{ \,28 \pi^3\,\alpha_w }{180} \Big[1+\frac{1}{2}\,\cos^2(\theta_w)\Big]\, \frac{y}{\tau^5} \,,  \label{reStotposhel} \ee adding the charged and neutral current contributions and including the helicity suppression factor we find
\be \frac{\Delta^+}{\xi} =  \frac{ 7\, \pi^3\,\alpha_w }{90} \, \Big[1+\frac{1}{2}\,\cos^2(\theta_w)\Big]\, \frac{1}{\tau^4} \,.  \label{deltotposghel} \ee

 \section{Effective mixing angles  and production rates}
 The effective mixing angles are given by (\ref{tetaeffpm}) and in the previous sections we obtained $\Delta^\pm/\xi$ and $\gamma^\pm/\xi$.

 \subsection{Negative helicity:}

  For negative helicity $\Delta^-/\xi$ is given by  (\ref{deloverchimin}) and $\gamma^-/\xi$ by (\ref{gaminchi}), therefore
 \be \theta^2_{eff}(\tau,y) =  \frac{\theta^2\,\Big(\frac{M_s}{M_W}\Big)^4}{\Bigg[\Big( \frac{M^2_s}{M^2_W} +J^-(\tau,y) \Big)^2+ \Big(I^-(\tau,y)\Big)^2 \Bigg]}\,.  \label{tetefneghel}\ee

 Figures (\ref{fig:Iminus},\ref{fig:Jmin} ) display  $I^-(\tau,y),J^-(\tau,y)$ as a function of $y$ for various values of $\tau$. These show important features:  $J^-$  vanishes and becomes negative at a value of $y^*(\tau)$ that increases monotonically with  $\tau$. This behavior implies that for $M_W >> M_s$ the vanishing of $J^-(\tau,y)$   implies an MSW resonance \emph{in absence of lepton asymmetry} for the effective mixing angle for \emph{negative helicity}. However, this resonance is ``screened'' by the contribution to the mixing angle from the imaginary part, inspection of both $J^-$ and $I^-$ (see figs.\ref{fig:Iminus},\ref{fig:Jmin}) and an exhaustive numerical study reveal that in the broader region $0< y^*(\tau) \lesssim y$    the imaginary part $I^-$  yields the  dominant contribution to the denominator of (\ref{tetefneghel})  and there is no substantial enhancement of the mixing angle as $y$ sweeps through the resonance for any $\tau$. In other words, the presence of the MSW resonance does not influence the effective mixing angle in a substantial manner.
  The numerical analysis shows that for $M_s/M_W \lesssim 10^{-2}$ the term $(M_s/M_W)^2$ in the denominator in (\ref{prodnegi}) can be safely neglected and the effective mixing angle is $\propto (M_s/M_W)^4$ reflecting a strong in-medium suppression, even when the denominator becomes large because $J^-, I^-$ are very small as shown in the figures. The smallness of the denominator is thus compensated by the large power of $M_s/M_W\ll 1$ in the numerator.

 In the low temperature limit $T\ll M_W$ the effective mixing angle in the medium is given by
  \be \theta_{eff}(\tau,y) =  \frac{\theta}{ \Big[ 1 +  \frac{ 28 \pi^3\,\alpha_w }{45} \, \Big(\frac{M_W}{M_s}\Big)^2\,\Big[1+\frac{1}{2}\,\cos^2(\theta_w)\Big]\, \frac{y^2}{\tau^6} \Big] } \,,  \label{louTtetaf}\ee the contribution from the imaginary part is subleading as it is suppressed by an extra power of $\alpha_w$ from two loop  contributions to the self-energy.

The production rate (see \ref{gamasterfin},\ref{neghelprodu} ) is
 \be  \Gamma^-_{prod}(\tau,y) =   \frac{ M_W\, \theta^2 \, \Big(\frac{M_s}{M_W}\Big)^4}{y\Big[e^y+1\Big]}\,\frac{\tau\,I^-(\tau,y)}{\Bigg[\Big( \frac{M^2_s}{M^2_W} +J^-(\tau,y) \Big)^2+ \Big(I^-(\tau,y)\Big)^2 \Bigg]}\, \,. \label{prodnegi}\ee Although both $J^-,I^-$ vanish as $y \rightarrow 0$, and the effective mixing angle reaches its maximum ($\theta_{eff} \rightarrow  \theta$) the imaginary part $I^-$ vanishes exponentially as $y \rightarrow 0$ and the production rate vanishes in this limit. The effect of the MSW resonance is ``screened'' by the imaginary part and the resulting production rate features a peak as a function of $y$ for fixed $\tau$ that is a result of the competition between the peak in $I^-(y,\tau)$ (see figs. (\ref{fig:Iminus})) and the \emph{increase} in $(J^-)^2$.

 We will analyze further the production rate in the following section within the context of cosmological expansion.

 \vspace{2mm}

 \subsection{Positive helicity:} For positive helicity the effective mixing angle is given by
 \be \theta^2_{eff}(\tau,y) =  \frac{\theta^2}{\Bigg[\Big( 1 +J^+(\tau,y) \Big)^2+ \Big(I^+(\tau,y)\Big)^2 \Bigg]}\,,  \label{tetefposhel}\ee  and the production rate
  \be  \Gamma^+_{prod}(\tau,y) =   \frac{ M_W\, \theta^2 \, \Big(\frac{M_s}{M_W}\Big)^2}{y\Big[e^y+1\Big]}\,\frac{\tau\,I^+(\tau,y)}{\Bigg[\Big( 1 +J^+(\tau,y) \Big)^2+ \Big(I^+(\tau,y)\Big)^2 \Bigg]}\, \,. \label{prodposi}\ee

 From the figures (\ref{fig:Iplus},\ref{fig:Jplus}) we see that $I^+,J^+ \ll 1$ in the region in which $I^+$ peaks, therefore we can set $J^+ \simeq 0~;~I^+ \simeq 0$ in the denominator in (\ref{prodposi}), in other words \emph{the in-medium contribution to the mixing angle is negligible}, namely $\theta_{eff} \simeq \theta$ and we can approximate the production rate of \emph{positive helicity} neutrinos as
  \be  \Gamma^+_{prod}(\tau,y) \simeq   { M_W\, \theta^2 \, \Big(\frac{M_s}{M_W}\Big)^2} \,\frac{\tau\,I^+(\tau,y)}{y\Big[e^y+1\Big]}\, \,. \label{prodposapp}\ee

  This is an important result: the helicity suppression entails a \emph{much weaker coupling to the medium} which in turn results in negligible in-medium correction to the mixing angles and the effective mixing angle is just the vacuum mixing angle.

   The positive helicity production rate is $\propto (M_s/M_W)^2$ as a consequence of helicity suppression, whereas the negative helicity production rate (\ref{prodnegi}) is $\propto  (M_s/M_W)^4$ as a consequence of the in-medium suppression of the mixing angle, but is enhanced by the small denominator. As it will be discussed below, there is a range of masses and temperatures for which the negative and positive helicity rates are \emph{comparable}: positive helicity states feature  helicity suppressed couplings but nearly vacuum mixing angles, whereas negative helicity states feature stronger coupling to the medium which in turn leads to strongly suppressed in-medium mixing angles, thus the competition. This aspect is studied below.

\section{Cosmological production:} We consider a spatially flat Friedmann-Robertson-Walker cosmology during a radiation dominated stage. The effect of cosmological expansion is included by replacing the momentum in Minkowski space time with the physical momentum in the expanding cosmology, namely \be q \rightarrow \frac{q}{a(t)} \label{physmom}\ee where $a(t)$ is the scale factor and now $q$ refers to the (constant) comoving momentum. As we focus on the production during the radiation dominated era, the physical temperature is
\be T(t) = \frac{T_0}{a(t)} \label{physT}\ee where $T_0$ is the temperature that the plasma would have \emph{today} as we normalize the scale factor to 1 today. We note that the variable
\be y = \frac{q(t)}{T(t)} = \frac{q}{T_0} \label{yconst}\ee is a \emph{constant} under cosmological expansion in terms of the comoving momentum and the temperature that the plasma would feature today.

In terms of the comoving wave vector $q$ and the invariant ratio $y=q/T$ the quantum kinetic equation in the expanding cosmology reads
\be \frac{d n^h_2(q;t)}{dt} = \Gamma^h_2(q,t) \Big[n_{eq}(q) -n^h_2(q;t)\Big]\,, \label{qukinexp} \ee where now $\Gamma^h_2(q,t) \equiv \Gamma^h_2(\tau(t),y)$ depends on time through $\tau(t) = M_W/T(t)= M_W a(t)/T_0$,

Under the assumption of $\theta \ll 1$  and a vanishing initial population,  we neglect the build up of the population and approximate the quantum kinetic equation as
\be \frac{d\,n^h_2(t)}{dt} = \Gamma^h_{prod}(\tau(t),y) \label{cosmoprod}\ee where $\Gamma^\mp_{prod}(\tau(t),y)$ are given by (\ref{prodnegi}),\ref{prodposi}). Since the production rate depends on time through $\tau(t)$ it is convenient to use this variable in the kinetic equation, with
\be \frac{dn^h_2(t)}{dt} =   \frac{dn^h_2(\tau,y)}{d\tau}\, H(t)\, \tau(t) \label{kinetau}\ee where during radiation domination
\be H(t) = \frac{\dot{a}(t)}{a(t)} = 1.66 \, g^{1/2}_{eff}(T) \,\frac{T^2(t)}{M_{pl}}~~;~~ M_{pl} = 1.22 \,\times \,10^{19}\,\mathrm{GeV} \,. \label{hubble} \ee $g_{eff}(T)$ is the effective number of relativistic degrees of freedom, $g_{eff} \simeq 100$, and  varies slowly in the temperature regime $1\,\mathrm{GeV} < T < 100\,\mathrm{GeV}$. We will approximate $g_{eff} \simeq 100$ and constant in this temperature range anticipating that freeze-out will occur at $T_f \simeq \mathrm{few}\,\mathrm{GeV}$.

\vspace{2mm}

\textbf{Negative helicity}

For negative helicity we find
\be  \frac{dn^-_2(\tau,y)}{d\tau}\simeq  \frac{0.92\,\times 10^{16}\,  \theta^2  \, \Big(\frac{M_s}{M_W}\Big)^4}{y\Big[e^y+1\Big]}\,\frac{\tau^2\,I^-(\tau,y)}{\Bigg[\Big( \frac{M^2_s}{M^2_W} +J^-(\tau,y) \Big)^2+ \Big(I^-(\tau,y)\Big)^2 \Bigg]}\,. \label{neghelqk}  \ee As both $J^-,I^-$ decrease as the temperature decreases (and $\tau$ increases) there are two competing effects: the damping rate $\propto I^-$ decreases but the effective mixing angle \emph{increases} as a result, for a fixed value of $y=q/T$ the production rate peaks as a function of $\tau$ and falls off sharply. We write (\ref{neghelqk}) as
\be \frac{dn^-_2(\tau,y)}{d\tau}\equiv     0.92\,\times 10^{16}\,  \theta^2  \, \Big(\frac{M_s}{M_W}\Big)^4~R^-(y,\tau)\,. \label{Rmindef}\ee As discussed above, for $M_s/M_W < 10^{-2} $ we find numerically that  $R^-(y,\tau)$ is nearly independent of $M_s$. The form (\ref{Rmindef}) separates the suppression factor from the effective mixing angle in terms of the prefactor   $\theta^2 M^4_s/M^4_w$, whereas for $M_s/M_W < 10^{-2}$ the function $R^-(y,\tau)$ is \emph{insensitive to the value of $M_s$} and only depends on standard model couplings and vector boson masses. As $\tau$ increases (temperature decreases) the effective mixing angle \emph{increases} whereas the damping rate $\gamma^-$ \emph{decreases}, therefore we expect that $R^-(y,\tau)$ for fixed values of $y$ to feature a peak as a function $\tau$. The analysis in the previous sections clarifies that for $\tau \gg \sqrt{y}$ the damping rate is exponentially suppressed (see \ref{gammamintot},\ref{Lsigcdefs}) whereas the real part (index of refraction) falls off as $1/\tau^5$ (see \ref{deltotneghel}) therefore the production rate is exponentially suppressed at large $\tau$ as the mixing angle grows much slower. This entails the \emph{freeze out} of the distribution function.

\begin{figure}[ht!]
\begin{center}
\includegraphics[height=4in,width=3.2in,keepaspectratio=true]{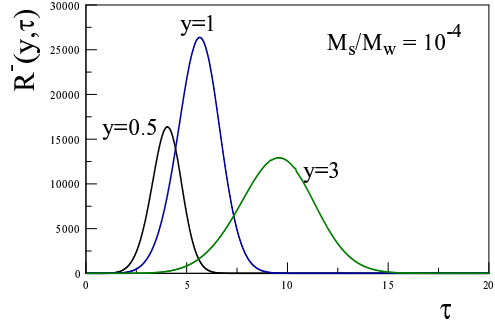}
\caption{ Rates  $R^-(y,\tau)$ (see eqns. (\ref{neghelqk}) and (\ref{Rmindef})) vs. $\tau$ for $y=0.5,1,3$ respectively for $M_s/M_W=10^{-4}$. }
\label{fig:rates}
\end{center}
\end{figure}

\begin{figure}[ht!]
\begin{center}
\includegraphics[height=4in,width=3.2in,keepaspectratio=true]{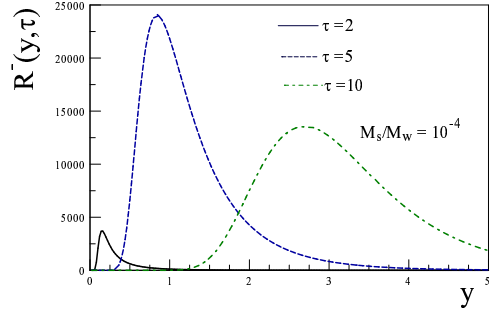}
\caption{ Rates  $R^-(y,\tau)$ vs. $y$ for $\tau=2,5,10$ respectively for $M_s/M_W=10^-4$. }
\label{fig:ratesneghel}
\end{center}
\end{figure}

\begin{figure}[ht!]
\begin{center}
\includegraphics[height=4in,width=3.2in,keepaspectratio=true]{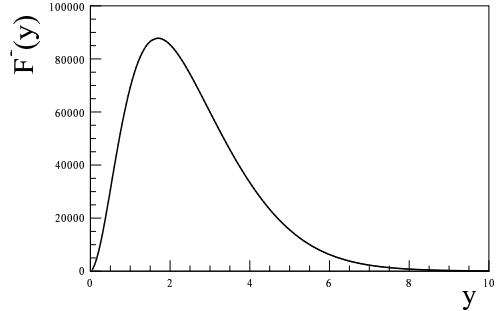}
\caption{ Asymptotic distribution function   $F^-(y)$ (see eqns. (\ref{asin2min}),(\ref{negheldist}))   vs. $y$. }
\label{fig:negheldist}
\end{center}
\end{figure}

This expectation is borne out from the numerical study, figs.(\ref{fig:rates},\ref{fig:ratesneghel}) shows $R^-(y,\tau)$ for various values of $y$ as a function of $\tau$ and as a function of $y$ for $\tau=2,5,10$. Numerically the case with $M_s/M_W = 10^{-4}$ is indistinguishable from that setting $M_s =0$ in the denominator of $R^-$ (see eqn. (\ref{neghelqk})). These figures clearly show  the ``freeze-out'' of the distribution as a function of $\tau$, as the rate vanishes for large $\tau$, larger values of $y$ freeze-out at larger $\tau$ but with much smaller amplitudes. This feature is expected, the vector bosons are suppressed at smaller temperatures  (larger $\tau$) and large values of $y$ are further suppressed by the detailed balance factor $1/(e^y +1)$. Fig. (\ref{fig:ratesneghel}) shows the ``filling'' of the different wave-vectors: as time evolves larger $y$ are populated but eventually larger values of $y$ are suppressed by the Fermi-Dirac factor $n_{eq}(y)$.
\emph{Assuming} that the initial population vanishes at $\tau \simeq 1$), the asymptotic distribution function is given by
\be n_2^-(y) = \int_1^\infty \frac{dn^-_2(\tau,y)}{d\tau}\,d\tau =  0.92\,\times 10^{16}\,  \theta^2  \, \Big(\frac{M_s}{M_W}\Big)^4 \, F^-(y)\,\label{asin2min}\ee
where we have defined the \emph{frozen} distribution
\be F^-(y)= \int_1^{\infty} R^-(y,\tau)\,d\tau \label{negheldist}\ee which  is shown in fig. (\ref{fig:negheldist}). Although we have set the lower limit $\tau =1$ for consistency in (\ref{negheldist}), we find that $R^-(y,\tau)$ vanishes as $\tau \rightarrow 0$ and the  lower limit can be effectively taken to $\tau = 0$.  A numerical study informs that the region $0< y \leq 10$ which features the largest contribution to the distribution function, freezes out at $\tau_f \simeq 15$ corresponding to a freeze-out temperature for negative helicity modes $T^-_f = M_W/\tau_f \simeq 5 \, \mathrm{GeV}$.

The vanishing of $F^-(y)$ as $y\rightarrow 0$ is a consequence of the vanishing of the imaginary part, and a direct consequence of the decay kinematics in the medium, as explained above the $y\rightarrow 0$ is dominated by the decay of vector bosons that are highly boosted in the rest frame of the plasma, and the population of these states is highly suppressed at $T \simeq M_W$. The broadening of the distribution as compared to the damping rate (compare figures (\ref{fig:Iminus},\ref{fig:negheldist}) is a consequence a longer freeze-out time resulting from  the competition between a decreasing damping rate $I^-$ and an increasing mixing angle, the modes with higher $y$ continue to populate as the mixing angle increases but eventually as modes with large values of $y$ are populated, their contribution is suppressed by the detailed balance factor $n_{eq}(y)$.
After freeze-out, the total number density of negative helicity neutrinos produced, (equal to the total number of positive helicity antineutrinos  in absence of a lepton asymmetry) is given by
\be \mathcal{N}^-_2 = \frac{T^3(t)}{2\pi^2} \int_0^\infty n^-_2(y)\,y^2\,dy\,, \label{totneg}\ee for which we need the result
\be \int_0^\infty y^2 F^-(y)\, dy \equiv N^- = 2.287\times 10^6 \,.\label{totint} \ee This integral is dominated by the region $0 < y \lesssim 10$, which freezes-out at $\tau \simeq 15$,  with the result that
\be \frac{ \int_{10}^\infty y^2 F^-(y)\, dy}{ \int_0^\infty y^2 F^-(y)\, dy} = 3.9 \times 10^{-3} \,. \label{ratF}\ee

Normalizing the number density to that of one degree of freedom of an active massless neutrino decoupled in equilibrium at the \emph{same} temperature, namely\be \mathcal{N}_\nu = \frac{T^3(t)}{2\pi^2} \int_0^\infty n_{eq}(y)\,y^2\,dy\,, \label{totnueq}\ee where $n_{eq}(y)=1/(e^y+1)$, we find
\be \frac{\mathcal{N}^-_2}{\mathcal{N}_\nu} = 285  ~\theta^2 \,\Big(\frac{M_s}{\mathrm{MeV}}\Big)^4\,, \label{noversneg}\ee this ratio is constant throughout the expansion history.   It is clear that this mechanism of production will \emph{not} yield a thermalized species for $M_s \,\mathrm{few}\, \mathrm{MeV}$  and $\theta \ll 10^{-4}$, under these conditions, we note that the approximation (\ref{cosmoprod}) of neglecting the build-up of the population in the quantum kinetic equation (\ref{qukinexp}) is consistent since the ratio $ {\mathcal{N}^-_2}/{\mathcal{N}_\nu} \ll 1$.

\vspace{2mm}

\begin{figure}[ht!]
\begin{center}
\includegraphics[height=4in,width=3.2in,keepaspectratio=true]{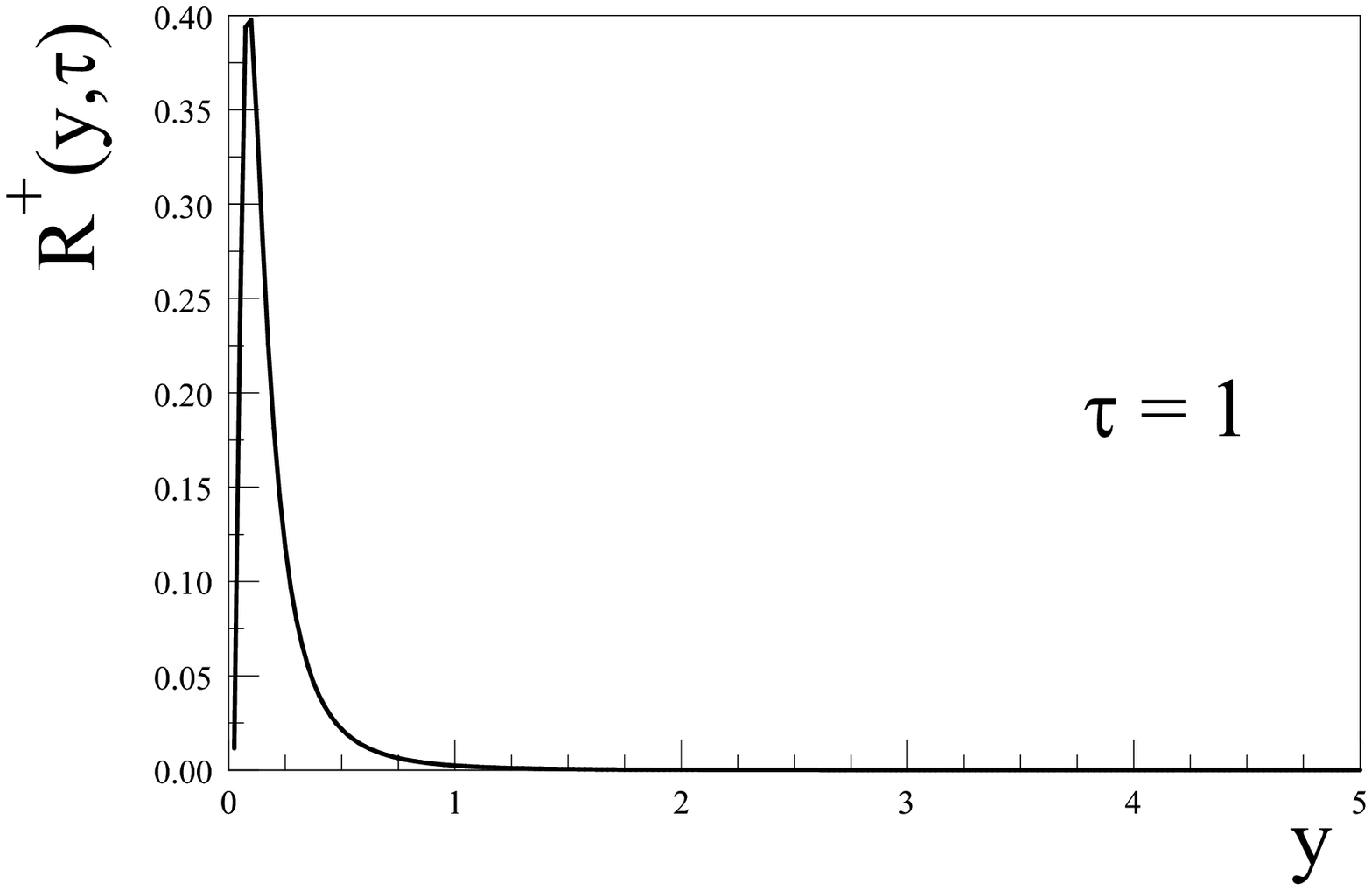}
\includegraphics[height=4in,width=3.2in,keepaspectratio=true]{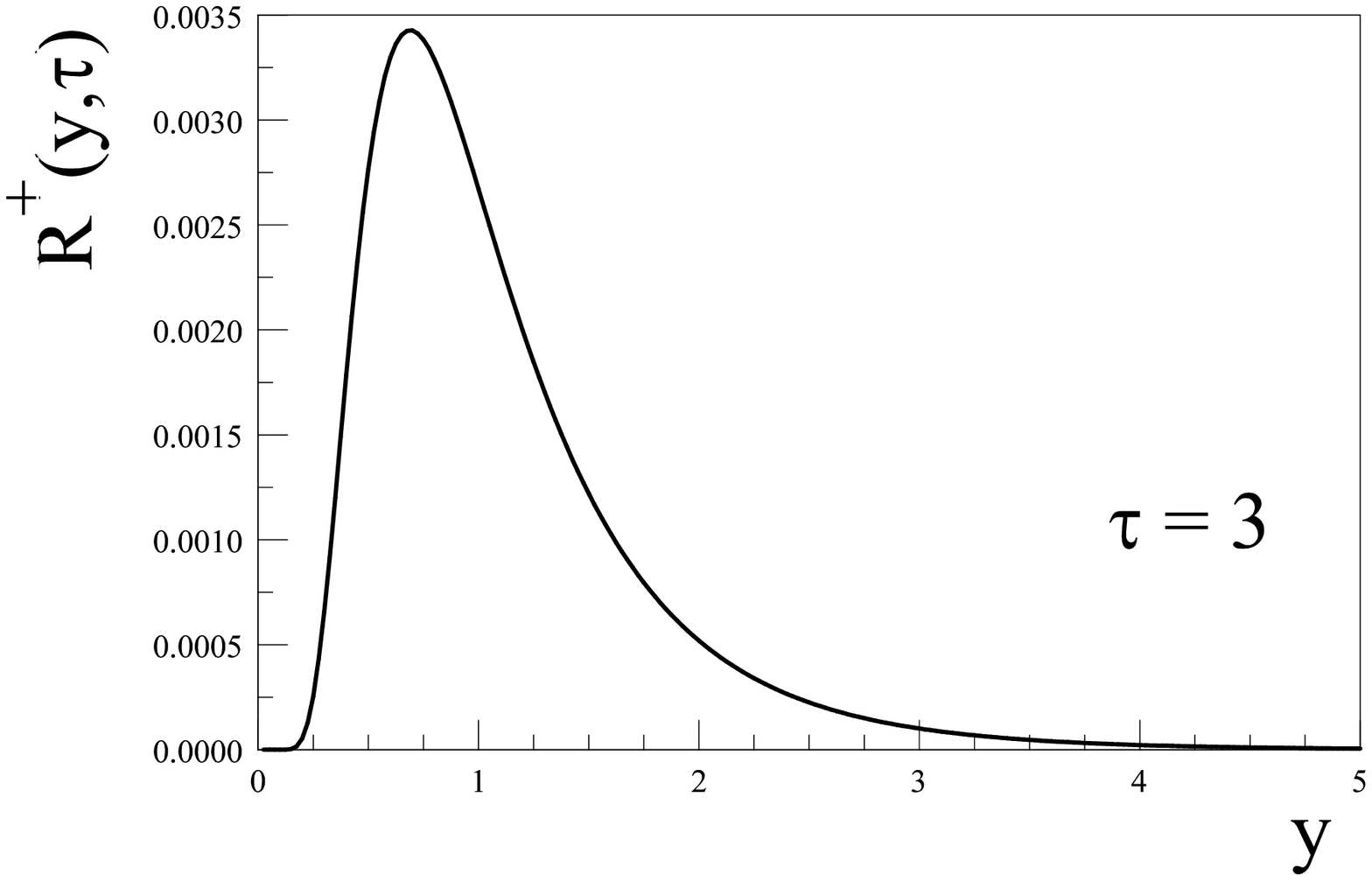}
\includegraphics[height=4in,width=3.2in,keepaspectratio=true]{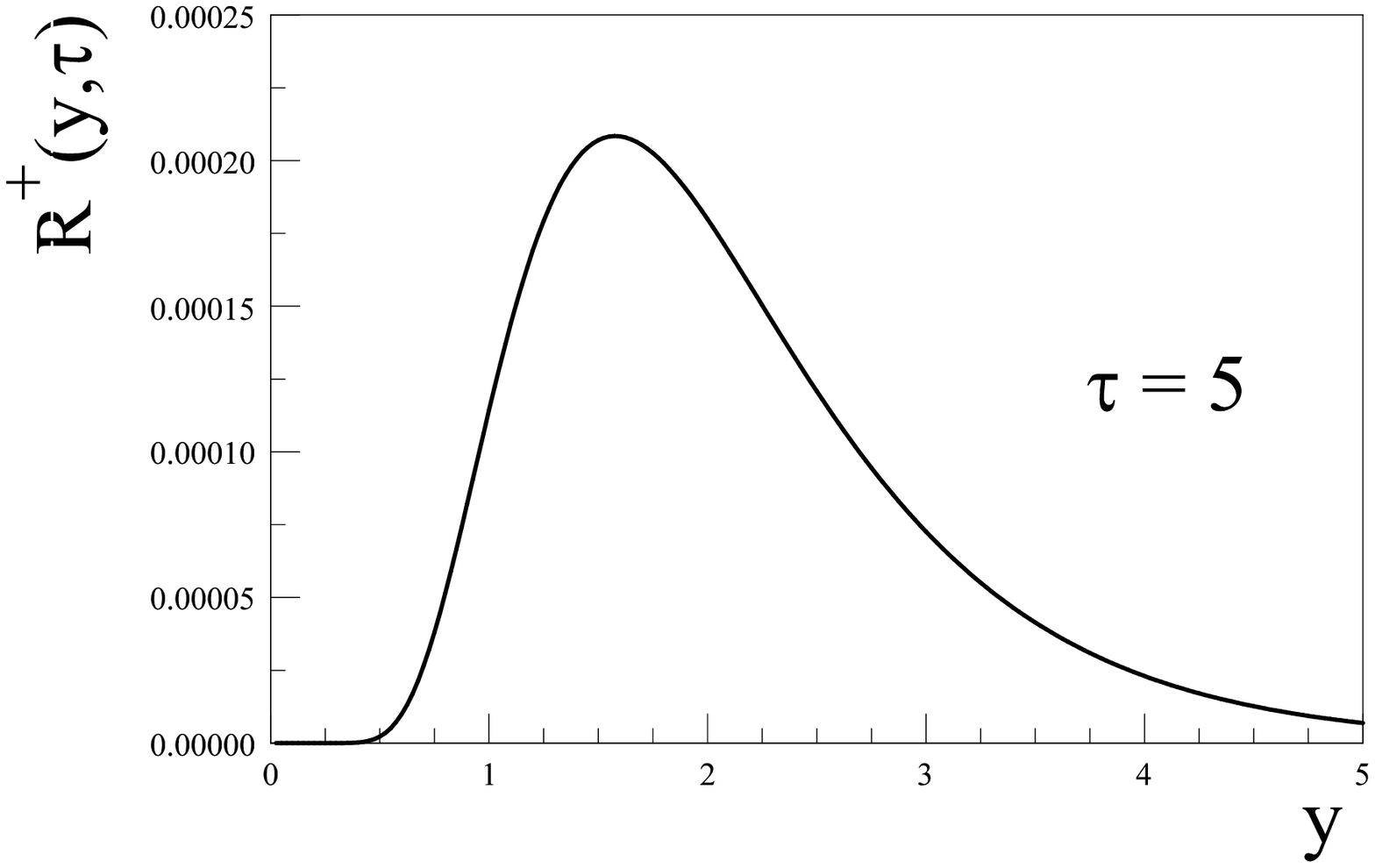}
\caption{  $R^+(\tau,y)$ eqn. (\ref{Rplusdef}) vs. $y=q/T$ for $\tau =  1,3,5$ respectively. }
\label{fig:rplustau}
\end{center}
\end{figure}

\begin{figure}[ht!]
\begin{center}
\includegraphics[height=4in,width=3.2in,keepaspectratio=true]{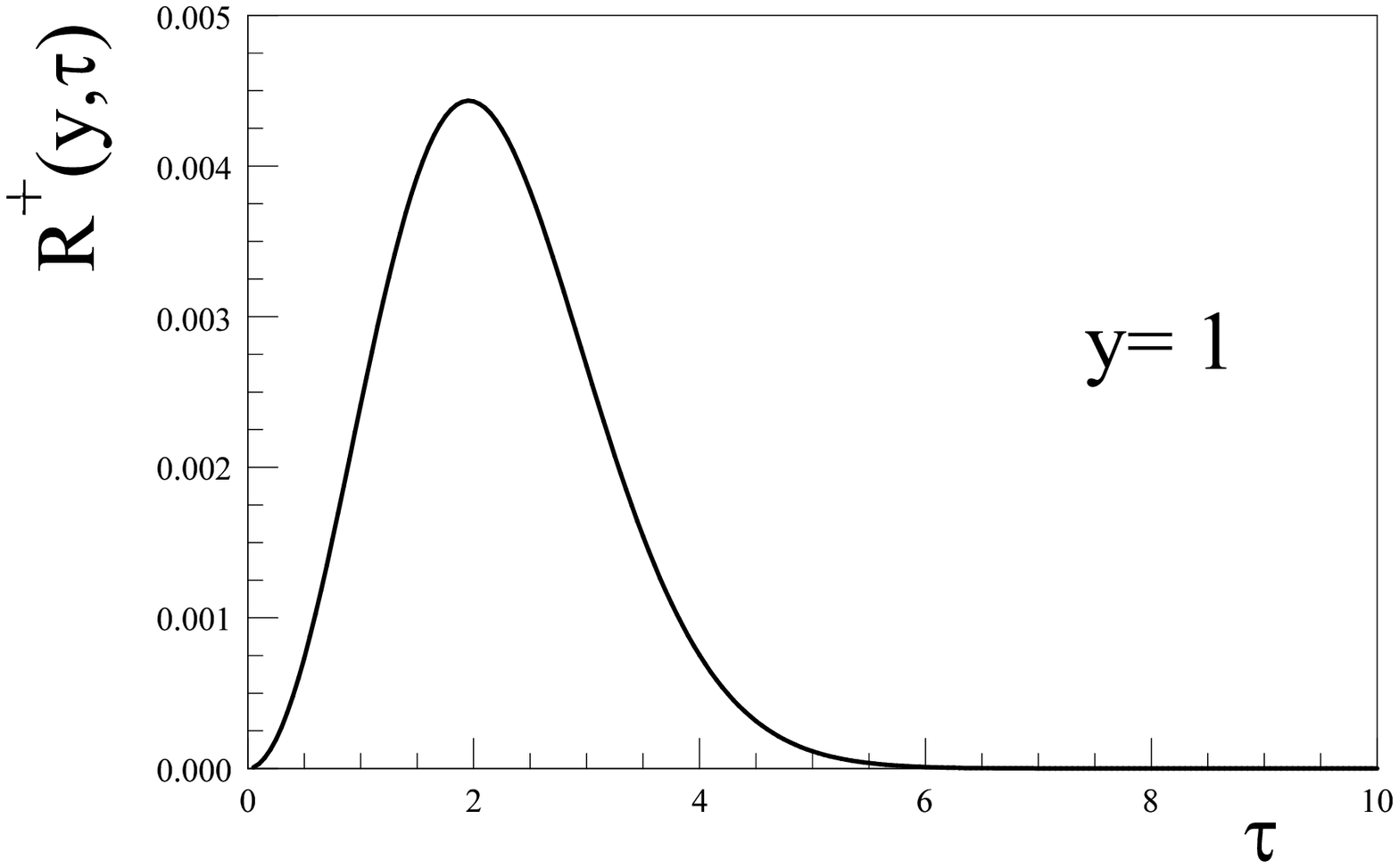}
\includegraphics[height=4in,width=3.2in,keepaspectratio=true]{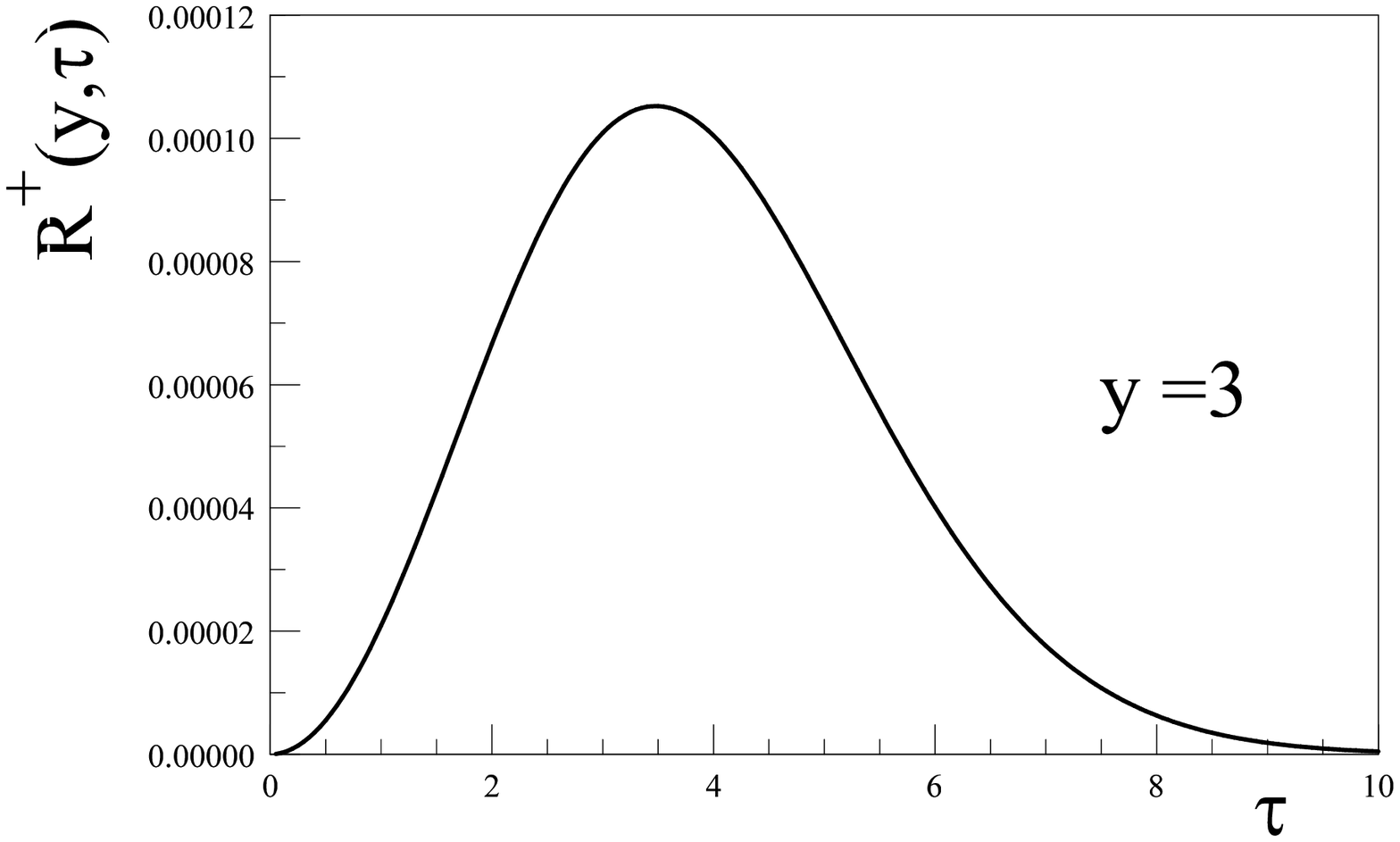}
\includegraphics[height=4in,width=3.2in,keepaspectratio=true]{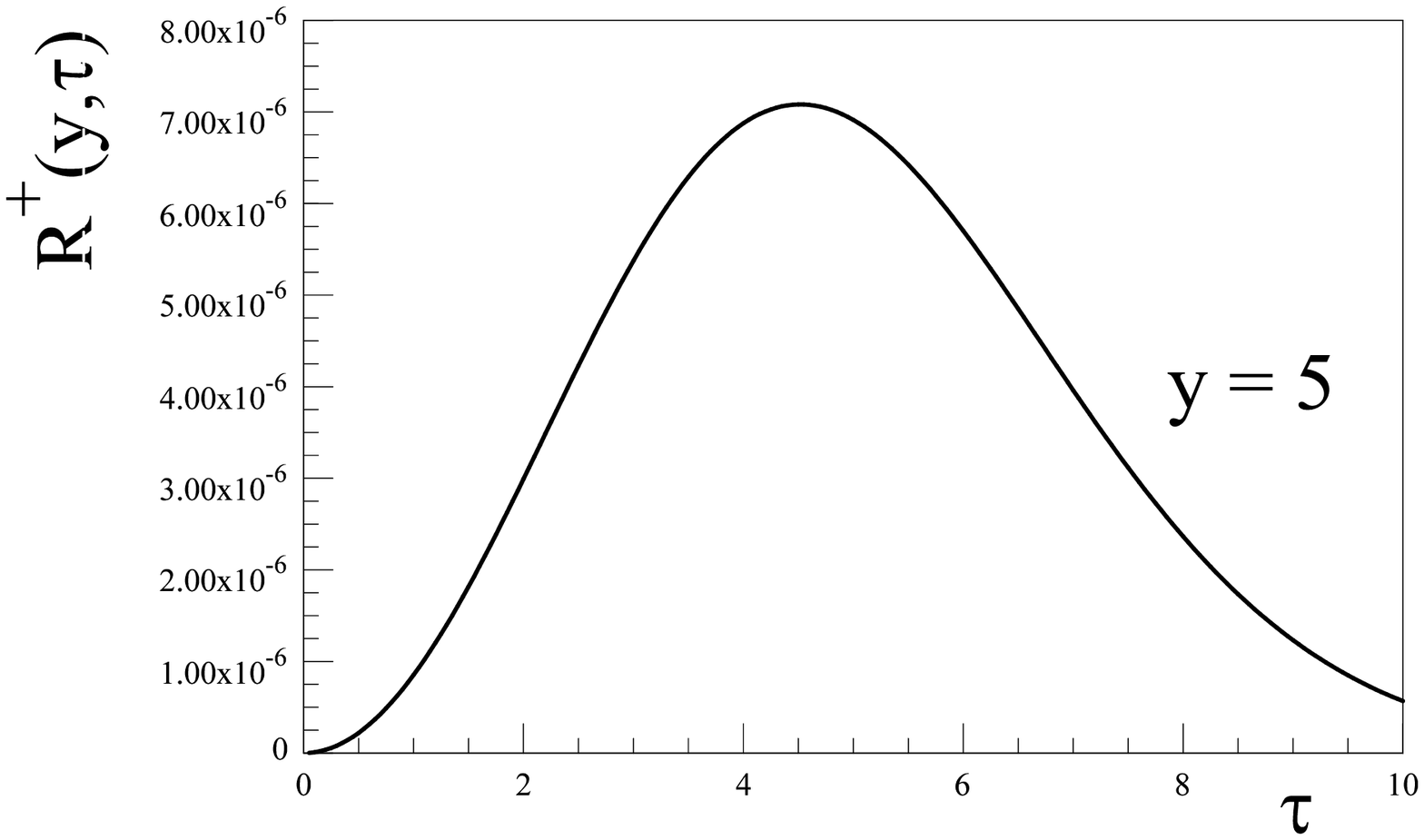}
\caption{  $R^+(y,\tau)$ vs. $\tau$ for $y =  1,3,5$ respectively. }
\label{fig:rplusy}
\end{center}
\end{figure}

\textbf{Positive helicity} for positive helicity we find

\be  \frac{dn^+_2(\tau,y)}{d\tau}\simeq   0.92\,\times 10^{16}\,  \theta^2  \, \Big( \frac{M_s}{M_W}\Big)^2 \,\frac{\tau^2\,I^+(\tau,y)}{y\Big[e^y+1\Big]}\,, \label{poshelqk}  \ee where as discussed above we have used the approximate rate (\ref{prodposapp}). As for the negative helicity case it proves convenient to write  (\ref{poshelqk}) as
\be \frac{dn^+_2(\tau,y)}{d\tau}\simeq   0.92\,\times 10^{16}\,  \theta^2  \, \Big( \frac{M_s}{M_W}\Big)^2 \, R^+(y,\tau)~~;~~ R^+(y,\tau)= \frac{\tau^2\,I^+(\tau,y)}{y\Big[e^y+1\Big]}\,, \label{Rplusdef}\ee where $R^+(\tau,y)$ is read off (\ref{poshelqk}) and does not depend on $M_s$. Fig.(\ref{fig:rplustau}) shows $R^+(\tau,y)$ vs. $y$ for $\tau = 1,3,5$ respectively and fig. (\ref{fig:rplusy}) shows $R^+(\tau,y)$ vs. $\tau$ for $y=1,3,5$ respectively. Together these figures show the ``filling'' of higher momentum modes as the temperature decreases and the freeze-out of the distribution function for different wavevectors. The larger values of $y$ take longer to be  populated and freeze out later, but their contribution is strongly suppressed by the detailed balance factor $1/(e^y +1)$.

\begin{figure}[ht!]
\begin{center}
\includegraphics[height=4in,width=4in,keepaspectratio=true]{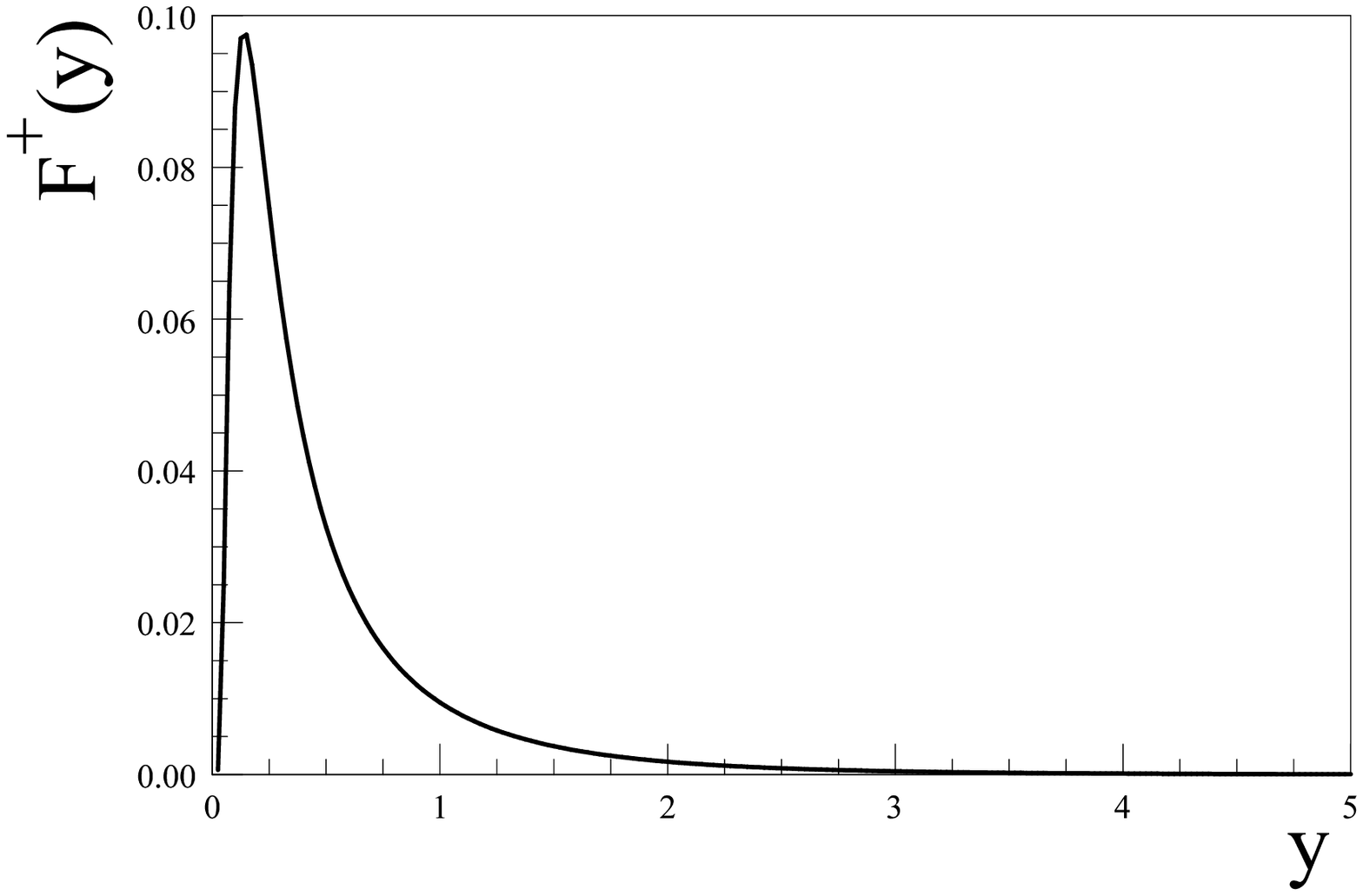}
\caption{ Asymptotic distribution function $F^+(y)$  eqn. (\ref{n2plu}) vs. $y=q/T$ . }
\label{fig:fplus}
\end{center}
\end{figure}

Similarly to the previous case, the asymptotic distribution function is
\be n^+_2(y) =  0.92\,\times 10^{16}\,  \theta^2  \, \Big( \frac{M_s}{M_W}\Big)^2\,F^+(y)~~;~~ F^+(y) = \int_1^\infty R^+(\tau,y)\,d\tau \label{n2plu}\ee
The asymptotic distribution function $F^+_2(y)$ is shown in fig. (\ref{fig:fplus}), it is dominated by the region $0< y \lesssim 8$ with
\be \frac{ \int_{8}^\infty y^2 F^+(y)\, dy}{ \int_0^\infty y^2 F^+(y)\, dy} = 3.3 \times 10^{-3} \,, \label{ratFplu}\ee
and freezes-out at $\tau \simeq 10$ corresponding to a freeze-out temperature $T^+_f \simeq 8 \,\mathrm{GeV}$.

 We note that the distribution function for the positive helicity component is sharply peaked at small momenta $y\approx 0.5$ as compared to that for the negative helicity component which is much broader and peaks at $y\simeq 2.5$, namely the positive helicity component yields a much \emph{colder} distribution  (compare figures (\ref{fig:negheldist} and \ref{fig:fplus})). The reason for this discrepancy is the fact that the production rate for the negative helicity component features a competition between a diminishing damping rate, but an increasing effective mixing angle as $\tau$ increases (temperature decreases). This competition   results in a longer freeze-out time allowing build-up in the population  of larger momentum modes as $\tau$ evolves as discussed above. It is remarkable that the distribution functions $F^\pm_2$ are very similar to those found from pion decay in ref.\cite{mixedlou,lellolightsterile}, the similarity is more striking for $F^-$. The physical reason for this similarity is actually simply the fact that low momentum modes are suppressed since the presence of low momentum sterile neutrinos in the decay of a much more massive particle implies that this ``mother'' particle must be highly boosted in the plasma.

The total population at asymptotically long time is given by
\be \mathcal{N}^+_2 = \frac{1}{2\pi^2} \int_0^\infty n^+_2(y)\,y^2\,dy\,, \label{totpos}\ee which
is determined by the integral
\be \int_0^\infty y^2\,F^+(y)\,dy \equiv N^+ = 0.025 \,.  \label{totingpos}\ee As in the negative helicity case, normalizing to the number density of relativistic neutrino decoupled in equilibrium at the same temperature (\ref{totnueq}) we find
\be \frac{\mathcal{N}^+_2}{\mathcal{N}_\nu} \simeq  2\times 10^4 ~\theta^2 \Big(\frac{M_s}{\mathrm{MeV}} \Big)^2\,. \label{noverspos}\ee Again, we see that for $M_s \simeq \,\mathrm{MeV}$ and $\theta \ll 10^{-2}$ the sterile-like species produced by vector boson decay
\emph{does not thermalize}.

\begin{figure}[ht!]
\begin{center}
\includegraphics[height=4in,width=4in,keepaspectratio=true]{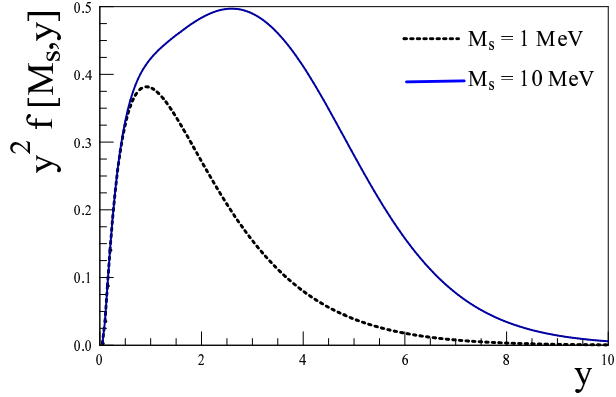}
\caption{ Total distribution function $f(M_s,y)$ eqn. (\ref{n2tota}) multiplied by $y^2$   vs. $y=q/T$ for $M_s = 1;10\,\mathrm{MeV}$. }
\label{fig:hump}
\end{center}
\end{figure}

We have studied the contributions of positive and negative helicity individually to highlight the different distribution functions and dependence on $M_s$, however each is simply a different decay channel for the production of sterile-like eigenstates from the decay of vector bosons and both channels contribute to the total abundance. Hence we combine both channels to give the total density
\be n_2(y) = n^+_2(y)+ n^-_2(y) =  3.6   ~\Bigg( \frac{\theta^2}{10^{-4}} \Bigg)\Big(\frac{M_s}{\mathrm{MeV}}\Big)^2\,f(M_s,y)~~;~~f(M_s,y) = \Bigg[\frac{F^+(y)}{N^+}+ \Big(\frac{M_s}{8.35\,\mathrm{MeV}}\Big)^2\,\frac{F^-(y)}{N^-}\Bigg] \,,\label{n2tota}   \ee where the normalization factors $N^\pm$ are given by (\ref{totint},\ref{totingpos}).  The effective distribution function multiplied by the phase space factor $y^2$ is shown in fig.(\ref{fig:hump}) for   $M_s =1\,\mathrm{MeV}$ where is dominated by the \emph{positive helicity component} and $M_s= 10\,\mathrm{MeV}$ where it is dominated by the \emph{negative helicity component}.

This figure clearly shows the strongly non-thermal total distribution function at freeze out, it also provides a specific example of the ``mixed dark matter'' nature\cite{mixedlou} when several different production channels with different kinematics and effective mixing angles contribute to the production of a sterile-like species. The ``hump'' in $f(M_s,y)$ for $M_s = 10 \,\mathrm{MeV}$ is a result of the competition between the two channels, the negative helicity channel is hotter since its distribution is peaked at larger momenta but becomes dominant at larger $M_s$, whereas that of the positive helicity is colder since it is peaked at lower momenta, but dominates for smaller $M_s$.
Accordingly, we find for the total abundance normalized to that of a single degree of freedom of a massless thermal neutrino, with $\mathcal{N}_2 = \mathcal{N}^+_2 + \mathcal{N}^-_2 $
\be \frac{\mathcal{N}_2}{\mathcal{N}_\nu} \simeq  2   ~\Bigg( \frac{\theta^2}{10^{-4}}\Bigg)~\Bigg(\frac{M_s}{\mathrm{MeV}} \Bigg)^2~\Bigg[1+ \Big(\frac{M_s}{8.35\,\mathrm{MeV}}\Big)^2\Bigg]\,. \label{totalabu}\ee The first term in the bracket is the contribution from the positive helicity states and the second from the negative helicity, both become comparable for $M_s \simeq 8.35\,\mathrm{MeV}$.

\emph{If} the sterile-like  neutrino is \emph{stable}, its comoving number density  would remain constant and upon becoming non-relativistic this species would contribute to dark matter a fraction given by\cite{mixedlou}
\be \mathcal{F}_2 = \frac{\Omega_{\nu_2}\,h^2}{\Omega_{DM}\,h^2} = \frac{M_s}{7.4\,\mathrm{eV}} \,\Big(\frac{g_{\nu_2}}{g_{d}} \Big) \, \int_0^\infty n_2(y)\,y^2\,dy \label{fracneg}\ee where  $g_{\nu_2}$ is the number of degrees of freedom for neutrinos of negative helicity, we will assume Dirac neutrinos in which case $g_{\nu_2}=2$   accounting for   neutrinos and  anti-neutrinos ($g_{\nu_2}=1$ for Majorana neutrinos) and $g_d \simeq 100$ is the number of ultrarelativistic degrees of freedom at decoupling (freeze-out) which occurs at $T_f\simeq 5-8\,\mathrm{GeV}$, yielding

\be  \mathcal{F}_2 = 0.97   \, \Big( \frac{\theta^2}{ 10^{-8}}\Big) \,\Big(\frac{M_s}{\mathrm{MeV}}\Big)^3~\Bigg[1+ \Big(\frac{M_s}{8.35\,\mathrm{MeV}}\Big)^2\Bigg]\,.   \label{fracDM}\ee The   terms in the bracket are the contribution from the positive helicity and negative helicity respectively, the latter dominates for $M_s \gg 8.35\,\mathrm{MeV}$.

 It is clear from this expression that the sterile neutrinos produced by vector boson decay \emph{cannot} yield a substantial   $\simeq \mathrm{KeV}$ warm dark matter component, since the X-ray data constrains such component to the mass range $\simeq \mathrm{few}\,\mathrm{KeV}$   and mixing angle $\theta^2 \lesssim 10^{-10}$\cite{casey,bulbul,boyarsky,kaplinghat2} which according to (\ref{fracDM}) would yield a negligible abundance of such species. However, accelerator and cosmological bounds\cite{kusbounds,gelmini} allow for \emph{heavy} sterile states with masses in the $\mathrm{MeV}$ range and mixing angles $\lesssim  10^{-5}$, in fact these are the bounds used in the recent analysis of $\mathrm{MeV}$ sterile neutrinos as possible solutions to the $^{7}Li$ problem\cite{ishida,salvati} which we discuss  further in section (\ref{sec:discussion}) below.

The results obtained above for the distribution and abundances constitute a \emph{lower bound}, this is because we have neglected any initial population and, as it will be discussed below, we expect other processes to yield sterile-like neutrinos at various stages of the thermal history.

\section{Discussion}\label{sec:discussion}
\subsection{Validity of approximations}
We have implemented several approximations to obtain the above results, which merit a discussion of their validity.
\begin{itemize}
\item {Ultrarelativistic neutrinos: an obvious approximation for the active-like mass eigenstates, for the sterile-like eigenstate this implies $M_s/q\ll 1$. In the expanding cosmology this inequality is in terms of the physical momentum $q_{phys}(t) = q/a(t)$ with $q$ being the comoving momentum. Since $y=q_{phys}(t)/T(t)$ is a constant and $T(t)=M_W/\tau$ hence $q_{phys}(t)= y M_W/\tau$. The inequality must be evaluated at freeze-out, therefore the condition for the validity of the ultrarelativistic limit for sterile-line neutrinos is
    \be \frac{y M_W}{M_s\tau_f} \gg 1,, \label{condiUR}\ee in the range of the distribution function with the largest support. With $\tau_f \simeq 15$ the condition (\ref{condiUR}) applies to $y \gtrsim M_s/6\,\mathrm{GeV}$, which is fulfilled for $y \gtrsim 10^{-3}$ for $M_s\simeq \mathrm{few}\,\mathrm{MeV}$. The distribution function is exponentially suppressed at small $y$ in both cases, therefore $M_s \simeq \mathrm{few}\,\mathrm{MeV}$ fulfills the criterion in almost the whole range but for extremely small values of $y$ which are suppressed both by the distribution and by phase space. }

    \item{$\theta \ll 1$, this approximation was used in expanding the square roots in (\ref{inverseS}) and extracting the dispersion relations (\ref{freqs},\ref{deloste},\ref{deloact}) and  effective mixing angles (\ref{tetaeff}). Assuming $\theta \ll1$, the actual approximation is $(1+(\Delta/\xi))^2+(\gamma/\xi)^2 \gg \theta^2$ or in fact that $\theta_{eff} \leq \theta$ which is fulfilled in both cases. As was discussed above in the negative helicity case   the (MSW) resonance   when $1+ \Delta_2/\xi \simeq 0$ is actually ``screened'' by the term $\gamma_2/\xi = (M^2_W/M^2_s)I^-$ which is actually $\gg 1$ for $M_W/M_s > 10^2$  suppressing the effective mixing angle $\theta_{eff}\ll \theta$. Therefore this approximation is consistent, namely assuming that the vacuum mixing angle is $\ll 1$ implies that the effective mixing angle is also $\ll 1$ and the corrections are such that $\theta/\theta_{eff} \leq 1$ . }

   \item{Active-like neutrinos in LTE: this approximation was invoked to obtain the neutral current contribution to the self-energy with thermalized neutrinos in the intermediate state. For $\theta\ll 1 \rightarrow \theta_{eff} \ll 1$ and $\cos(\theta_{eff}) \simeq 1$ implying that the interaction vertices of active-like neutrinos are the usual standard model ones. This, in turn implies the validity of the usual argument that leads to conclude that active neutrinos are in LTE down to $T\simeq \mathrm{MeV}$ which is much smaller than the freeze-out temperature of sterile-like neutrinos $T_f \simeq \mathrm{few}\,\mathrm{GeV}$. Therefore this approximation is valid all throughout the region of production via vector boson decay and even much lower temperatures down to the usual decoupling temperature $\simeq \mathrm{MeV}$ for weak interactions.  }

   \item{Perturbative expansion: the validity of perturbation theory in describing sterile-like production and freeze-out relies on \emph{two} small dimensionless parameters: $\alpha_w \simeq 1/32$ and $\theta \ll 1$. Inspection of the ratio $\delta \omega^h_2/\omega_2(q)$ (see eqns. (\ref{freqs},\ref{deloste}) clearly shows that this ratio is $\ll 1$ for $\theta_{eff}\ll 1$, $\alpha_w \ll 1$ and the ultrarelativistic limit, confirming the validity of the perturbative expansion for the description of production and freeze-out of sterile-like neutrinos. }

\end{itemize}

\vspace{2mm}

\subsection{Other contributions and higher orders}
Production of sterile-like neutrinos from vector boson decay is the \emph{dominant} process at $T \simeq M_{W,Z}$, and is of order $\alpha_w$ as clearly exhibited by the results obtained above. This is the leading contribution to the self-energy in this temperature range, namely the one-loop contributions depicted in fig. (\ref{fig:self}). In the same temperature regime there are several other processes that contribute to the imaginary part of the self-energy, hence to the production rate: heavy quark and lepton decays   via charged   current interactions, $q \rightarrow q\, \overline{\nu}\,\nu$, $\tau,\mu \rightarrow \overline{\nu}\,\nu \, l$  charged lepton annihilation $l^+ l^- \rightarrow \overline{\nu}\nu$  (via neutral currents) and several other processes (for a more detailed discussion see\cite{mixedlou}). These processes contribute to the imaginary part of the self-energy at \emph{two loops}, therefore are of order $\alpha^2_w$. Furthermore at $T \ll M_{W,Z}$ these are further suppressed by a vector boson propagator, therefore their contribution to the imaginary part is generically $C G^2_F T^4 q$ typically with $C \simeq 1 $. As the temperature diminishes through the cosmological expansion, the damping rate from vector boson decay will become of the same order as the contribution to the imaginary part from these higher order processes which must  then be taken into account if the available energy is larger than the threshold for sterile-like production. Therefore, the results obtained in the previous section provide a \emph{lower bound} to the abundance of sterile-like neutrinos, as processes that are of higher order but dominate at lower temperatures increase the abundance.

The index of refraction, namely the real part of the self-energy is dominated by the one-loop result, which for low temperatures (large $\tau$) is given by the low temperature limits (\ref{reStotneghel},\ref{reStotposhel}). This observation is important: at $T\ll M_{W}$ on dimensional grounds the two loop processes yield real and imaginary parts of the  self-energy $\propto G^2_F T^4 q$ since this limit is well describe by the local Fermi theory, therefore compared to the low temperature limit of the \emph{one loop} contribution (\ref{reStotneghel},\ref{reStotposhel}) the two-loop contribution to the real part  is suppressed by a power of $\alpha_w$.  Therefore for $T \ll M_W$ the leading contribution to real part or index of refraction is given by the one loop results (\ref{reStotneghel},\ref{reStotposhel}), whereas the imaginary part (damping rate) is determined by the two loop diagrams and are $\propto G^2_F T^4 q$ the proportionality constant determined by the nature and number of degrees of freedom (leptons, quarks) that enter in the processes. Therefore, in principle a complete description of production and freeze out should include all possible processes at one and two loops in the self-energy. The real part is dominated by the one loop term, but the imaginary part will receive contributions from both one and two loops, the relevance of each will depend on the temperature regime. For the mixing angle, both the real and imaginary part (damping rate) are needed, however the imaginary part is of the same order than the real part only for the one-loop contribution, namely at temperatures of the order of $M_{W,Z}$ (or larger), however, at much lower temperatures, the corrections to the mixing angle are dominated by one loop contribution to  the real part given by (\ref{deltotneghel},\ref{deltotposghel})  and the two loop contributions to both the real and imaginary part can be safely neglected  in agreement with the results of ref.\cite{notzold,dolgovhansen,dolgovenqvist}. Ref.\cite{laine2} discusses further
contributions at $T >> M_{W}$ which merit further study but are outside the scope of this article.

 In the temperature regime considered here $T \simeq M_{W,Z}$, self-energy corrections to the propagators of massive vector bosons (quark  and  lepton loops) are perturbative and suppressed by another power of $\alpha_w$. Self-energy corrections to the charged lepton propagators are the same as the hard thermal loop corrections. In section (\ref{sec:wdecay}) we discussed the possible influence of hard thermal loop corrections to the lepton propagators arguing that these are subleading for kinematic reasons. The numerical results obtained in the previous sections confirm this argument: the production rate and final distribution functions receive the largest support in the region of momenta $q/T \lesssim 2-3$. In this region of phase space, kinematically the charged leptons produced via vector boson decay feature momenta $q \simeq T$ and in this region of momenta hard thermal loop corrections can be safely neglected. However, the large momentum tail of the sterile neutrino distribution, although strongly suppressed as shown numerically, will be affected by these corrections. A study of these is beyond the scope of this article and is relegated to a future study.

\vspace{2mm}

\subsection{Lifetime constraints:} Massive sterile-like neutrinos can decay in various leptonic  channels\cite{shrock,barger}. Consider the simpler case of one sterile-like $\nu_2$ and one active-like $\nu_1$ neutrino with $\theta \ll 1$, the  charged current channel  $\nu_2 \rightarrow e^+e^-\nu_1$, is available for $M_2\simeq M_s > 1\,\mathrm{MeV}$ and the ``invisible'' neutral current $\nu_2 \rightarrow 3\nu_1$ channel which is available for any $M_s$ of cosmological relevance for WDM or CDM, and the radiative channel $\nu_2 \rightarrow \gamma\,\nu_1$ which is suppressed by one power of $\alpha_{em}$. The decay widths for these channels have been obtained\cite{shrock,barger,gorbunov}
\be
\Gamma(\nu_2 \rightarrow e^+ e^- \nu_1) \simeq  {3.5}\times {10^{-5}} \, {\theta^2} \, \left(\frac{M_{s}}{MeV}\right)^5  ~K\Bigg[\frac{m^2_e}{M^2_s}\Bigg]\,\Big(\frac{1}{s}\Big)\label{nuhdec1}
\ee where the function $K \rightarrow 0$ for $M_s \rightarrow 2 m_e$ and $K \rightarrow 1$ for $M_s \gg m_e$\cite{shrock}. For other leptonic channels similar expressions were obtained in ref.\cite{gorbunov}.

 The  decay rate  into active-like neutrinos mediated by neutral currents (not GIM (Glashow-Iliopoulos-Maiani)) suppressed with sterile-like heavy neutrinos) is given by   (see \cite{shrock,barger})
  \be
\Gamma(\nu_2 \rightarrow 3\nu_1  ) \simeq     {3.5}\times {10^{-5}} \, {\theta^2}\,  \left(\frac{M_{s}}{MeV}\right)^5  \,\Big(\frac{1}{s}\Big)\label{nuh3nus}  \ee and the radiative decay width\cite{palwolf,barger}
\be \Gamma(\nu_2 \rightarrow \gamma \nu_1) \simeq   10^{-7} \theta^2 \,\Big(\frac{M_s}{\mathrm{MeV}}\Big)^5 \, \,\Big(\frac{1}{s}\Big)\,.\label{gamahfot}\ee

Recent results for a lower bound  on the lifetime of Dark Matter yields $t_b \simeq 160\,\mathrm{Gyr}$\cite{audren}, a similar bound but in terms of  the fraction of cold dark matter is given in ref.\cite{poul}.  Adding both leptonic channels assuming that $M_s \gtrsim \mathrm{MeV}$ and taking both of the same order, the condition that the sterile species would be a suitable dark matter candidate implies that its lifetime is longer than or equal  to this lower bound, namely $\Gamma_{tot}\, t_b  \leq 1$, implying that
  \be {\theta^2}\,  \left(\frac{M_{s}}{MeV}\right)^5 \lesssim 10^{-14}\,.  \label{lifecon}\ee
  Combining this bound with the fractional abundance (\ref{fracDM})   we find that
    \be \frac{\mathcal{F}_2~\Big(\frac{M_s}{\mathrm{MeV}}\Big)^2}{\Big[1+ \Big(\frac{M_s}{8.35\,\mathrm{MeV}}\Big)^2\Big]} \lesssim 10^{-6} \label{stab}\ee
     which could yield $\mathcal{F}_2 \simeq 1$ for $M_s \simeq \mathrm{few}\,\mathrm{KeV}$, which, however would require a very large mixing angle $\theta \simeq 10^{-2}$ which is ruled out by cosmological X-ray bounds\cite{casey,bulbul,boyarsky,kaplinghat2}. Hence, we conclude that sterile-like neutrinos produced via vector boson decay cannot be suitable dark matter candidates.

     However, if there is a hierarchy of sterile-like neutrinos  heavy neutrinos with $M_s \, \mathrm{few}\, \mathrm{MeV}$ and mixing angles $\theta^2 \gg 10^{-13}$, these may decay into lighter $\simeq \mathrm{KeV}$ sterile-like states that could contribute to the dark matter abundance. This possibility of cascade decay merits further study and is clearly beyond the scope of this article.

\vspace{2mm}

\subsection{Comparison to other results}\label{sub:compa}
The expressions for the quantum kinetic equation (\ref{qukinfin}) and the effective mixing angle in the medium (\ref{tetaeff}) that we obtained   are  \emph{exact} to all orders in standard model couplings and to leading order in $\theta^2\ll 1$. In ref.\cite{volkas,abacir} an expression for the effective mixing angle that includes both the real part (index of refraction) and imaginary part (damping rate) of the self-energy (from standard model interactions) has been proposed that seems to be valid for arbitrary vacuum mixing angle. Our result is only valid for $\theta \ll1$ where we can extract unambiguously the mixing angle from the position of the complex poles in the propagator.   It is not straightforward to define or extract a real mixing angle in the case of large $\theta$, the subtleties   are discussed in ref.\cite{boywu}.   The final form of the quantum kinetic equation (\ref{qukinfin}) is similar to that used in \cite{dolgovhansen,dodwid}, although in these articles the mixing angle only includes the index of refraction (real part), valid in the temperature regime of interest in those articles. Crucially, our analysis   shows the    importance of the positive helicity component:  at high temperature the in-medium suppression of the mixing angle for negative helicity is much larger than that for positive helicity since the latter interacts with the medium with a coupling that is helicity suppressed. However the contributions from the in-medium suppressed negative helicity and the helicity suppressed positive helicity may be competitive within a range of masses and temperatures. This is an aspect that has not been discussed previously.

Ref.\cite{laine2} considers the production
of $\simeq \mathrm{GeV}$ sterile neutrinos in the
temperature range $5\,\mathrm{GeV} < T < 160 \,\mathrm{GeV}$,
focusing on the washout of lepton densities, our results are in broad agreement with those of this reference for the processes associated with vector boson decay for negative helicity states in the regions where a comparison is meaningful.
However, in contrast, we focus on the mass scale $\simeq \,\mathrm{MeV}$ motivated by the possible solution to the  $^{7}Li$ problem, and concentrate on the role which helicity plays
in the effective mixing angles in the medium and   the damping
and production rates. Furthermore, we  study the full dynamics of production and freeze-out including cosmological expansion for both helicity states obtaining the abundance at freeze-out, and establishing the freeze-out temperature and time scales for each helicity.

The role that helicity plays is an immediate consequence
of the  $V-A$ coupling in the standard model. The interaction of positive helicity states with the medium
is helicity suppressed, relative to those with negative helicity.
As we have shown,
the  mixing angle for positive helicity states is hardly modified
by their  interaction with the medium, whereas that for negative
helicity states is strongly suppressed, \emph{positive helicity} states dominate the production in a wide range of masses. Furthermore,
we find an MSW resonance for negative helicity states
\emph{in absence of a lepton asymmetry}, which is screened by the
damping rate and does not enhance the production. We study the full
dynamics of production with cosmological expansion analyzing the
freeze-out of both helicity states, establishing the freeze-out
temperature for each channel and obtaining the frozen distributions.
The final distribution  exhibits the contribution from both helicity channels,
one colder than the other and with both distinctly non-thermal.
We also analyze the various accelerator, lifetime  and cosmological
constraints both for dark matter as well as the possibility that
$\simeq \mathrm{MeV}$ sterile neutrinos might solve the $^7Li$ problem
(see the discussion below).

\vspace{2mm}

\subsection{Thermalization?}\label{sub:thermalization}

The result for the total abundance of sterile neutrinos produced via vector boson decay, compared to that of a thermal species given by  equation (\ref{totalabu})  shows that for $\theta \ll 10^{-2}, M_s \,\mathrm{few}\,\mathrm{MeV}$ sterile neutrinos produced via this mechanism do \emph{not} thermalize.
If the lifetime of the sterile-like neutrino is (much) shorter than the age of the Universe, it means that at some time in the past history of the Universe  the \emph{rate} $dn_s(t)/dt <0$ since   if such species is present today its population is decaying in time. We have argued that the quantum kinetic equation (\ref{qukinfin}) is \emph{exact} to all orders in standard model couplings to leading order in $\theta^2 \ll 1$. In fact as per the discussion leading up to (\ref{qukinfin}) the production term is completely determined by the damping rate, the mixing angle (in the combination $\Gamma^h_2$) and detailed balance, which if the sterile-like mass eigenstate is \emph{not} relativistic entails that the correct form of the quantum kinetic equation is
 \be \frac{d n^h_2(q;t)}{dt} = \Gamma^h_2(q)  \Big[n_{LTE}(E) -n^h_2(q;t)\Big]\,, \label{qukinLTE} \ee where
 \be n_{LTE}(E) = \frac{1}{e^{\frac{E}{T}}+1} ~~;~~ \frac{E}{T} = \sqrt{y^2+\frac{M^2_2}{T^2}}\,  \label{nlte}\ee and $\Gamma^h_2(q)$ are the damping rates in terms of the   imaginary part of the self-energy (\ref{gamasterfin}).  In Minkowski space time, however small the mixing angle (hence $\Gamma_2$) sterile-like mass eigenstates will \emph{always} thermalize, the longer the thermalization time scale the smaller $\Gamma_2$. With cosmological expansion freeze out occurs when $\Gamma_2 /H(t) \ll 1$ (for a more detailed discussion see ref.\cite{mixedlou}). In the original form (gain-loss) of the kinetic equation (\ref{quakin}) the gain term always involves the annihilation of one or several species (vector bosons, leptons,quarks) that by assumption are in (LTE) in the plasma, as the Universe expands and cools the abundances of these species diminishes and the  gain  contributions diminish accordingly. The loss terms that involve the annihilation of one or more species in LTE also diminish under cosmological expansion, however, if the sterile neutrino can decay into other species, this decay contribution only entails the \emph{creation} of the decay products, and these contributions \emph{do not} vanish as the temperature diminishes. Three processes that contribute to the loss term and survive in the low temperature limit are precisely the decay channels (\ref{nuhdec1},\ref{nuh3nus},\ref{gamahfot}). Therefore if the sterile-like mass eigenstate decays with a lifetime smaller than the age of the Universe, these loss terms dominate the quantum kinetic equation at some late time and the rate becomes negative before today. Then the form (\ref{qukinLTE})  implies that at some time in the past the sterile-like neutrino has \emph{thermalized}, since the production term (gain) dominates initially but the decay (loss) dominates at late times, the rate must have passed through zero in between, namely the distribution reached (LTE) and started to decay after this point. This discussion becomes relevant with the possibility that $M_s ~\mathrm{few}\,\mathrm{MeV}$ and lifetime $\simeq 10^5 \,(s)$ could provide a solution to the $^{7}Li$ problem as suggested in refs.\cite{ishida,poulin,salvati}, as discussed below.

 In ref.\cite{hernandez} a study of low scale type I seesaw models suggests that processes that are dominant at much lower temperature result in the thermalization of heavy states. Therefore, it is possible that other mechanisms which dominate at lower temperature and not considered in our study may lead to the thermalization of the heavy sterile states produced via vector boson decay. An important aspect to be understood is whether thermalization occurs before or after the time scale required to solve the $^{7} Li $ problem\cite{salvati} as this will impact the destruction of $^{7}Be$ in the nucleosynthesis chain.

 The study of such possibility is definitely important but  beyond the scope of this article.

\subsection{Solution to the $^{7}Li$ problem?:} \label{sub:Li7}
In ref.\cite{salvati} the authors performed an exhaustive analysis of the parameter space within which the decay of sterile neutrinos of $M_s \simeq \mathrm{few}\,\mathrm{MeV}$ could yield a solution of the $^{7}Li$ problem as previously advocated in refs.\cite{ishida,poulin}. The analysis of ref.\cite{salvati} included the most recent data on (CMB) anisotropies and concluded that a heavy sterile neutrino with $M_s \simeq 4.35\,\mathrm{MeV}$ and lifetime $\Gamma^{-1} \simeq 1.8\,\times \,10^5\,(s)$ would provide a suitable solution. However, the parameter space also bounds the ratio\footnote{This reference actually bounds $\mathcal{N}_s/\mathcal{N}_{cmb}$ which differs by a factor 3/4.} $\mathcal{N}_s/\mathcal{N}_\nu \simeq 10^{-4}$ and the mixing angle $\theta^2 \simeq 10^{-4}$. These values are in significant tension both with the results that we obtained above and the bounds of ref.\cite{aaron}. In particular with $M_s \,\mathrm{few}\,\mathrm{MeV}$ and such large mixing angle our result (\ref{totalabu}) indicates that $\mathcal{N}_s/\mathcal{N}_\nu \simeq 1$ suggesting full thermalization, furthermore, as is discussed above, our results provide a \emph{lower bound} for the abundance of sterile-like heavy neutrinos. Both the region of abundance and   mixing angles found in ref.\cite{salvati} seem in strong tension with the bounds in\cite{aaron}, both caveats are recognized in\cite{salvati} which suggests, as possible alternative, a low reheating temperature\cite{gelmini}. Of course our results rely on (LTE) at the electroweak scale, therefore they are not applicable to such scenario. Hence, although the production mechanism of sterile-like neutrinos studied in this article which is the leading order in standard model couplings and provides a  lower bound to the abundance, offers a compelling mechanism for production of heavy sterile-like neutrinos with the possibility to solve the $^{7}Li$ problem, significant tension arises between the parameter range of the solution established in ref.\cite{salvati}, our result as a lower bound on the abundance  and the cosmological bounds obtained in ref.\cite{aaron}. The resolution of this tension merits a deeper study, well beyond the scope of this article.

\vspace{2mm}

\subsection{WDM from cascade decay:}\label{sub:cascade}
The analysis of the solution of the $^{7}Li$ problem suggested in in refs.\cite{ishida,poulin,salvati} is a specific example of a cascade decay mechanism: heavy ($M_s \sim \,\mathrm{few}\,\mathrm{MeV}$)  sterile-like neutrinos produced at a (high) scale that eventually decay into several channels with the daughter particles influencing  important physical processes during  cosmological expansion. If there is a hierarchy of sterile-like massive neutrinos that include $\mathrm{MeV}$   and $\mathrm{KeV}$ scales, the heavier mass states may be produced at a high temperature, such as explored in this article, and the decay of this heavy state on a time scale $\simeq 10^5\,s$ to solve the $^{7}Li$ problem (if the caveats discussed above can be overcome), can also lead to the production of the lighter mass states that can be suitable WDM candidates. While this lifetime is interesting within the context of the $^{7}Li$ problem, a heavy neutrino with $M_s \simeq \mathrm{MeV}$ and $\theta\simeq 10^{-7}$  would feature a lifetime $\simeq 10^{12}\,(s)$ therefore decaying into a WDM candidate just after matter radiation equality.   This possibility emerges naturally  by writing the weak interaction vertices in mass eigenstates, then the process $\nu_2 \rightarrow 3\nu_1$ yields a contribution $\nu_2 \rightarrow 2\nu_1\nu_m$ with $\nu_2,\nu_m$ the heavier ($\simeq \mathrm{MeV}$) and lighter ($\simeq \mathrm{KeV}$) mass eigenstates respectively. The branching ratio for such process  is $\propto \theta^2_m$ where $\theta^2_m$ is the mixing angle of the active (flavor) neutrinos with the sterile-like lighter neutrino $\nu_m$. This  mechanism of production of WDM candidates is a tantalizing possibility that would be a natural scenario in extensions beyond the standard model that posit the existence of  several sterile neutrinos   merits further study  clearly beyond the scope of this article.

\vspace{2mm}

\section{Summary of results, conclusions, further questions}

Our goals in this article are two-fold: i) to obtain the general form of the quantum kinetic equations and effective mixing angles in the medium to describe production and freeze-out of sterile-like (mass eigenstates) neutrinos in a broad range of temperature and under a minimal set of assumptions. Our study departs from previous ones (see the recent review\cite{revster}) in several important aspects: we focus on the different helicity contributions, and we systematically include the absorptive part of the self-energy in the in-medium modification of the mixing angle.  ii) To apply the kinetic equations thus found to study the production to leading order in standard model couplings from vector boson decay at $T\simeq M_W$.

We obtained the effective mixing angles in the medium directly from the equations of motion in the case of mixing of one sterile with one active neutrino via a see-saw mass matrix with standard model interactions for the active (flavor) neutrino valid when the vacuum mixing angle $\theta \ll 1$ but  to \emph{all orders in standard model couplings}. Assuming that all standard model degrees of freedom are in (LTE) in the relevant temperature range we obtained the quantum kinetic equation that describes the production, evolution and freeze-out of sterile-like mass eigenstates. The mixing angles in the medium and the production rate are determined by the real and imaginary parts of the self-energy on the mass shell of the sterile-like mass eigenstate, and depend on helicity. The full quantum kinetic equation to leading order in $\theta \ll 1$ is
\be \frac{d n^h_2(q;t)}{dt} = \Gamma^h_2(q)  \Big[n_{LTE}(q) -n^h_2(q;t)\Big]\,, \nonumber \ee where $h=\pm$ correspond to helicity states and  $\Gamma^\mp_2(q)$ are given by (\ref{gamasterfin}) with (\ref{gamneghel},\ref{gamneghel}), and $n_{LTE}$ is the Fermi-Dirac distribution function in (LTE). The full expression for the mixing angles in the medium, valid to all orders in standard model couplings and to leading order in $\theta \ll 1$ is given   in the relativistic limit by
\be \theta^h_{eff}(q) = \frac{\theta}{\Bigg[\Big( 1+\frac{\Delta^h(q)}{\xi}\Big)^2+\Big(\frac{\gamma^h(q)}{\xi} \Big)^2\Bigg]^{1/2}}\,, \nonumber \ee where $\Delta,\gamma,\xi$ are given by (\ref{gamneghel}-\ref{delposhel},\ref{chi}) respectively in terms of the real ($\Delta$) and imaginary ($\gamma$) part of the \emph{active neutrino self-energy on the mass shell of the sterile-like eigenstate}.

We implemented the quantum kinetic equation to obtain the production of sterile-like neutrinos from vector boson decay at $T \simeq M_W$ including  cosmological expansion. For negative helicity neutrinos (and positive helicity anti-neutrinos) the effective mixing angle   is strongly suppressed by the medium, however for positive helicity neutrinos (and negative helicity anti-neutrinos) the medium corrections are negligible because the interaction with the medium is helicity suppressed. We find that there is a region of masses for which the production of both species is comparable.

 It is noteworthy that the mixing angle for negative helicity neutrinos features an MSW resonance \emph{in absence of lepton asymmetry}, which, however, is screened by the imaginary part of the self-energy. This is an important aspect that has not been previously addressed before: the absorptive (imaginary) part of the self-energy \emph{also} contributes to the mixing angle in the medium.

 Negative helicity neutrinos freeze-out at $T^-_f \simeq 5\,\mathrm{GeV}$ with a broader distribution as a consequence of a competition between a diminishing damping rate $\gamma$ and an \emph{increasing} effective mixing angle as temperature diminishes. Positive helicity neutrinos  freeze-out temperature is $T^+_f \simeq 8\,\mathrm{GeV}$ with a distribution that peaks at much smaller momenta, describing a  colder species. Accounting for both channels we find that the distribution function of sterile-like neutrinos of mass $M_2 \simeq M_s$ is given by
\be n_2(y) =    3.6   ~\Bigg( \frac{\theta^2}{10^{-4}} \Bigg)\Big(\frac{M_s}{\mathrm{MeV}}\Big)^2\,f(M_s,y)  \,,\nonumber   \ee where  $y = q/T$ and $y^2 f(M_s,y)$ is strongly non-thermal and is displayed in fig.(\ref{fig:hump}) revealing the competition between the colder (positive helicity) and hotter (negative helicity) components. The total abundance normalized to that of one relativistic degree of freedom in thermal equilibrium ($\mathcal{N_\nu}$) is
\be \frac{\mathcal{N}_2}{\mathcal{N}_\nu} \simeq  2   ~\Bigg( \frac{\theta^2}{10^{-4}}\Bigg)~\Bigg(\frac{M_s}{\mathrm{MeV}} \Bigg)^2~\Bigg[1+ \Big(\frac{M_s}{8.35\,\mathrm{MeV}}\Big)^2\Bigg]\,. \label{totalabufin} \ee The first term in the bracket is the contribution from the positive helicity states and the second from the negative helicity, both become comparable for $M_s \simeq 8.35\,\mathrm{MeV}$. We argue that this expression is a \emph{lower bound} on the abundance of sterile-like neutrinos.

 The fractional abundance of dark matter contributed by both helicity components is given by (\ref{fracDM}). Constraints from X-ray data on masses and mixing angles suggest that sterile-like neutrinos produced by vector boson decay  \emph{cannot} yield a substantial $M_s \simeq \mathrm{KeV}$ warm dark matter component. However, this production mechanism yield a substantial abundance of $M_s \simeq \,\mathrm{MeV}$ \emph{heavy} sterile-like neutrinos with $\theta^2 < 10^{-4}$ consistent with accelerator constraints. Therefore this production mechanism may yield the heavy neutrinos recently invoked to solve the $^{7}Li$ problem\cite{ishida,poulin,salvati}. However, the parameter range determined in \cite{salvati} also bounds $\mathcal{N}/\mathcal{N_\nu} \simeq 10^{-4}$ which is incompatible with the result (\ref{totalabufin}) for the range of mass and mixing angles reported in this reference, and is also in conflict with recent bounds reported in\cite{aaron}. The possibility that heavy  $\simeq \mathrm{MeV}$ sterile-like neutrinos   decaying   after BBN injecting energy in the medium providing a solution of the $^{7}Li$ problem as suggested also in refs.\cite{ishida,poulin,salvati} merits a deeper study both of the production mechanism as well as the cosmological impact of this heavy neutrino species.

 \vspace{2mm}

 \textbf{Further questions:}
 The study of production at $T\gg M_{W,Z}$ requires a deeper understanding of the finite temperature corrections to the dispersion relations of the vector bosons near the electroweak crossover regime, a worthy study  which is beyond the scope of this article.
 We have suggested several other processes that contribute to the production throughout the thermal history of the Universe, while these are higher order (two loops) processes   they may comparable to the leading order processes or even dominate at temperature $T \ll M_{W}$. However the medium corrections to the mixing angles are completely determined by the one loop contribution and the effective mixing angle is given by eqn. (\ref{louTtetaf}). Further study of these processes is clearly warranted, they can be competitive near the freeze-out temperature of the leading one-loop contribution for heavy sterile neutrinos and, crucially contribute to the production of lighter mass eigenstates. We have also argued that  sterile neutrinos with lifetimes shorter than the age of the Universe that are decaying today must necessarily have thermalized at some time in the past. Studying this thermalization process is of fundamental interest since most of the calculations of production neglect this possibility by neglecting the loss term in the kinetic equation, yet thermalization of heavy sterile neutrinos may   have  important cosmological consequence for the expansion history. In ref.\cite{hernandez} a study of low scale type I seesaw models suggest that heavy sterile neutrinos do thermalize through collision processes dominant at much lower temperatures motivating further studies of thermalization.
 We have also suggested that several compelling extensions beyond the standard model posit a hierarchy of sterile neutrino masses, and this opens the possibility that heavier sterile-states may be produced at high temperature, as analyzed here, and decay well after BBN or near the time of matter-radiation equality into lighter sterile states that may be suitable WDM candidates. This mechanism of cascade decay, which is fundamentally similar to that advocated for the solution of the $^{7}Li$ problem as an energy injection mechanism, is worthy of  study.

\acknowledgments The authors thank Mikko Laine and Goran Senjanovic for bringing their work to the author's attention. D. B. and L.L. gratefully acknowledge support
from NSF through grant PHY-1506912. L. L. and R. D. P.
thank the U.S. Department of Energy for their support
under contract DE-SC0012704.
L. L. is partially supported by the U.S. Department of Energy,
Office of Science, Office of Workforce Development for
Teachers and Scientists, Office of Science Graduate Student
Research (SCGSR) program. The SCGSR program is administered by the
Oak Ridge Institute for Science and Education for the DOE under
contract number DE-AC05-06OR23100. L.L. thanks the theory group of
Brookhaven National Laboratory for their hospitality under this program.

\appendix

\section{Spectral density.}\label{app:specdens}
The spectral densities are obtained for a generic vector boson mass $M$, with a straightforward application for either charged or neutral current  cases.

We need to identify the regions in which the product of delta functions in (\ref{imapmbfin}) with (\ref{rhobb})  have support in the interval  $W^- \leq W \leq W^+$.

Using the identities:
\be n_f(-k) = 1-n_f(k) ~~;~~ N_b(-p^0) = -(1+N_b(p^0)) \label{identists}\ee the product
\be     \Big[\delta(k_0-k)-\delta(k_0+k) \Big] ~ \Big[\delta(q_0-k^0-W_p)-\delta(q_0-k^0+W_p)\Big]~\Big[1-n_{f}(k_0)+N_b(p_0)\Big]~~;~~p^0 = q^0 - k^0 \ee  is gathered into four different terms, keeping the finite temperature contributions only these are:
\be \mathbf{1}): ~~\delta(k^0-k)\delta(q^0-k-W)\Big[-n_f(k)+N_b(q^0-k)\Big]\,. \label{one}\ee This contribution describes the \emph{decay} of the sterile neutrino and its inverse process $\nu_s  \leftrightarrow l W $. The delta function $\delta(q^0-k-W)$ has support for $W^- \leq q^0-k \leq W^+$ as a function of $k$ this constraint implies
\be k^- \leq k \leq k^+ ~~;~~ k^{\pm} = \frac{(q^0)^2-q^2-M^2}{2(q^0 \mp q)} ~~\mathrm{for}~~q^0 \geq \sqrt{q^2+M^2} \label{constone}\ee

\vspace{2mm}

\be \mathbf{2}): ~~ \delta(k^0+k)\delta(q^0+k+W)\Big[1-n_f(-k)+N_b(q^0+k)\Big]\,,\label{twoa}  \ee
For this term it must be that $q^0 +k <0$ since $W^-\leq W \leq W^+$ and because $k>0$, it follows that $q^0 <0$. Therefore using the identity (\ref{identists}) and keeping solely the finite temperature contributions this term yields
\be \mathbf{2}): ~~ -\delta(k^0+k)\delta(|q^0|-k-W)\Big[ -n_f(k)+N_b(|q^0|-k)\Big]~~;~~ |q^0|-k >0 \,,\label{two} \ee which is similar to the previous case. The region of support is $q^0 <0$ and for $k$ is
\be k^- \leq k \leq k^+ ~~;~~ k^{\pm} = \frac{(q^0)^2-q^2-M^2}{2(|q^0| \mp q)} ~~\mathrm{for}~~|q^0| \geq \sqrt{q^2+M^2} \label{consttwo}\ee

\vspace{2mm}

\be \mathbf{3}):  -\delta(k^0-k)\delta(q^0-k+W)\Big[1-n_f(k)+N_b(q^0-k)\Big] \label{threea} \ee this term has support for $k-q^0 >0$, using (\ref{identists})
\be \mathbf{3}):   \delta(k^0-k)\delta(k-q^0-W)\Big[n_f(k)+N_b(k-q^0)\Big]\,. \label{three} \ee For $q_0 >0$ this contribution describes the \emph{decay} of the charged lepton and its inverse $l \leftrightarrow \nu_s W$, whereas for $q_0<0$ it describes the \emph{decay} of the W into a charged lepton and a sterile \emph{anti-neutrino} and its inverse, namely $W \leftrightarrow \overline{\nu}_s l$. The regions of support are:

\bea i:) && 0 \leq q^0 \leq q ~;~ k^- \leq k < \infty~~;~~ k^- = \frac{q^2+M^2-(q^0)^2}{2(q-q^0)} \label{threei}\\
ii:) && 0>q^0>-q ~~;~~ k^+=\infty ~~;~~ k^-= \frac{q^2+M^2-(q^0)^2}{2(q-q^0)} \label{threeii}\\
iii:) && q^0 <0 ~;~ q \leq |q^0|\leq \sqrt{q^2+M^2}~~;~~ k^- \leq k \leq k^+~~;~~ k^\pm = \frac{q^2+M^2-(q^0)^2}{2(|q^0|\mp q)} \label{threeiii}\eea

\vspace{2mm}

\be \mathbf{4}): -\delta(k^0+k)\delta(q^0+k-W)\Big[n_f(k)+N_b(q^0+k)\Big]\,. \label{four} \ee For $q_0 >0$ the delta function $\delta(q^0+k-W)$ describes the decay of the W into a sterile neutrino and a charged lepton. The regions of support are:
\bea && i:)~ 0 \leq q^0 \leq q: ~~;~~ k^- \leq k < \infty ~~;~~ k^- = \frac{q^2+M^2-(q^0)^2}{2(q+q^0)} \label{fouri}\\
&& ii:) ~q\leq q^0 \leq \sqrt{q^2+M^2}: ~~;~~ k^- \leq k \leq k^+ ~~;~~ k^\pm = \frac{q^2+M^2-(q^0)^2}{2(q \mp q^0)} \label{fourii} \\
&& iii:)~ -q \leq q^0 \leq 0 : ~~;~~ k^- \leq k < \infty ~~;~~ k^- =  \frac{q^2+M^2-(q^0)^2}{2(q+q^0)} \label{fouriii}\eea

The regions i) and iii) describe the process of Landau damping\cite{robLD}, with support below the light cone, corresponding to emission and reabsorption of thermal excitations similarly to the case of the self-energy of quarks interacting with gluons or electrons with photons. For the case of the massive vector boson the Landau damping cut is continued above the light cone, corresponding to the region ii:). This term is the only one that contributes to the imaginary part on the mass shell of the sterile-like neutrino.

\vspace{2mm}

 The integrals over $k^0$ can be done straightforwardly, the contributions from $L[Q,k]\cdot P[p]$ (see (\ref{LdotPplus},\ref{LdotPmin})  yield the following terms for negative and positive helicity respectively:

 \vspace{2mm}

\begin{center}
\textbf{$k^0$ integrals: Negative helicity}
\end{center}

\begin{itemize}
\item{~~~~ for $\mathbf{1}: -\frac{M^2}{q}\Big[ F_1(q_0,q)+ k\,F_2(q_0,q) \Big]$  }
\item{~~~~ for $\mathbf{2}: -\frac{M^2}{q}\Big[ F_1(q_0,q)- k\,F_2(q_0,q) \Big]$  }
\item{~~~~ for $\mathbf{3}: -\frac{M^2}{q}\Big[ F_1(q_0,q)+ k\,F_2(q_0,q) \Big]$  }
\item{~~~~ for $\mathbf{4}: -\frac{M^2}{q}\Big[ F_1(q_0,q)- k\,F_2(q_0,q) \Big]$  }
\end{itemize}

\vspace{2mm}

\begin{center}
\textbf{$k^0$ integrals: Positive helicity}
\end{center}

\begin{itemize}
\item{~~~~ for $\mathbf{1}: \frac{M^2}{q}\Big[ G_1(q_0,q)+ k\,G_2(q_0,q) \Big]$  }
\item{~~~~ for $\mathbf{2}: \frac{M^2}{q}\Big[ G_1(q_0,q)- k\,G_2(q_0,q) \Big]$  }
\item{~~~~ for $\mathbf{3}: \frac{M^2}{q}\Big[ G_1(q_0,q)+ k\,G_2(q_0,q) \Big]$  }
\item{~~~~ for $\mathbf{4}: \frac{M^2}{q}\Big[ G_1(q_0,q)-k\,G_2(q_0,q) \Big]$  }
\end{itemize}

 \begin{center}

\textbf{$k$ Integrals}

\end{center}

The next step is to calculate the $k-$ integrals, this is facilitated by the following identities:
\be n_f(k) = -T \frac{d}{dk} \ln[1+e^{-k/T}]  \, \label{nfder}\ee
\be N_b(q^0-k) = -T \frac{d}{dk} \ln[1-e^{k/T}e^{-q^0/T}] \label{Nbder}\ee and a similar identity for $N_b(k-q^0);N_b(q^0+k)$. With these identities we find
\be \int^{k^+}_{k^-} n_f(k)~ dk = -T\,\ln\Bigg[\frac{1+e^{-k^+/T}}{1+e^{-k^-/T}}\Bigg] \label{intnf}\ee
\be \int^{k^+}_{k^-}N_b(q^0-k)~ dk = -T\,\ln\Bigg[\frac{1-e^{k^+/T}e^{-q^0/T}}{1-e^{k^-/T}e^{-q^0/T}}\Bigg] \label{intNb1}\ee
\be \int^{k^+}_{k^-}N_b(k-q^0)~ dk = T\,\ln\Bigg[\frac{1-e^{-k^+/T}e^{q^0/T}}{1-e^{-k^-/T}e^{q^0/T}}\Bigg] \label{intNb2}\ee
\be \int^{k^+}_{k^-}k~ n_f(k)~ dk = -T^2\,\Bigg\{k^+\ln\Big[ {1+e^{-k^+/T}}\Big] -k^-\ln\Big[{1+e^{-k^-/T}}\Big] - \sum_{n=1}^{\infty} \frac{(-1)^n}{n^2}\Big[e^{-nk^+/T}-e^{-nk^-/T} \Big] \Bigg\} \label{intknf}\ee
\be \int^{k^+}_{k^-}k~N_b(q^0-k)~ dk = -T^2\,\Bigg\{k^+\ln\Big[ {1-e^{k^+/T}e^{-q^0/T}}\Big] -k^-\ln\Big[{1-e^{k^-/T}e^{-q^0/T}}\Big] + \sum_{n=1}^{\infty} \frac{e^{-nq^0/T}}{n^2}\Big[e^{nk^+/T}-e^{nk^-/T} \Big] \Bigg\} \label{intkNb1}\ee
\be \int^{k^+}_{k^-}k~N_b(k-q^0)~ dk = -T^2\,\Bigg\{-k^+\ln\Big[ {1-e^{-k^+/T}e^{q^0/T}}\Big] +k^-\ln\Big[{1-e^{-k^-/T}e^{q^0/T}}\Big] + \sum_{n=1}^{\infty} \frac{e^{nq^0/T}}{n^2}\Big[e^{-nk^+/T}-e^{-nk^-/T} \Big] \Bigg\} \label{intkNb2}\ee
In the integrals (\ref{intknf}-\ref{intkNb2}) we have used the identities (\ref{nfder},\ref{Nbder}), integrated by parts, expanded the logarithms in power series and integrated term by term. The infinite sums can be expressed in terms of di-logarithmic (Spence's) functions but such form is not particularly useful.

\vspace{2mm}

\textbf{Numerical Implementation for the real part:}

The numerical implementation for the real parts from the dispersive form (\ref{resigma}) is best achieved in a ``modular form'' which is facilitated by introducing
\bea \int^{k^+}_{k^-} \Big[N_b(q^0-k)-n_f(k) \Big] \, dk & \equiv & -T\, D1(q^0,q) \label{D1}\\
\int^{k^+}_{k^-} \Big[N_b(q^0-k)-n_f(k) \Big]\,k\,dk & \equiv & -T^2\,Dk1(q^0,q) \label{Dk1}\\
\int^{k^+}_{k^-} \Big[N_b(k-q^0)+n_f(k) \Big] \, dk & \equiv & -T\, D2(q^0,q) \label{D2}\\
\int^{k^+}_{k^-} \Big[N_b(k-q^0)+n_f(k) \Big] \,k\, dk & \equiv & -T^2\, Dk2(q^0,q)\,, \label{Dk2}\eea where the respective integrals are given above. In terms of these quantities and $F_{1,2}(q^0,q)$ defined by eqns. (\ref{F1},\ref{F2}) (and $G_{1,2}$ defined by (\ref{G1},\ref{G2}) for positive helicity) we find for negative helicity
the following contributions to the imaginary parts in the different regions $1-4$ of $q^0$ defined by the support of the corresponding delta functions described above:
\be \mathrm{Im}\Sigma^{(1)}= \frac{g^2 M^2  T}{16\pi\,q^2} \Bigg[F_1(q^0,q)\,D1(q^0,q) + T\, F_2(q^0,q)\,Dk1(q^0,q) \Bigg]~~;~~q^0 \geq \sqrt{q^2+M^2} \label{imsig1}\ee
\be \mathrm{Im}\Sigma^{(2)}= -\frac{g^2 M^2  T}{16\pi\,q^2} \Bigg[F_1(q^0,q)\,D1(|q^0|,q) - T\,F_2(q^0,q)\,Dk1(|q^0|,q) \Bigg]~~;~~q^0 <0 ~~,~~ |q^0|\geq  \sqrt{q^2+M^2} \label{imsig2}\ee
\bea \mathrm{Im}\Sigma^{(3)} & = &  \frac{g^2 M^2  T}{16\pi\,q^2} \Bigg\{\Big[ F_1(q^0,q)\,D2(q^0,q) + T\, F_2(q^0,q)\,Dk2(q^0,q)\Big] (0<q^0<q) \nonumber \\
& + & \Big[ F_1(q^0,q)\,D2(q^0,q) + T\, F_2(q^0,q)\,Dk2(q^0,q)\Big]( 0>q^0>-q) \nonumber \\
& + & \Big[ F_1(q^0,q)\,D2(q^0,q) + T\, F_2(q^0,q)\,Dk2(q^0,q)\Big]( -q>q^0>-\sqrt{k^2+M^2})\Bigg\} \label{imsig3} \eea in the first two terms $k^+ = \infty \rightarrow D2(k^+=\infty)=Dk2(k^+=\infty) =0$ and only the lower limit with $k^-$ corresponding to the case (3) above contributes, the two limits $k^\pm$ contribute to the last term.

\bea \mathrm{Im}\Sigma^{(4)} & = &  -\frac{g^2 M^2  T}{16\pi\,q^2} \Bigg\{\Big[ F_1(q^0,q)\,D2(-q^0,q) - T\, F_2(q^0,q)\,Dk2(-q^0,q)\Big] (0<q^0<q) \nonumber \\
& + & \Big[ F_1(q^0,q)\,D2(-q^0,q) - T\, F_2(q^0,q)\,Dk2(-q^0,q)\Big]( q<q^0<\sqrt{k^2+M^2}) \nonumber \\
& + & \Big[ F_1(q^0,q)\,D2(-q^0,q) - T\, F_2(q^0,q)\,Dk2(-q^0,q)\Big]( 0>q^0>-q)\Bigg\}\,. \label{imsig4} \eea Now the real part of the self energy is calculated with the dispersive form (\ref{resigma}) with
\be  \mathrm{Im}\Sigma(q^0,q)= \mathrm{Im}\Sigma^{(1)}(q^0,q)+\mathrm{Im}\Sigma^{(2)}(q^0,q)+
\mathrm{Im}\Sigma^{(3)}(q^0,q)+\mathrm{Im}\Sigma^{(4)}(q^0,q)\,. \label{imsigtot}\ee

In each region in $q^0$ the values of $k^\pm$ are given by the different cases analyzed above. The principal part is obtained by excising an interval of width $2 \epsilon$ around $q^0 = q$ with $\epsilon \ll 1$.

For positive helicity the same analysis above holds with the following modifications:
\begin{itemize}
\item \textbf{i:)} $M^2 \rightarrow -M^2$ \emph{only in the pre-factor}
\item \textbf{ii:)} $F_{1,2}(q^0,q) \rightarrow G_{1,2}(q^0,q) $ where $F_{1,2},G_{1,2}$ are given by eqns. (\ref{F1},\ref{F2}) and  (\ref{G1},\ref{G2}) respectively. Note that $F_{1,2}$ and $G_{1,2}$ obey the relation (\ref{relaFG}).
\end{itemize}

Finally, we introduce the dimensionless variables:
\be z= q^0/T ~~;~~ y=q/T ~~;~~ \tau=M_W/T \label{alldims}\ee where as discussed in the text we use $M_W$ as the baseline scale. The integrals over $q^0$ are then rendered dimensionless in terms of these variables. The dispersive integrals over the dimensionless variable $z$ are carried out numerically and the final results for the real part of the self energy are generically of the form
\be \mathrm{Re}\Sigma^\pm(q) = \frac{g^2 M^2  T}{16\pi^2\,q^2}\,\mathcal{K}^\pm[\tau,y]\,, \label{generalformresig}\ee where $\mathcal{K}^\pm[\tau,y]$ are dimensionless functions of $\tau,y$ that are obtained numerically with the procedure detailed above. For charged currents
$g^2 = g^2_w/2, M=M_W$, for neutral currents $g^2=g^2_w/(2c)^2, M= M_W/c, c=\cos(\theta_w) \simeq 0.88$ and for neutral currents $\tau \rightarrow \tau/c$ in the argument of $\mathcal{K}[\tau,y]$.

\vspace{2mm}

\textbf{Low Temperature limit.}

In the low temperature limit $M_W \gg T$ and keeping only the finite temperature contributions and using the identities (\ref{identists}) we can neglect $N_b$. In (\ref{imapmb})  the product of delta functions and distribution functions becomes
\be  n_f(k) \Bigg\{ \delta(k^0-k) \Big[\delta(q^0-k+W_p)-\delta(q^0-k-W_p) \Big]   +    \delta(k^0+k) \Big[ \delta(q^0+k+W_p)-\delta(q^0+k-W_p)\Big] \Bigg\} \,. \label{ltsf} \ee Now it is more convenient to integrate over $q^0$ and $k^0$ in the dispersive integral (\ref{resigma}), leaving only the integrals in $k$, with
\be d^3k = (2\pi)k^2 dk d(\cos(\varphi)) \ee with $\varphi$ the angle between $\vec{q}$ and $\vec{k}$.

 For negative helicity we find
 \be L_{\mu \nu}[Q^+,k]P^{\mu \nu}[p] = k(1-\cos(\varphi)) + \frac{2k}{M^2}(q^0-q \cos(\varphi))\Big(q^0-k-q +k\cos(\varphi)  \Big)~~;~~\mathrm{for}~~k^0=k \label{Lpnegplu} \ee

\be L_{\mu \nu}[Q^+,k]P^{\mu \nu}[p] = -k(1+\cos(\varphi)) - \frac{2k}{M^2}(q^0+q \cos(\varphi))\Big(q^0+k-q +k\cos(\varphi)  \Big)~~;~~\mathrm{for}~~k^0=-k \,.\label{Lpnegmen} \ee Integrating over $q^0$ implementing the delta functions and expanding the numerator and denominator in powers of $k/M, q/M$, integrating over $\cos(\varphi)$,  keeping only the leading order terms (proportional to $1/M^4$) and using
\be \int^\infty_0 k^3\,n_f(k)\,dk  = \frac{7\, \pi^4 \, T^4 }{120} \label{k3int}\ee we find for negative helicity
\be \mathrm{Re}\Sigma^-(q) = \frac{14 \pi^2}{90}\, g^2 \,\Bigg(\frac{T}{M}\Bigg)^4\,q \label{lowTreSneghelapp}\ee

For positive helicity we follow the same steps, with
\be L_{\mu \nu}[Q^-,k]P^{\mu \nu}[p] = k(1+\cos(\varphi)) + \frac{2k}{M^2}(q^0-q \cos(\varphi))\Big(q^0-k+q -k\cos(\varphi)  \Big)~~;~~\mathrm{for}~~k^0=k \label{Lpposplu} \ee

\be L_{\mu \nu}[Q^-,k]P^{\mu \nu}[p] = -k(1-\cos(\varphi)) - \frac{2k}{M^2}(q^0+q \cos(\varphi))\Big(q^0+k+q -k\cos(\varphi)  \Big)~~;~~\mathrm{for}~~k^0=-k \,.\label{Lpposmen} \ee Following the same steps as for negative helicity, we find
\be \mathrm{Re}\Sigma^+(q) = \frac{14 \pi^2}{180}\, g^2 \,\Bigg(\frac{T}{M}\Bigg)^4\,q \,.\label{lowTreSposhelapp}\ee

\newpage

U.S. Department of Energy Office of Nuclear Physics or High Energy Physics

{\it Notice:}
This manuscript has been co-authored by employees of Brookhaven
Science Associates, LLC under Contract No. DE-SC0012704 with
the U.S. Department of Energy. The publisher by accepting the manuscript for
publication acknowledges that the United States Government retains a
non-exclusive, paid-up, irrevocable, world-wide license to publish or
reproduce the published form of this manuscript, or allow others to do so,
for United States Government purposes.
This preprint is intended for publication in a journal or proceedings.
Since changes may be made before publication, it may not be cited or
reproduced without the author’s permission.
{\it DISCLAIMER}:
This report was prepared as an account of work sponsored by an agency of the
United States Government.  Neither the United States Government nor any
agency thereof, nor any of their employees, nor any of their contractors,
subcontractors, or their employees, makes any warranty, express or implied,
or assumes any legal liability or responsibility for the accuracy,
completeness, or any third party’s use or the results of such use of any
information, apparatus, product, or process disclosed, or represents that
its use would not infringe privately owned rights. Reference herein to any
specific commercial product, process, or service by trade name, trademark,
manufacturer, or otherwise, does not necessarily constitute or imply its
endorsement, recommendation, or favoring by the United States Government or
any agency thereof or its contractors or subcontractors.  The views and
opinions of authors expressed herein do not necessarily state or reflect
those of the United States Government or any agency thereof.

\end{document}